\documentclass[
	reprint,
	superscriptaddress,
	showkeys,
	aps,
	pra,
	longbibliography,
	nofootinbib
]{revtex4-2}

\usepackage[utf8]{inputenc}
\usepackage{xcolor}

	\definecolor{fsu-blue}{cmyk}{1.0,0.7,0.1,0.5}			% university
	\definecolor{fsu-gold}{cmyk}{0.2,0.25,0.6,0.25}			% award
	\definecolor{fsu-violet}{cmyk}{0.8,0.85,0.0,0.0}		% theology
	\definecolor{fsu-purple}{cmyk}{0.5,1.0,0.0,0.1}			% philosophy
	\definecolor{fsu-magenta}{cmyk}{0.15,1.00,0.25,0.15}	% law
	\definecolor{fsu-red}{cmyk}{0.0,0.95,0.8,0.15}			% medicine
	\definecolor{fsu-orange}{cmyk}{0.1,0.7,1.0,0.0}			% physics
	\definecolor{fsu-yellow}{cmyk}{0.05,0.4,1.0,0.1}		% economics
	\definecolor{fsu-light-green}{cmyk}{0.7,0.1,1.0,0.05}	% chemistry
	\definecolor{fsu-green}{cmyk}{0.9,0.3,1.0,0.1}			% life-science
	\definecolor{fsu-turquoise}{cmyk}{1.0,0.25,0.4,0.1}		% mathematics
	\definecolor{fsu-light-blue}{cmyk}{0.7,0.2,0.0,0.2}		% social

\usepackage[caption=false]{subfig}

\usepackage{amsmath} % AMS mathematical facilities for LaTeX
\usepackage{amssymb} % Additional math symbols
\usepackage{bm} % Access bold symbols in maths mode -- \mathbb{R}, \mathbb{Z} etc.
\usepackage{upgreek} % Upright Greek letters
\usepackage{slashed} % Feynman slash notation

\usepackage{comment} %Selectively include/exclude portions of text -- defines comment environment
\usepackage{graphicx} % Enhanced support for graphics -- include figure files
\usepackage{placeins} % Control float placement -- defines \FloatBarrier

\usepackage{booktabs} % Publication quality tables in LaTeX

\allowdisplaybreaks% Allow breaking of formulas across multiple pages

\newcommand{\Z}{\mathcal{Z}}
\newcommand{\dd}{\mathrm{d}}
\newcommand{\ee}{\mathrm{e}}
\newcommand{\ii}{\mathrm{i}}

\newcommand{\vdistance}{\vphantom{\bigg(\bigg)}}
\newcommand{\Vdistance}{\vphantom{\Bigg(\Bigg)}}

\newcommand{\B}{\mathcal{B}}
\newcommand{\F}{\mathcal{F}}
\newcommand{\N}{\mathcal{N}}
\newcommand{\W}{\mathcal{W}}

\newcommand{\boson}{\mathrm{b}}
\newcommand{\fermion}{\mathrm{f}}

\newcommand{\barvartheta}{\tilde{\vartheta}}

\renewcommand{\S}{\mathcal{S}}

\DeclareMathOperator{\STr}{\mathrm{STr}}

\newcommand{\GbInv}{\B^{-1}}%{r_\boson + U^{\prime \prime}}
\newcommand{\Gb}{\B}%{\frac{1}{r_\boson + U^{\prime \prime}}}

\newcommand{\GfInv}{\F^{-1}}%{r_\fermion + H}
\newcommand{\Gf}{\F}%{\frac{1}{r_\fermion + H}}

\usepackage[
	colorlinks,
	pdfpagelabels,
	breaklinks,
	pdfstartview=FitH,
	bookmarksopen=true,
	bookmarksnumbered=true,
	bookmarksopenlevel=2,
	plainpages=false,
	hypertexnames=false,
	citecolor=fsu-magenta,
	linkcolor=fsu-blue,
	urlcolor=fsu-magenta, % //TOTO: adopt colors
	pdftitle={{Functional Renormalization Group flows as diffusive Hamilton-Jacobi-type equations}},
	pdfauthor={{Adrian Koenigstein, Martin J. Steil, and Stefan Floerchinger}},
	pdfkeywords={Functional Renormalization Group, Hamilton-Jacobi equation, conservation laws, fluid-dynamics, numeric, partial differential equation, Kurganov-Tadmor}
]{hyperref}
\usepackage{bookmark} % A new bookmark (outline) organization for hyperref

\usepackage{orcidlink}

\usepackage[capitalise,sort,compress]{cleveref} % Intelligent cross-referencing used primarily unless text and/or linebreaking requires manual \ref, \cite, or \eqref
\usepackage[acronym]{glossaries-extra}
\glssetcategoryattribute{acronym}{nohyperfirst}{true}
\setabbreviationstyle[acronym]{long-short}

\newacronym{wrt}{w.r.t.}{with respect to}
\newacronym{rhs}{r.h.s.}{right hand side}
\newacronym{lhs}{l.h.s.}{left hand side}

\newacronym[plural=PDEs,longplural=partial differential equations]{pde}{PDE}{partial differential equation}
\newacronym[plural=ODEs,longplural=ordinary differential equations]{ode}{ODE}{ordinary differential equation}

\newacronym[plural=QFTs,longplural=quantum field theories]{qft}{QFT}{quantum field theory}
\newacronym{qcd}{QCD}{Quantum Chromodynamics}
\newacronym[plural=LEFTs,longplural=low-energy effective theories]{left}{LEFT}{low-energy effective theory}
\newacronym{frg}{FRG}{functional renormalization group}
\newacronym{grg}{(F)RG}{(functional) renormalization group}
\newacronym{ea}{EAA}{effective action}
\newacronym{eaa}{EAA}{effective average action}
\newacronym{rg}{RG}{renormalization group}
\newacronym{uv}{UV}{ultraviolet}
\newacronym{ir}{IR}{infrared}
\newacronym{lpa}{LPA}{local potential approximation}
\newacronym{lpap}{LPA${}^\prime$}{local potential approximation prime}
\newacronym{sclpa}{scLPA}{self-consistent local potential approximation}
\newacronym{ipi}{1PI}{one-particle irreducible}
\newacronym{tpi}{2PI}{two-particle irreducible}
\newacronym{npi}{\texorpdfstring{$n$PI}{nPI}}{$n$-particle irreducible}
\newacronym{kt}{KT}{Kurganov-Tadmor}
\newacronym{kthj}{KT-HJ}{Kurganov-Tadmor-Hamilton-Jacobi}
\newacronym{hj}{HJ}{Hamilton-Jacobi}
\newacronym{hjb}{HJB}{Hamilton-Jacobi-Bellman}
\newacronym{muscl}{MUSCL}{Monotonic Upstream-Centered Schemes for Conservation Laws}
\newacronym{wlog}{w.l.o.g.}{without loss of generality}
\newacronym{gny}{GNY}{Gross-Neveu-Yukawa}

\newacronym{gn}{GN}{Gross-Neveu}
\newacronym{qm}{QM}{quark-meson}

\newacronym{cfd}{CFD}{computational fluid dynamics}
\newacronym[plural=ICs,longplural=initial conditions]{ic}{IC}{initial condition}
\newacronym[plural=BCs,longplural=boundary conditions]{bc}{BC}{boundary condition}

\newcommand{\ie}{\textit{i.e.}} % id est: "That is (to say)" -- usage ..., \ie{}, ...
\newcommand{\eg}{\textit{e.g.}} % exempli gratia: "for example", "for instance" -- usage ..., \eg{}, ...
\newcommand{\cf}{\textit{cf.}} % confer: "bring together" and hence "compare" -- usage ..., \cf{} ...
\newcommand{\etc}{\textit{etc.}} % et cetera: "and the others"
\newcommand{\apriori}{\textit{a priori}} % "independent of experience"
\newcommand{\aposteriori}{\textit{a posteriori}} % "based on experience"
\begin{document}

%\preprint{}

\title{
	Functional Renormalization Group flows as diffusive Hamilton-Jacobi-type equations
}

\author{Adrian Koenigstein \orcidlink{0000-0001-7482-2195}}
	%\email{info@adrian-koenigstein.com}
	\affiliation{
		Theoretisch-Physikalisches Institut,
		Friedrich-Schiller Universität,
		D-07743 Jena,
		Germany.
	}

\author{Martin J. Steil \orcidlink{0000-0001-8465-9803}}
	%\email{martinjsteil@gmail.com}
	\affiliation{
		Institut f\"ur Kernphysik, Theoriezentrum,
		Technische Universit\"at Darmstadt,
		D-64289 Darmstadt,
		Germany
	}

\author{Stefan Floerchinger \orcidlink{0000-0002-3428-4625}}
	\affiliation{
		Theoretisch-Physikalisches Institut,
		Friedrich-Schiller Universität,
		D-07743 Jena,
		Germany.
	}

\date{\today}

\begin{abstract}
	In order to find reliable and efficient numerical approximation schemes, we suggest to identify the Functional Renormalization Group flow equations of one-particle irreducible two-point functions as Hamilton-Jacobi(-Bellman)-type partial differential equations.
	Based on this reformulation and reinterpretation we adopt a numerical scheme for the solution of field-dependent flow equations as nonlinear partial differential equations.
	
	We demonstrate this novel approach by first applying it to a simple fermion-boson system in zero spacetime dimensions -- which itself presents as an interesting playground for method development.
	Afterwards, we show, how the gained insights can be transferred to more interesting problems:
	One is the bosonic $\mathbb{Z}_2$-symmetric model in three Euclidean dimensions within a truncation that involves the field-dependent effective potential and field-dependent wave-function renormalization.
	The other example is the $(1 + 1)$-dimensional Gross-Neveu model within a truncation that involves a field-dependent potential and a field-dependent fermion mass/Yukawa coupling at nonzero temperature, chemical potential, and finite fermion number.
\end{abstract}

 %COMMENT: Use showkeys class option if keyword display desired
\maketitle

\tableofcontents%

\section{Introduction}%
\label{sec:introduction}

	Within the last decades, the \gls{frg} has become a powerful tool for the study of models from statistical physics and \gls{qft}.
	It was applied successfully to a wide range of problems, such as the study of phase transitions and critical phenomena in the context of particle physics, gravity, and condensed matter physics, see, \eg{}, Ref.~\cite{Dupuis:2020fhh} for a review.
	A strength of the \gls{frg} is the possibility to access and evolve theories in the full range of energy/length scales: from high-energy/microphysical scales in the \gls{uv} to low energy/macrophysical scales in the \gls{ir}, while resolving field space in a single nonperturbative framework without the need of explicitly solving complicated high-dimensional functional integrals.

\subsection{Research issue}
	However, also the \gls{frg} has its own shortcomings and practical limitations.
	One central problem is the issue of truncations, which are usually necessary to make the explicit solution of \gls{frg} flow equations, derived from the central governing Wetterich equation,  tangible.
	Still, also within a given truncation, the \gls{frg} flow equations are often highly nonlinear coupled \glspl{pde} and/or \glspl{ode} and their solution is \apriori{} an involved task.
	Prime examples for such systems and common truncations are \glspl{left} for \gls{qcd} or systems from solid state theory, both in their high-density regimes, where the \gls{frg} is used to study -- among other things -- the scale-dependence of the effective potential, field-dependent wave-function renormalizations, and the flow of field-dependent Yukawa couplings or fermion masses (depending on the perspective).
	The question arises, how reliable the results of such truncations and the numerical solutions of the corresponding flow equations are, because so far, there seems little consensus on the mathematical structure of the problem and its numerical treatment.
	One way to address this question is to compare the results from the \gls{frg} to results obtained with other methods or exact results from solvable models.\footnote{We use the term solvable model in the sense of a model, where the expectation values, correlation functions, and vertex functions can be calculated exactly or numerically to arbitrary precision via some other method without approximations.}
	Solvable models are rare or they are trivial, especially in the context of \glspl{qft}.
	Another approach is to perform selfconvergence tests, where the \glspl{pde} are solved with different discretization schemes and resolutions.
	Hence, within this work, we address different closely related topics:
	\begin{enumerate}
		\item We search for toy models that are simple enough to be exactly solvable and rich enough to serve as benchmark tests that mimic many complications of \gls{frg}(-model) studies whose truncations involve a field-dependent potential, field-dependent wave-function renormalizations, and/or field-dependent Yukawa couplings.
		
		\item We aim at a reformulation of the \gls{frg} flow equations in a form that allows a better application of numerical methods and algorithms and at the same time provides deeper insights into the mathematical structure of the Wetterich equation and its truncated field-dependent flow equations.
		
		\item We want to explicitly find, modify, apply, and test numeric schemes to solve the corresponding \glspl{pde}.
	\end{enumerate}

\subsection{Scope, structure and conception of this work}%
\label{subsec:scope}

During its conception and early stages\footnote{%
	The present work started as an extension of the series~\cite{Koenigstein:2021syz,Koenigstein:2021rxj,Steil:2021cbu} and is mentioned in it as \textit{``Part IV: A fermion-boson model.''}
	We initially set out to study a zero-dimensional model with four fermions and three scalars with a $SU(2)$ symmetry.
	We soon realized that such a model is already diagrammatically and in terms of the resulting differential equations very involved to the point of being ill-suited for a first study of Grassmann-valued degrees of freedom in zero dimensions, see Section~3.3 of Ref.~\cite{Steil:2023sfd} for a discussion.
	Hence, we decided to consider the simpler -- in a sense for our purposes minimal -- model discussed in this work.
	During the preparation of this manuscript, we decided against structuring and framing this work as a direct extension of the series~\cite{Koenigstein:2021syz,Koenigstein:2021rxj,Steil:2021cbu} in form of a part IV in favor of a more independent publication with a broader focus.
}, this work was primarily focused on zero-dimensional fermion-boson models and their study within the \gls{frg}. 
The idea was to construct and analyze such models as exactly solvable testbeds for exploring the fermion-boson, \ie{} Grassmann-scalar, systems in zero dimensions with a focus on the development of numerical methods for coupled flow equations.
As the project developed, however, it became apparent that many of the challenges and insights encountered in the zero-dimensional setting pointed towards more general structural questions about the \gls{frg} itself. 
This realization led us to broaden the scope and structuring of the present work: before discussing our zero-dimensional fermion-boson model, we first discuss the Wetterich equation and its functional derivatives at a conceptual level.
Building on this foundation, we then return to the construction and treatment of the zero-dimensional model, adapt suitable numerical schemes based on our findings, and finally apply these schemes to our zero-dimensional model as well as higher-dimensional systems from statistical physics and \gls{qft}.

Concretely, we begin in \cref{sec:wetterich}, by discussing the Wetterich equation and its functional derivatives, with a particular focus on the flow equation of the two-point function in \cref{subsec:wetterich_ii}. 
We demonstrate that the flow equation of the two-point functions takes the form of a functional (infinite-dimensional), viscous \gls{hjb} equation.
This structure/form is also present in the explicit \glspl{pde} derived from this equation -- which in fact prompted us to make the identification on the functional level.
When working with the flow equation for the two-point function as a \gls{hjb} equation, the advective and diffusive nature of \gls{frg} flows become directly apparent.
Furthermore this form allows a more rigorous treatment of nonconservative terms, which arise in \gls{cfd} formulations based on the flow equation of the one-point function, \cf{} \cref{subsec:wetterich_i}.
Apart from practical implications for numerical schemes for the derived \glspl{pde} the identification of a \gls{hjb} structure opens questions of a more general nature, such as the existence and uniqueness of solutions, their stability, convergence, as well as the relation of the \gls{frg} to other fields like optimal transport theory, \gls{cfd}, and infinite-dimensional stochastic optimal control problems.
We elaborate on some of this questions and relations in \cref{subsec:frgHJB} and \cref{subsec:frgHJB0d}.

Following the general discussion of \cref{sec:wetterich}, we turn to the construction and introduction of the zero-dimensional fermion-boson model in \cref{sec:the_model}.
The presented model is both minimal and solvable, yet structurally rich enough to be a suitable testing ground for the following numerical developments.
Its \gls{frg} formulation, presented in \cref{sec:the_frg_approach}, ties in directly into the conceptual observations of \cref{sec:wetterich} and extends on it using the explicit zero-dimensional setting.
It aligns with our previous works~\cite{Koenigstein:2021syz,Koenigstein:2021rxj,Steil:2021cbu,Zorbach:2024rre} on purely bosonic, zero-dimensional models with $N \in \mathbb{N}$ scalar degrees of freedom, which are also called vector models with or without $O(N)$ symmetry, see, \eg{}, Refs.~\cite{Bessis:1980ss,Zinn-Justin:1998hwu,DiVecchia:1990ce,Hikami:1978ya,Nishigaki:1990sk,Schelstraete:1994sc,Catalano:2019,Fl_rchinger_2010,Keitel:2011pn,SkinnerScript,Moroz:2011thesis,Pawlowski:talk,Strocchi:2013awa,Kemler:2013yka,Rosa:2016czs,Millington:2019nkw,Millington:2020Talk,Millington:2021ftp}.

In \cref{sec:numerical_scheme} we adapt and combine a numerical scheme~\cite{KT2000:HamiltonJacobi} developed for the solution of viscous \gls{hjb} equations with the finite volume methods~\cite{KTO2-0} used in our previous works~\cite{Koenigstein:2021syz,Koenigstein:2021rxj,Steil:2021cbu,Zorbach:2024rre}.

In \cref{sec:zero_dim_exp}, we apply this scheme to a set of test(case)s constructed within the zero-dimensional fermion-boson model. 
This model turned out to be a rather rich setting to explore the interplay between fermionic and bosonic fluctuations in a coupled system of field-dependent, flowing couplings.
Besides discussing the convergence, stability, and accuracy of our numerical scheme we also discuss how a ``sign-problem'' -- the problem of probability distributions with ``negative probability'' -- can manifest itself in \gls{frg} flow equations of the model under consideration.
The related nonanalytic structures in field-space -- \ie{} Yang-Lee zeros -- are also discussed in the context of the construction of a reference solution for the \gls{frg} flow.

Finally, in \cref{sec:higher_dim_exp}, we extend the discussion to two higher-dimensional models in order to illustrate the broader applicability of our approach.
Of particular interest for us are higher-dimensional models and \glspl{left} of \gls{qcd} in the high-density regime, where a good and numerically stable resolution of the dynamics in field-space, for example the field-dependent flow of Yukawa couplings, wave-function renormalization, and the effective potential, are crucial~\cite{Grossi:2021ksl,Ihssen:2023xlp}.
To this end, we investigate the flow of field-dependent wave-function renormalizations in a $\mathbb{Z}_2$-symmetric scalar theory in \cref{sec:field-dependent-wave-function-renormalization} in three dimensions and the Gross-Neveu-Yukawa model in $1+1$ dimensions at nonzero temperature and density in \cref{subsec:gny_model}. 
These examples show the applicability of the formulation based on the two-point function/\gls{hjb} equations and the adapted numerical schemes to physically relevant systems.

In summary, the scope of this work is to connect conceptual advances with concrete applications: beginning with the identification of \gls{hjb} structures in derivatives of the  Wetterich equation, moving to the construction and analysis of a zero-dimensional fermion-boson model as a testbed, and finally adapting and testing numerical schemes both in the toy-model and in selected higher-dimensional applications.

\section{The Wetterich equation and its functional derivatives}%
\label{sec:wetterich}

	In this section, we discuss the Wetterich equation and its functional derivatives, which play a central role in the \gls{frg} approach.
	The following discussion is focused on the functional dependencies in the Wetterich equation and their consequences and implications. 
	This section does not include a derivation of the Wetterich equation, which can be found in the literature~\cite{Wetterich:1991be,Wetterich:1992yh,Reuter:1993kw,Morris:1993qb,Tetradis:1993ts,Ellwanger:1993mw}.
	For a more detailed discussion and introduction of the \gls{frg}, we refer the interested reader to the nonexhaustive list of Refs.~\cite{Berges:2000ew,Wetterich:2001kra,Pawlowski:2005xe,Gies:2006wv,Kopietz:2010zz,Rosten:2010vm,Delamotte:2007pf,Dupuis:2020fhh}.

	The Wetterich equation for the \gls{rg}-time-dependent \gls{eaa} $\bar{\Gamma} ( t, \Phi )$ is given by
		\begin{align}
			&	\partial_t \bar{\Gamma} ( t, \Phi ) = \, &	\vdistance	\label{eq:wetterich_equation}
			\\
			= \, & \STr \big[ \big( \tfrac{1}{2} \, \partial_t R ( t ) \big) \, \big( \bar{\Gamma}^{(2)} ( t, \Phi ) + R ( t ) \big)^{- 1} \big] \, ,	\vdistance	\nonumber
			\\
			\equiv \, & \STr \big[\big( \tfrac{1}{2} \, \partial_t R ( t )\big) \, G( t, \Phi ) \big] \, .	\vdistance	\label{eq:wetterich_equation_G}
		\end{align}
	The generic multi-field $\Phi$, collects the field content of the theory under consideration, while \gls{rg}-time ${t \equiv -\ln(k/\Lambda) \in [ 0, \infty )}$ parametrizes the \gls{rg}-scale $k$ with respect to a \gls{uv} reference scale $\Lambda$.
	$R ( t )$ is a matrix-valued regulator and $\STr$ denotes the supertrace in the conventions of Appendix~C of Ref.~\cite{Koenigstein:2023wso}.
	In \cref{eq:wetterich_equation_G}, we have introduced the notation $G( t, \Phi ) \equiv \big( \bar{\Gamma}^{(2)} ( t, \Phi ) + R ( t ) \big)^{- 1}$ for the inverse of the second functional derivative of the effective action with respect to the fields $\Phi$ -- the full scale-dependent propagator of the theory. 
	At this point we may note, that the \gls{rhs}\ of the Wetterich equation \labelcref{eq:wetterich_equation} is a functional of the second functional derivative of the \gls{eaa} with respect to the fields, $\bar{\Gamma}^{(2)} ( t, \Phi )$, and the regulator function $R ( t )$, motivating the form
		\begin{align}
			&	\partial_t \bar{\Gamma} ( t, \Phi ) = \, \mathcal{F}_R( t, \Phi )\big[\bar{\Gamma}^{(2)}\big] \, .	\label{eq:wetterich_equation_F}
		\end{align}
	The \gls{rhs}\ does notably not depend on the \gls{eaa} $\bar{\Gamma} ( t, \Phi )$ itself, but only on its functional derivatives in a highly nonlinear fashion.
	This is a well-known feature shared among implementation of the \gls{rg} and is a manifestation of the fact, that the vacuum energy (zero-point function) does not affect correlation functions and its absolute value is not relevant\footnote{%
		Note, however, that in thermodynamic computations $\lim_{t\rightarrow 0}\bar{\Gamma} ( t, \Phi )$ acts as a grand potential and thus differences in the zero-point functions are linked to the pressure.%
	} for the dynamics of the theory and its renormalization, \cf{} Refs.~\cite{Zinn-Justin:2010,Peskin:1995ev}.
	At this point we can also see the typical $n, n + 2$ problem posed within flow equations in the \gls{frg}: a flow equation for the $n$-th moment of the \gls{eaa} (here explicitly the zeroth moment) depend on the \gls{rhs} on moments up to order $n+2$ (here explicitly on the two-point function).
	This structure is rooted in the inherently nonperturbative nature of the \gls{frg} which on the level of \cref{eq:wetterich_equation_G} manifests itself by the appearance of the full scale-dependent propagator $G(t,\Phi)$ on the \gls{rhs}.

\subsection{Functional flow equation for the one-point function}%
\label{subsec:wetterich_i}

	Motivated by the fact, that the Wetterich equation \labelcref{eq:wetterich_equation_F} for the \gls{eaa} depends only on derivatives of the latter, we consider the first functional derivative of \cref{eq:wetterich_equation_F} with respect to the field $\Phi$\footnote{For compactness and illustrative purposes, we opted for a very compact notation for the functional derivatives and involved signs and permutations, where we hide complexity in the implicit field-space matrix notation and the supertrace $\STr$. For a complete derivation we refer the interested reader to App.~C.2.4. of Ref.~\cite{Koenigstein:2023wso}.}:
		\begin{align}
			& \partial_t \bar{\Gamma}^{(1)} ( t, \Phi ) =	\Vdistance	\nonumber
			\\
			= \, & \frac{\delta}{\delta \Phi} \, \mathcal{F}_R ( t, \Phi ) \big[ \bar{\Gamma}^{(2)} \big] =	\Vdistance	\label{eq:wetterich_equation_Fi}
			\\
			= \, & \STr \big[ \big( -\tfrac{1}{2} \, \partial_t R ( t )\big) \, G ( t, \Phi ) \, \bar{\Gamma}^{(3)} ( t, \Phi ) \, G( t, \Phi )\big] \, .	\Vdistance	\label{eq:wetterich_equation_Gi}
		\end{align}
	\Cref{eq:wetterich_equation_Fi} takes the form of a functional flow/convection\footnote{In the following we use the term convection in the fluid-dynamical sense as a phenomenon including both advection and diffusion.} equation for the one-point function $\bar{\Gamma}^{(1)} ( t, \Phi )$ and diagrammatically reads
	\begin{align}
		& \partial_t \Big(
		\begin{gathered}
			\includegraphics{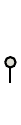}
		\end{gathered} \Big) +
		\frac{\delta}{\delta \Phi} \, \Bigg(
		\begin{gathered}
			\includegraphics{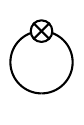}
		\end{gathered} \Bigg) = 0 \, .	\label{eq:wetterich_equation_Fi_diagrammatic}
	\end{align}
	This form of the Wetterich equation is central to the works~\cite{Grossi:2019urj,Wink:2020tnu,Grossi:2021ksl,Koenigstein:2021syz,Koenigstein:2021rxj,Steil:2021cbu,Stoll:2021ori,Koenigstein:2023wso,Batini:2023nan,Murgana:2023xrq,Steil:2023sfd,Ihssen:2023dmx,Zorbach:2024rre,Zorbach:2024zjx,Ihssen:2024miv,Zorbach:2025drj,Sattler:2025hcg,Ihssen:2022xjv,Ihssen:2022xkr,Ihssen:2023qaq,Ihssen:2023xlp,Jeong:2024rst,Sattler:2024ozv,Batini:2025ftv} considering (truncations of) the Wetterich equation as conservative equations.
	Note, that here and in the following, we absorb a minus sign in the symbolic representation of the regulator insertion, $\otimes = - \frac{1}{2} \, \partial_t R ( t )$ in order to better visualize the relation to \gls{cfd} and \gls{hjb} equations.

	The highly nonlinear dependence of the propagator $G( t, \Phi )$ on the Hessian $\bar{\Gamma}^{(2)} ( t, \Phi )$ of the \gls{eaa} can manifest in advective, diffusive, and most challengingly nonconservative contributions to the flow of the one-point function.
	Nonconservative contributions are very difficult to handle in explicit numerical schemes, without a deeper understanding of the underlying mathematical structure.
	To gain further insight into the nature of such contributions and the Wetterich equation itself, we propose to consider the second functional derivative of the Wetterich equation \labelcref{eq:wetterich_equation} with respect to the field $\Phi$.

\subsection{Functional flow equation for the two-point function}
\label{subsec:wetterich_ii}

	Taking another functional derivative of \cref{eq:wetterich_equation_Fi} using the explicit form \labelcref{eq:wetterich_equation_Gi} yields the functional flow equation for the two-point function:
	\begin{widetext}
		\begin{align}
			& \partial_t \bar{\Gamma}^{(2)} ( t, \Phi ) + \STr \big[ \big( -\tfrac{1}{2} \, \partial_t R ( t ) \big) \, G( t, \Phi ) \, \bar{\Gamma}^{(3)} ( t, \Phi ) \, G( t, \Phi ) \, \bar{\Gamma}^{(3)} ( t \,, \Phi ) \, G( t, \Phi )\big] =	\vdistance	\label{eq:wetterich_equation_Gii}
			\\
			= \, & \STr \big[ \big( -\tfrac{1}{2} \, \partial_t R ( t ) \big) \, G( t, \Phi ) \, \bar{\Gamma}^{(4)} ( t, \Phi ) \, G( t, \Phi ) \big] \, ,	\vdistance	\nonumber
		\end{align}
	\end{widetext}
	which takes the form of a functional (infinite-dimensional), viscous \glsfirst{hjb} equation, see Refs.~\cite{Bellman1954Apr,Yong1999,Fleming2006,Fabbri2017,Tran2020,KT2000:HamiltonJacobi,Lizana:2015hqb,Ivanov:2020zkm,Becchi:2002kj} for a non-exhaustive list of references on \gls{hjb} equations, their widespread use, and first applications to \gls{rg} flow equations. 
	Diagrammatically \cref{eq:wetterich_equation_Gii} can be symbolized as
	%\begin{widetext}
		\begin{align}
		& \partial_t \Big(
		\begin{gathered}
			\includegraphics{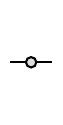}
		\end{gathered} \Big) +
		\begin{gathered}
			\includegraphics{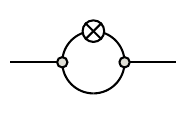}
		\end{gathered} =	\Vdistance
		\\
		= \, &
		\begin{gathered}
			\includegraphics{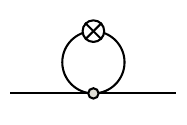}
		\end{gathered}\, .	\Vdistance	\nonumber
		\end{align}
	%\end{widetext}
	The scale-dependent two-point function $\bar{\Gamma}^{(2)} ( t, \Phi )$ acts as primitive variable and thus \cref{eq:wetterich_equation_Gii} may be formulated as
		\begin{align}
			& \partial_t \bar{\Gamma}^{(2)} ( t, \Phi ) + \mathcal{H}_R( t, \Phi ) \big[ \bar{\Gamma}^{(2)}, \bar{\Gamma}^{(3)}\big] =	\Vdistance	\label{eq:wetterich_equation_Fii}
			\\
			= \, & \varepsilon_R ( t, \Phi ) \big[ \bar{\Gamma}^{(2)}, \bar{\Gamma}^{(4)} \big] \, ,	\Vdistance	\nonumber
		\end{align}
	The tadpole diagram manifests as a nonlinear, viscous diffusion term $\varepsilon_R ( t, \Phi ) \big[ \bar{\Gamma}^{(2)}, \bar{\Gamma}^{(4)} \big]$, where
		\begin{align}
			G( t, \Phi ) \, \big( - \tfrac{1}{2} \, \partial_t R ( t ) \big) \, G( t, \Phi )
		\end{align}
	acts as a nonlinear diffusion coefficient, while the bubble diagrams manifest as a nonlinear advective/drift contribution, encoded by the Hamiltonian $\mathcal{H}_R ( t, \Phi ) \big[ \bar{\Gamma}^{(2)}, \bar{\Gamma}^{(3)} \big]$.
	The dependence on the primitive variable/the two-point function remains highly nonlinear through the products of propagators in the super-traces, while the higher moments of $\bar{\Gamma}^{(2)}$ -- namely the three- and four-point functions $\bar{\Gamma}^{(3)}$ and $\bar{\Gamma}^{(4)}$ appear only squared and linearly, respectively.
	The $n,n+2$ problem (in this subsection explicitly with $n=2$) is therefore directly apparent.
	In the present context it may be seen not as a technical obstacle, but as a structural feature that invites reformulation and reinterpretation.
	The \gls{hjb} equation is nonconservative but allows for a more rigorous classification of the flow equation, even on the functional level.
	Both \gls{cfd} and \gls{hjb} formulations provide ways to work the $n,n+2$ problem by providing frameworks on how to deal with the higher derivatives, without the explicit need for a truncation in the canonical vertex-expansion sense or ad-hoc approaches to compute the higher derivatives numerically, which both can be very limited due to properties of $\bar{\Gamma}$ like convexity (in the \gls{ir}), phase transitions, and/or the emergence of nonanalytic points.
	From the \gls{hjb} perspective, entropy solutions -- central in the conservative/\gls{cfd} formulations -- may find a more natural and robust generalization within the viscosity solution framework of \gls{hjb} theory~\cite{Fleming2006,Kappen2008Jul}. This could open new directions for both the analysis and numerical treatment of \gls{frg} equations, suggesting (numerical) approximation schemes and concepts that are better aligned with the fundamental nature of the \gls{frg}.

	The classification of \cref{eq:wetterich_equation_Gii} in \cref{eq:wetterich_equation_Fii} as a functional \gls{hjb} equation is of central importance for this work and might have even further reaching consequences beyond the scope of this work.

	For the remainder of this subsection, we outline the impact of this identification and formulation on the present work, while in the next \cref{subsec:frgHJB} we elaborate and, to some extent, speculate on further conceptual implications of identifying a functional \gls{hjb} equation as governing equation of the \gls{frg}.
	In \cref{sec:the_frg_approach} we will discover a direct manifestation of \cref{eq:wetterich_equation_Fii} as a system of \glspl{pde} describing the zero-dimensional model of \cref{sec:the_model}, while in \cref{sec:higher_dim_exp} \glspl{pde} of \gls{hjb}-type arise in truncations of higher-dimensional models and projections involving the two-point function. 
	Considering the \glspl{pde} as \gls{hjb} equations allows us to combine at this point established finite volume methods used in the \gls{cfd}-based formulations of the Wetterich equation with methods developed for \gls{hjb} equations.
	We use a \gls{hjb}-approach for nonconservative contributions while maintaining a finite volume formulation for conservative contributions.
	The reason for this split is based \aposteriori{} on our explicit numerical tests in \cref{sec:zero_dim_exp}, and especially in \cref{sec:higher_dim_exp}, rather than on conceptual arguments.

\subsection{The FRG as an infinite dimensional stochastic optimal control problem}%
\label{subsec:frgHJB}

	An intriguing perspective for future research emerges from the observation that the \gls{frg} flow equation~\eqref{eq:wetterich_equation_Gii} for the two-point function manifests as a functional \gls{hjb} equation, which naturally decomposes into drift and diffusion terms.
	Considering $G( t, \Phi ) ( -\tfrac{1}{2} \, \partial_t R ( t ) ) \, G( t, \Phi )$ as a diffusion coefficient of a McKean-Vlasov-type stochastic process, it might be possible to interpret the \gls{frg} as an infinite dimensional stochastic optimal control problem~\cite{Fabbri2017,Fleming2006,Kappen2008Jul}. 
	In this context, the nonlinear but instantaneous dependence of the diffusion coefficient on the evolving two-point function reflects the endogeneity of the system, where noise and control are governed by the current effective description, highlighting the semigroup and nonperturbative nature of the \gls{frg} flow: all relevant information is encoded in the present, not in explicit memory of the past.
	Wilson's RG approach of integrating out fluctuations momentum shell by momentum shell might be interpreted as an implementation of Chapman-Kolmogorov/Bellman's principle in the stochastic optimal control/dynamic programming framework.

	In the context of the \gls{rg} ideas in this/similar directions are not new, \cf{} Refs.~\cite{Carosso:2019qpb,Ivanov:2020zkm,Cotler:2022fze}, but to our knowledge an attempt of a formulation/strong link for/between the \gls{frg} formulated in terms of the \gls{eaa} governed by the Wetterich equation~\eqref{eq:wetterich_equation_G} and stochastic optimal control has not been made yet.
	Making such a link and studying connections to optimal transport, dynamic programming, and related concepts could yield a deeper understanding of the mathematical structure of \gls{grg}\footnote{
		We use the abbreviation (F)RG in statements that apply both to the functional renormalization group and to the nonperturbative renormalization group in general.
	} flows and inspire new conceptual insights as well as more powerful numerical and analytical methods.
	Further work and remarks to this end are beyond the scope of the present study, which ultimately focuses on numerical methods, but we believe the connections sketched above warrant closer examination in future studies.
	
\section{Testbed: A zero-dimensional fermion-boson model}%	
\label{sec:the_model}

	In this section, we introduce a toy model that is simple enough to allow for a detailed comparison of numeric calculations with analytical solutions, while being rich enough to serve as a nontrivial test for \gls{rg} dynamics in field space with fermions.
	Furthermore, it allows us to study the aforementioned mathematical structure of the \gls{frg} flow equations in a concrete setting.

	We consider a zero-dimensional \gls{qft} with a single bosonic and two fermionic degrees of freedom, which are coupled to each other.
	The (moment) generating function of the model under consideration reads
		\begin{align}
			\Z ( J, \tilde{\eta}, \eta ) = \, & \mathcal{N} \int_{- \infty}^{+\infty} \dd \phi \int \dd \tilde{\theta} \, \dd \theta \times	\Vdistance	\label{eq:partition_function}
			\\
			& \times \exp \big( - \S ( \phi, \tilde{\theta}, \theta ) + J \, \phi + \tilde{\eta} \, \theta + \tilde{\theta} \, \eta \big) \, ,	\Vdistance	\nonumber
		\end{align}
	where $\phi$ is the bosonic field, $\tilde{\theta}$ and $\theta$ are the fermionic fields, and $J$, $\tilde{\eta}$, and $\eta$ are the sources for the bosonic and fermionic fields, respectively, and $\mathcal{N}$ is the normalization factor.
	Of course, the term ``field'' is a bit misleading in this context, because bosonic fields are simply real numbers, while the fermionic fields are two different ordinary Grassmann numbers ($\tilde{\theta}$ and $\theta$ are not complex conjugate to each other).
	In our following discussion we will however maintain the term field even though mathematically we are just discussing numbers.
	The Grassmann numbers $\tilde{\theta}$ and $\theta$ obey
		\begin{align}
			&	\theta^2 = \tilde{\theta}^2 = 0 \, ,	&&	\theta \, \tilde{\theta} + \tilde{\theta} \, \theta = 0 \, .
		\end{align}
	with the usual conventions, see, \eg{}, the textbooks~\cite{Berezin1966,Greiner:1996zu,Peskin:1995ev}, for Berezin integration~\cite{Berezin1966} over Grassmann-variables, \eg{},
		\begin{align}
			&	\int \dd \theta_a \, 1 = 0 \, ,&& \int \dd \theta_a \, \theta_a = 1 \, ,	&&	\theta_a \in \{ \theta, \tilde{\theta} \} \, .	\label{eq:BerezinDef}
		\end{align}
	The most general action, that is at least quadratic in the fields, reads
		\begin{align}
			\S ( \phi, \tilde{\theta}, \theta ) = \tilde{\theta} \, H ( \phi ) \, \theta + U ( \phi ) \, ,	\label{eq:classical_action_ansatz}
		\end{align}
	where $H ( \phi )$ and $U ( \phi )$ are arbitrary functions of $\phi$ that are supposed to be invariant under the transformation $\phi \to - \phi$.
	This model shares some similarities with the zero-dimensional supersymmetric model in Ref.~\cite{SkinnerScript} but is in the present form and context novel to our knowledge.
	We will also assume that $H ( \phi ) \neq 0$ at least for some values of $\phi$ -- otherwise the partition function would vanish and the model would be trivial.
	Note that for constant $H ( \phi)$ fermions and bosons decouple and we are left with a bosonic $\mathbb{Z}_2$-symmetric model.
	We studied such bosonic $\mathbb{Z}_2$-symmetric models in detail in our previous works~\cite{Koenigstein:2021syz,Koenigstein:2021rxj,Steil:2021cbu} and we refer to these works for an in-depth discussion of this limit/sector of the present model.
    To ensure convergence of the integrals in the partition function \labelcref{eq:partition_function} and the related expectation values, \cf{} \cref{subsec:exp_c}, $U ( \phi )$ should asymptotically at least grow like $\phi^2$ and $H ( \phi )$ should not grow faster than polynomial.
	However, the functions $U ( \phi )$ and $H ( \phi )$ do not need to be analytic.
	In the following we denote $H ( \phi )$ as a (generalized) Yukawa coupling that can involve a mass term for the fermions as well as couplings to the bosonic field.\footnote{Note, that for the sake of simplicity and in contrast to Yukawa couplings in higher spacetime dimensions our field-dependent Yukawa coupling is supposed to be an even function of $\phi$.
	Furthermore due to the field-dependency $H ( \phi )$ does not act as a simple two-fermion-one-boson coupling like the name would canonically suggest. Nonvanishing derivatives $\partial_\phi^n H ( \phi )$ correspond to two-fermion-$n$-boson couplings, cf.\ \cref{eq:flow_H_1}.}
	The function $U ( \phi )$ is the (self-interaction) potential for the bosonic field.
	To resolve the entire field-space $\phi$ one can introduce an explicit term in the potential which breaks $\phi \to - \phi$ symmetry or work with nonzero sources $J$ to sample different values of $\phi$. 
	For this work we practically choose to do the latter.
	However, this is mentioned explicitly.

\subsection{Expectation values and generating functions}%
\label{subsec:expval_genfunc}

	Next, we turn to the calculation of expectation values and generating functions for the model.
	We start with the normalization factor, which is given by
		\begin{align}
			\N^{-1} = \, & \int_{- \infty}^{\infty} \dd \phi \, H ( \phi ) \exp \big( - U ( \phi ) \big) \, ,	\label{eq:normalization}
		\end{align}
	where we applied the rules~\eqref{eq:BerezinDef} for integration over Grassmann numbers (Berezin integrals).
	The remaining $\phi$-integration may be performed by hand for specific choices of $H ( \phi )$ and $U ( \phi )$.
	Otherwise, the integral can be evaluated to preferred precision using standard methods for numerical integration.

	Now, we study two classes of correlation functions.
	First, we consider purely bosonic expectation values.
	The expectation value of some function $f ( \phi )$ is given by
		\begin{align}
			\langle f ( \phi ) \rangle = \N \int_{- \infty}^{\infty} \dd \phi \, H ( \phi ) \, f ( \phi ) \exp \big( - U ( \phi ) \big) \, ,	\label{eq:expvalf}
		\end{align}
	where we already integrated out the fermionic fields.
	Second, we consider expectation values involving fermionic fields.
	Since there are only two fermionic fields, the only nonvanishing type of expectation value is the product of a function $g ( \phi )$ times $\tilde{\theta} \, \theta$,
		\begin{align}
			\langle g ( \phi ) \, \tilde{\theta} \, \theta \rangle = - \N \int_{- \infty}^{\infty} \dd \phi \, g ( \phi ) \exp \big( - U ( \phi ) \big) \, .	\label{eq:expvalg}
		\end{align}
	This also includes the fermionic two-point correlation  function%
		\footnote{Again, we use the names from standard \gls{qft} even though the term ``two-point'' does not really make sense in a model that lives in a single spacetime point.}, 
	for which $g ( \phi ) = 1$.
	Again, the remaining $\phi$-integrations are easily calculated for specific choices of $H ( \phi )$ and $U ( \phi )$ either analytically or numerically.
	Note, that the expectation value of a single fermionic field as well as higher-order fermionic expectation values vanish due to the Berezin integration, \ie{}, the Grassmann-nature of $ \tilde{\theta}$ and $\theta$.
	On the other hand, the functions $f$ and $g$ can be arbitrary functions of $\phi$, however, expectation values that are odd in $\phi$ vanish for vanishing source $J$ -- without the additional term $- c \, \phi$ in the potential.

	Before we continue, let us mention that on the level of the above integrals also functions with $H ( \phi ) < 0$ for some values of $\phi$ are allowed and may lead to negative expectation values.
	From the perspective of a probability distribution and partition function this seems to be problematic and can be considered as a simple model for the ``sign-problem'' in \gls{qft} and statistical physics, which we will also analyze from the \gls{frg} perspective.

	Furthermore, it is interesting to remark the following:
	As long as one is solely interested in purely bosonic expectation values, one may integrate out the fermionic fields right from the start and obtain an effective action\footnote{This is only directly possible, where $H ( \phi ) > 0$, while where $H ( \phi ) < 0$ one may use $\ln ( | H ( \phi ) | ) + \ii \uppi$.}
		\begin{align}
			\tilde{\S} ( \phi ) = U ( \phi ) - \ln \big( H ( \phi ) \big)
		\end{align}
	for the bosonic field and a partition function
		\begin{align}
			\tilde{\Z} ( J ) = \int_{- \infty}^{\infty} \dd \phi \, \exp \big( - \tilde{\S} ( \phi ) + J \, \phi \big) \, .
		\end{align}
	This is also standard in \gls{qft} problems, while it is usually not possible to further evaluate the remaining $\phi$-integrals because of the general spacetime dependence of the fields and derivative-terms in the fermion determinant.

	Introducing the Schwinger functional (in the present context a simple function)
		\begin{align}
			\W ( J, \tilde{\eta}, \eta ) = \ln \Z ( J, \tilde{\eta}, \eta ) \, ,	\label{eq:schwinger_functional}
		\end{align}
	one can calculate connected correlation functions by taking functional derivatives of $\W$ with respect to the sources.
	Explicit expressions for the first few relevant connected correlation functions can be found in App.~\labelcref{subsec:exp_c}.

	Since the \gls{frg} is formulated in terms of the \gls{eaa}, we are also interested in the vertex functions of the model.
	Otherwise, direct comparison between the \gls{frg} and the exact results is not possible.
	First, we introduce the effective action as the Legendre transform of the Schwinger functional,
		\begin{align}
			\Gamma ( \varphi, \tilde{\vartheta}, \vartheta ) = \underset{J, \tilde{\eta}, \eta}{\mathrm{sup}} \big\{ J \, \varphi + \tilde{\eta} \, \vartheta + \tilde{\vartheta} \, \eta - \W ( J, \tilde{\eta}, \eta ) \big\} \, .	\label{eq:effective_action}
		\end{align}
	Here, $\varphi$, $\tilde{\vartheta}$, and $\vartheta$ are the mean-fields.
	Then, the vertex functions are defined as the functional derivatives of the effective action with respect to the fields,
		\begin{align}
			\Gamma^{( l, m, n )} ( \varphi, \tilde{\vartheta}, \vartheta ) = \frac{\delta^{(l+m+n)} \Gamma ( \varphi, \tilde{\vartheta}, \vartheta )}{\delta \varphi^l \, \delta \vartheta^m \, \delta \tilde{\vartheta}^n} \, , \label{eq:effective_action_dIII}
		\end{align}
	which need to be evaluated at the physical point -- here $\varphi = 0 = \tilde{\vartheta} = \vartheta$ -- corresponding to the minimum of the effective action and is in agreement with the expectation values of $\langle \phi \rangle = 0 = \langle \tilde{\theta} \rangle = \langle \theta \rangle$ in the absence of explicit symmetry breaking terms.
	Explicit expressions for the first few relevant vertex functions can be found in App.~\labelcref{subsec:exp_vf}.
	
\subsection{Full field-dependence of the two-point vertex functions}%
\label{sec:field-dependent-vertex-functions}

	Factually, during our \gls{frg} calculations we will not only obtain the vertex functions at the physical point, but we even have to resolve the full field-dependence of the vertex functions for solving the corresponding \glspl{pde}.
	Oftentimes, the full field-dependence of the vertex functions is of minor interest, because the physical point is the only relevant one, and it only becomes important with explicit symmetry breaking terms \etc.
	However, we could also compare the solution of the \gls{rg} flows globally, at every point in field space, to the exact results to benchmark our methods.
	
	In order to do so, we first need the relation between $\varphi$ and $J$ from the Legendre transformation \labelcref{eq:effective_action},
		\begin{align}
			\varphi ( J ) = \, & \frac{\delta \W ( J, 0, 0 )}{\delta J} =	\Vdistance
			\\
			= \, & \frac{1}{\mathcal{Z} ( J, 0, 0 )} \, \frac{\delta \mathcal{Z} ( J, 0, 0 )}{\delta J} = \langle \phi \rangle_J \, .\Vdistance	\nonumber
		\end{align}
	This quantity can be computed numerically for arbitrary $J$ from \cref{eq:partition_function}.
	(For $J = 0$ we recover the expectation value of the bosonic field $\langle \phi \rangle$ that vanishes for symmetric $U(\phi)$ and $H(\phi)$.)
	The inversion of this dependence is $J ( \varphi )$, which can be computed (numerically) from $\varphi ( J )$.
	However, from the Legendre transformation \labelcref{eq:effective_action} we also have
		\begin{align}
			J ( \varphi ) = \, & \frac{\delta \Gamma ( \varphi, 0, 0 )}{\delta \varphi} \, .	\label{eq:j_of_phi}
		\end{align}
	Hence, we also directly obtain the full field-dependence of $J$ as the field-dependent bosonic one-point vertex function.
	This already allows for a global comparison of the \gls{frg} results to the exact results.

	However, we can also go a step further and consider the field-dependence of the two-point vertex functions.
	For the bosonic one, we have
		\begin{align}
			\Gamma^{\varphi \varphi} ( \varphi ) = \, & \frac{\delta^2 \Gamma ( \varphi, 0, 0 )}{\delta \varphi^2} = \label{eq:gamma2_of_phi}	\Vdistance
			\\
			= \, & \bigg( \frac{\delta^2 \mathcal{W} ( J, 0, 0 )}{\delta J^2} \bigg)^{- 1} \bigg|_{J = J ( \varphi )} =	\Vdistance	\nonumber
			\\
			= \, & \bigg( \frac{1}{\mathcal{Z} ( J, 0, 0 )} \, \frac{\delta^2 \mathcal{Z} ( J, 0, 0 )}{\delta J^2} +	\Vdistance	\nonumber
			\\
			& \quad - \bigg( \underbrace{\frac{1}{\mathcal{Z} ( J, 0, 0 )} \, \frac{\delta \mathcal{Z} ( J, 0, 0 )}{\delta J}}_{\varphi} \bigg)^2 \bigg)^{- 1} \bigg|_{J = J ( \varphi )} \, ,	\Vdistance	\nonumber
		\end{align}
	which reduces to \cref{eq:vertex_phi_phi} at the physical point.
	It can be calculated from the numerical solution of the \gls{rg} flow equations for all discretization points in field space directly.
	The reference values are obtained by calculating the respective $J ( \varphi )$ with \cref{eq:j_of_phi} and then evaluating the \gls{rhs} of \cref{eq:gamma2_of_phi} numerically with \cref{eq:partition_function}.

	Lastly, we find the full field-dependence of the fermionic two-point function
		\begin{align}
			\Gamma^{\tilde{\vartheta} \vartheta} ( \varphi ) = \, & \frac{\delta^2 \Gamma ( \varphi, 0, 0 )}{\delta \tilde{\vartheta} \, \delta \vartheta} =	\label{eq:gamma_theta_theta_of_phi}	\Vdistance
			\\
			= \, & - \bigg( \frac{\delta^2 \mathcal{W} ( J, 0, 0 )}{\delta ( - \eta ) \, \delta \tilde{\eta}} \bigg)^{- 1} \bigg|_{J = J ( \varphi )} \, .	\Vdistance	\nonumber
		\end{align}
	Again, we can calculate this quantity for all discretization points in field space from the numeric solution of the \gls{pde} of the \gls{rg} flow equations and compare it to the reference values.
	The latter are obtained the same way as for the bosonic two-point vertex function.
	Note, that we directly dropped vanishing terms in all of the above formulae for the sake of the readability.

	Finally, let us remark that one can use exactly the same formulae to extract the exact field-dependent two-point functions during the \gls{rg} flow at nonzero regulator to compare them to our numerical results or simply study their scale-dependence.
	Here, one merely has to add the regulator insertion (see next paragraphs and \cref{eq:DeltaSdef}) to the classical action $\S$ in \cref{eq:partition_function} and repeat the above steps for finite \gls{rg} time $t$.

	Having these quantities at hand, we are now well equipped to turn to the \gls{frg} approach and compare the results of the \gls{frg} to the exact results of the toy model.

\section{The FRG approach to the zero-dimensional model}%
\label{sec:the_frg_approach}

	In this section, we turn to the \gls{frg} formulation of the toy model.
	Here, we shall however not repeat the general derivation of the Wetterich equation, which was presented numerous times in the literature.
	The same applies to the above discussion on the flow-equation for the two-point vertex function and its relation to the \gls{hjb} equation.
	Still, because we are working in the unconventional context of zero spacetime dimensions, we discuss the main differences to the standard approach.
	Let us therefore summarize the fields in the field-space vector $\Phi = ( \varphi, \vartheta, \tilde{\vartheta} )^T$.
	Then, the \gls{frg} flow equation~\eqref{eq:wetterich_equation} reads
		\begin{align}
			&	\partial_t \bar{\Gamma} ( t, \Phi ) = \, &	\vdistance	\label{eq:wetterich_equation_zeroD}
			\\
			= \, & \STr \big[ \big( \tfrac{1}{2} \, \partial_t R ( t ) \big) \big( \bar{\Gamma}^{(2)} ( t, \Phi ) + R ( t ) \big)^{- 1} \big] \, ,	\vdistance	\nonumber
		\end{align}
	where $R ( t )$ is the matrix-valued regulator function and $\STr$ denotes the supertrace in the conventions of Appendix C of Ref.~\cite{Koenigstein:2023wso}.
	Furthermore, $\bar{\Gamma}^{(2)} ( t, \Phi )$ is the matrix of second derivatives of the \gls{eaa} with respect to the fields.
	Remember, that the \gls{eaa} is defined as a shift of the scale-dependent effective action by the regulator insertion,
		\begin{align}
			\bar{\Gamma} ( t, \Phi ) = \Gamma ( t, \Phi ) - \Delta \S ( t, \Phi ) \, ,	\label{eq:def-gamma_bar_def}
		\end{align}
	which both approach the full effective action \cref{eq:effective_action} in the limit $t \to \infty$.
	The regulator insertion for our zero-dimensional model is given by
		\begin{align}
			\Delta \S ( t, \Phi )= \, & \tfrac{1}{2} \, r_\boson ( t ) \, \varphi^2 + \tilde{\vartheta} \, r_\fermion ( t )\, \vartheta \, .	\vdistance\label{eq:DeltaSdef}
		\end{align}
	The mathematical nature of the regulator is, where one main difference to \gls{frg} in nonzero dimension arises.
	In our zero-dimensional setup, the regulator shape functions $r_\boson ( t )$ and $r_\fermion ( r )$ are plain monotonically decreasing functions of the \gls{rg} time $t \in [ 0, \infty)$ with % chktex 9
		\begin{align}
			&	r_{\boson/\fermion} ( t = 0 ) = \Lambda \, ,	&&	\lim_{t \to \infty} r_{\boson/\fermion} ( t ) = 0 \, ,
		\end{align}
	where $\Lambda$ is the ultraviolet cutoff that is supposed to be much larger%
		\footnote{Actually one should send $\Lambda \to \infty$, which is not possible in practice for numerical calculations.} 
	than any other scales in the problem.
	In general, regulator shape functions also depend on the momentum and in some setups even on the fields themselves, see, \eg{}, Refs.~\cite{Pawlowski:2005xe,Zorbach:2024zjx,Pawlowski:2015mlf,Braun:2020bhy,Rosten:2010vm,DePolsi:2022wyb} for a general discussion.
	In our zero-dimensional model, momenta do not exist and we do not consider regulator insertions with additional field-dependencies beyond \cref{eq:DeltaSdef}.
	Additionally, the \gls{rg} scale is usually of energy dimension, while in our case it is dimensionless.
	One can simply think of the regulator shape functions as scale-/\gls{rg}-time-dependent (dimensionless) mass terms for the fields -- similar to the Callan-Symanzik-type regulators in higher dimensions~\cite{Symanzik:1971cu,Symanzik:1971vw,Alexandre:2000eg,Callan:1970yg,Symanzik:1970rt,Pawlowski:2005xe,Braun:2022mgx,Oevermann:2024edu}.
	Otherwise and structurally, the \gls{frg} flow equation is the same as in the higher-dimensional case (within a truncation), especially \gls{wrt} its structure as a \gls{pde} system in field space.
	To be specific, we use
		\begin{align}
			r_{\boson/\fermion} ( t ) = \Lambda \, \ee^{- \gamma_{\boson/\fermion} \, t} \, .
		\end{align}
	It is also possible to use other monotonically decreasing functions, due to reparameterization invariance of the \gls{frg} in zero dimensions.
	In fact, we use the same regulator shape functions for fermions and bosons ${r_{\boson} ( t ) = r_{\fermion} ( t ) }$ with $\gamma_{\boson/\fermion} = 1$.
	However, it is also possible to shift the regulators against each other, which we also did during testing by setting $\gamma_\mathrm{b} = 1$ and $\gamma_\mathrm{f} \neq 1$.
	Similar shifts between the fermionic and bosonic \gls{rg}-scale/-time have recently gained some attention in \gls{frg} studies of \glspl{left} of \gls{qcd}, see, \eg{}, Ref.~\cite{Ihssen:2023xlp}.
	A shift of dynamics between fermions and bosons is \apriori{} difficult to motivate in the present zero-dimensional study and after semi-successful tests, we decided to keep the focus on other aspects of our novel approach.
	Still, it might be an interesting aspect from a numerics perspective for further studies beyond the scope of this work.

	Finally, let us remark that $\Gamma ( t, \Phi )$ from the \gls{rhs}\ of \cref{eq:def-gamma_bar_def}, as being the Legendre transform of the scale-dependent Schwinger functional $\W ( t, \mathcal{J} )$, has to be a convex function for all $t$, such that $\Gamma^{(2)} = \bar{\Gamma}^{(2)} + R$ has to be positive for finite $t$ and positive or zero for $t \to \infty$.
	The former ensures invertibility in the Wetterich equation \labelcref{eq:wetterich_equation_zeroD}.
	For finite \gls{uv} cutoff $\Lambda$ this of course restricts the space of admissible initial actions $\S ( \Phi )$ to those that yield a convex  $\Gamma ( t, \Phi )$ at $t = 0$.
	During the flow, the Wetterich equation is self-healing and always ensures convexity of $\Gamma ( t, \Phi )$ for $t > 0$, if it is convex at $t = 0$.
	However, also note that in contrast to higher-dimensional examples, where it is not always possible to enlarge the \gls{uv} cutoff trivially, this can indeed be done in our zero-dimensional models, such that here, we do not have to worry about this restriction in practice and simply sufficiently enlarge $\Lambda$.

\subsection{An exact truncation}

	Normally it is not possible to solve the Wetterich equation \labelcref{eq:wetterich_equation} exactly and one has to resort to truncation schemes, such as the derivative, vertex or operator expansions.
	This is usually done by specifying a set of relevant fields and interactions and then truncating the \gls{eaa} $\bar{\Gamma} ( t, \Phi )$ at a certain order in the fields or derivatives.
	Within this approximation, a particular subset of these couplings/vertices is treated as scale-dependent.
	Thereby one has to ensure, that $\bar{\Gamma} ( t, \Phi )$ matches the classical action $\S ( \Phi )$ (or the one-loop correction to $\S$) at the ultraviolet cutoff scale $\Lambda$ for $\Lambda \to \infty$.
	However, zero-dimensional models are special in this respect, because all possible interactions can be included in the \gls{eaa} via a simple ansatz.
	It is therefore theoretically possible to solve the Wetterich equation exactly for the toy model.
	The particular ansatz is of similar shape as the most general form for the classical action \labelcref{eq:classical_action_ansatz},
		\begin{align}
			\bar{\Gamma} ( t, \Phi ) = \tilde{\vartheta} \, H ( t, \varphi ) \, \vartheta + U ( t, \varphi ) \, .	\label{eq:frg_ansatz}
		\end{align}
	The difference is, that we promote the functions $H ( \phi )$ and $U ( \phi )$ to scale-dependent functions $H ( t, \varphi )$ and $U ( t, \varphi )$ and exchange the fluctuating quantum fields $\phi$, $\tilde{\theta}$, $\theta$ from the (functional) integral \labelcref{eq:partition_function} with mean-fields $\varphi$, $\tilde{\vartheta}$, $\vartheta$ from the Legendre transformation \labelcref{eq:effective_action}.
	Note again that reference values for $H(t, \varphi)$ and $U(t, \varphi)$ can be calculated directly for all $t$ from \cref{eq:gamma2_of_phi,eq:gamma_theta_theta_of_phi,eq:def-gamma_bar_def} by adding the regulator insertion \cref{eq:DeltaSdef} to the classical action in \cref{eq:partition_function} and repeating the steps from \cref{sec:field-dependent-vertex-functions}.

	In the next step, one would have to insert the ansatz \labelcref{eq:frg_ansatz} into the Wetterich equation \labelcref{eq:wetterich_equation} and solve the \gls{pde} with the initial condition $\bar{\Gamma} ( t = 0, \Phi ) = \S ( \Phi )$.
	Of course, this is a nontrivial task, because the Wetterich equation is a nonlinear \gls{pde} in field space and especially the exact inversion of the full matrix-valued and field-dependent two-point function is rarely possible.
	In practice, one projects the flow equation onto the couplings/vertices contained in the truncation, here $H ( t, \varphi )$ and $U ( t, \varphi )$, and then solves the resulting system of coupled \glspl{pde} and/or \glspl{ode}.
	Usually, this is done by taking derivatives of \cref{eq:wetterich_equation} with respect to the fields and then evaluating the resulting expressions on a particular field configuration.
	We will also follow this approach.
	However, due to the momentum-independence of the model, we can also extract the flow equations for $H ( t, \varphi )$ and $U ( t, \varphi )$ directly from the Wetterich equation.
	Indeed, in zero dimensions, it is possible to exactly invert the full two-point function and afterwards to evaluate the trace on the right hand side of the Wetterich equation exactly.
	Both approaches are presented in detail in the Appendices.~\ref{app:exact_inversion} and~\ref{app:projection}, respectively.
	Hence, a clear advantage of this zero-dimensional model is that a truncation of the Wetterich equation is not necessary but still possible, which allows for a direct comparison of the results of the exact and truncated solutions as well as the numerical solution to the exact results.
	Within this work, we solely focus on the numerical solution of the exact flow equations without testing artificial truncations.
	This might be subject of future work.
	(Nonetheless, first unsystematic tests already suggest that significant truncation artifacts can be expected for most interesting test cases and applications.)
	A surplus is that it also allows to easily analyse the mathematical structure of the flow equations as coupled \glspl{pde}.

\subsection{FRG flow equations}

	Applying either the exact inversion approach, see Appendix~\ref{app:exact_inversion}, or the common projection approach, see Appendix~\ref{app:projection}, we obtain two coupled \glspl{pde} as \gls{rg} flow equations for our model from the Wetterich equation \labelcref{eq:wetterich_equation}.
	These are the flow equation of the (scale-dependent) effective potential,
		\begin{align}
			\partial_t U = \, & -\big( -\tfrac{1}{2} \, \partial_t r_\boson \big) \, \tfrac{1}{r_\boson + \partial_\varphi^2 U} + \big( -\partial_t r_\fermion \big) \, \tfrac{1}{r_\fermion + H} =	\label{eq:flow_u}
			\\
			= \, & -
			\begin{gathered}
				\includegraphics{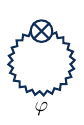}
			\end{gathered}
			+
			\begin{gathered}
				\includegraphics{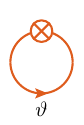}
			\end{gathered} \, ,	\nonumber
		\end{align}
	and the flow equation of the (scale-dependent) (generalized) Yukawa coupling
			\begin{align}
				\partial_t H = \, &
				-2 \, \big( -\tfrac{1}{2} \, \partial_t r_\boson \big) \, \tfrac{1}{( r_\boson + \partial_\varphi^2 U )^2} \, \tfrac{1}{r_\fermion + H} \, ( \partial_\varphi H )^2 +	\Vdistance	\label{eq:flow_H_1}
				\\
				&  +\big( -\tfrac{1}{2} \, \partial_t r_\boson \big) \, \tfrac{1}{( r_\boson + \partial_\varphi^2 U )^2} \, \partial_\varphi^2 H +	\Vdistance	\nonumber
				\\
				& - 2 \, \big( -\tfrac{1}{2} \, \partial_t r_\fermion \big) \, \tfrac{1}{r_\boson + \partial_\varphi^2 U} \, \tfrac{1}{( r_\fermion + H )^2} \, ( \partial_\varphi H )^2 =	\Vdistance	\nonumber
				\\
				= \, & -
				\begin{gathered}
					\includegraphics{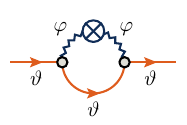}
				\end{gathered}
				+
				\begin{gathered}
					\includegraphics{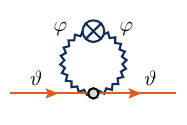}
				\end{gathered} +	\Vdistance	\nonumber
				\\
				& -
				\begin{gathered}
					\includegraphics{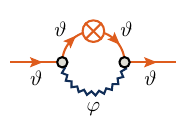}
				\end{gathered} \, .	\Vdistance	\nonumber
			\end{align}
	Here, we used the abbreviated form ${U \equiv U ( t, \varphi )}$, ${H \equiv H ( t, \varphi )}$ as well as ${r_\boson \equiv r_\boson ( t )}$ and ${r_\fermion \equiv r_\fermion ( t )}$ and introduced a common diagrammatic notation for the flow equations in terms of Feynman diagrams:
	Regulator insertions, $- \tfrac{1}{2} \, \partial_t r_\mathrm{b/f}$, are represented by crossed circles, bosonic/fermionic propagators,
		\begin{align}
			&	\tfrac{1}{r_\boson + \partial_\varphi^2 U} \, ,	&&	\tfrac{1}{r_\fermion + H} \, ,
		\end{align}
	by zigzag/solid blue/orange lines, respectively, and vertices, $\partial_\varphi^2 H$ and $\partial_\varphi H$, by dots.
	Solving this system of \glspl{pde} with the initial conditions $U ( t = 0, \varphi ) = U ( \varphi )$ and $H ( t = 0, \varphi ) = H ( \varphi )$ to $t \to \infty$ is indeed mathematically equivalent to calculating the vertex functions from the partition function.\footnote{Hereby, the vertex functions have to be extracted from $U ( t \to \infty, \varphi)$ and $H ( t \to \infty, \varphi)$ by calculating derivatives \gls{wrt} $\varphi$ at the minimum of the entire action, which is without explicit symmetry breaking at $\varphi = 0 = \tilde{\vartheta} = \vartheta$.}
	However, as is directly apparent from the flow equations, the system is highly nonlinear and a solution nontrivial, while the (numerical) calculation of the vertex functions from the partition function is in contrast straightforward.
	In realistic \glspl{qft} in higher dimensions, this is not the case and the direct calculation of expectation values and vertex functions from the functional integral is challenging.
	In this case, solving nonlinear \gls{pde} systems might be a reasonable alternative, which is why the \gls{frg} was developed in the first place and why we are testing adequate numerical methods for the solution of the flow equations in this work.

\subsection{Comments on the FRG flow equations and their nonconservative form}

	Before we continue with specific solution methods for the \gls{pde} system, we want to make some remarks on its form.
	\begin{enumerate}
		\item	The flow equations have almost exactly the same structure as the flow equations in higher spacetime dimensions for systems of bosons and fermions that are interacting via a local, field-dependent Yukawa coupling and bosonic self-interactions.
		In higher dimensions, however, considering solely a field-dependent Yukawa coupling (all quark-antiquark-$n$-boson scattering processes) and the field-dependent bosonic potential (all $n$-boson scatterings) constitutes a truncation -- the lowest order in the derivative expansion.\footnote{In Ref.~\cite{Ihssen:2023xlp} referred to as \gls{sclpa} in the context of the \gls{qm} model.}
		Still, working within this truncation requires a nontrivial numerical solution of the flow equations.
		Examples for studies that work within this approximation are Refs.~\cite{Ihssen:2023xlp,Fejos:2020lli,Grossi:2021ksl}.
		Some of these face sever challenges in the solution of the \gls{pde} system, especially in the presence of a chemical potential.
		At the end of this work, we also apply our developments to a particular higher-dimensional model -- the Gross-Neveu-Yukawa model -- within this truncation to demonstrate the applicability.

		\item	The system is of first order in $t$- and up to second order in $\varphi$-derivatives.
		Since $t$ serves as the evolution parameter its interpretation and denotation as a time $t \in [ 0, \infty )$ in the context of the \gls{pde} system is rather natural and we call it the \gls{rg} time. % chktex 9
		On the other hand, this implies that $\varphi \in ( - \infty, \infty )$ can be seen as the spatial domain, which has, however, to be mapped or truncated to a finite domain for numerical purposes.

		\item	It was already pointed out in previous works~\cite{Grossi:2019urj,Grossi:2021ksl,Koenigstein:2021syz,Koenigstein:2021rxj,Steil:2021cbu,Stoll:2021ori,Zorbach:2024zjx,Zorbach:2024rre}, that, as a consequence of the latter, the bosonic contribution to \cref{eq:flow_u} can indeed be interpreted as a (advection-)diffusion equation.
		The most simple way to see this is by taking another $\varphi$-derivative of \cref{eq:flow_u}, see also \cref{eq:wetterich_equation_Fi_diagrammatic},
				\begin{align}
					\mkern6mu\qquad \partial_t u
					= \, & \tfrac{\dd}{\dd \varphi} \Big[ \big( \tfrac{1}{2} \, \partial_t r_\boson \big) \, \tfrac{1}{r_\boson + \partial_\varphi u} - \big( \partial_t r_\fermion \big) \, \tfrac{1}{r_\fermion + H} \Big] =	\Vdistance	\label{eq:flow_du}
					\\
					= \, & - \big( \tfrac{1}{2} \, \partial_t r_\boson \big) \, \tfrac{1}{( r_\boson + \partial_\varphi u )^2} \, \partial_\varphi^2 u +	\Vdistance	\nonumber
					\\
					& + \big( \partial_t r_\fermion \big) \, \tfrac{1}{( r_\fermion + H )^2} \, \partial_\varphi H \, ,	\Vdistance	\nonumber
				\end{align}
		and define $u ( t, \varphi ) \equiv \partial_\varphi U ( t, \varphi )$.
		The first term on the right hand side of \cref{eq:flow_du} is a nonlinear diffusion term with a manifestly positive nonlinear diffusion coefficient that itself is a function of $t$ and $\partial_\varphi u$.
		The fermionic contribution, however, is of source type in the time evolution of $u$.
		However, let us also recapitulate that using the $\varphi$-derivative without explicitly evaluating it, \cref{eq:flow_du} can also be seen as rewriting the flow equation for $u$ in a conservative form, which is a common technique in the context of diffusion-advection equations, which was leveraged in the previously mentioned works.
		
		Another way of dealing with the flow equation for $U$ is to even go one derivative higher and work with the flow of the field dependent mass (squared)\footnote{Note, despite calling it a mass, $M$ can be negative and is not necessarily the mass of a particle in the sense of a physical interpretation.} $M ( t, \varphi ) = \partial_\varphi^2 U ( t, \varphi )$, as it was for example done in Refs.~\cite{Ihssen:2023qaq,Ihssen:2022xkr,Ihssen:2023xlp}.
		Then, the flow equation for $M$ reads
			\begin{widetext}
				\begin{align}	\label{eq:flow_m2}
					\partial_t M = \, & \tfrac{\dd}{\dd \varphi} \Big( - \big( \tfrac{1}{2} \, \partial_t r_\boson \big) \, \tfrac{1}{( r_\boson + M )^2} \, \partial_\varphi M + \big( \partial_t r_\fermion \big) \, \tfrac{1}{( r_\fermion + H )^2} \, \partial_\varphi H \Big) =	\Vdistance
					\\
					= \, & - \big( \tfrac{1}{2} \, \partial_t r_\boson \big) \, \tfrac{1}{( r_\boson + M )^2} \, \partial_\varphi^2 M +%
					2 \,\big( \tfrac{1}{2} \, \partial_t r_\boson \big) \, \tfrac{1}{( r_\boson + M )^3} \, ( \partial_\varphi M )^2 %
					+ \big( \partial_t r_\fermion \big) \, \tfrac{1}{( r_\fermion + H )^2} \, \partial_\varphi^2 H - 2 \, \big( \partial_t r_\fermion \big) \, \tfrac{1}{( r_\fermion + H )^3} \, ( \partial_\varphi H )^2 \, .	\Vdistance	\nonumber
				\end{align}
			\end{widetext}
		The first line is again the conservative form of the flow equation for $M$, while the second line has the structure of an advection-diffusion equation for $M$ with nonlinear coefficients.
		Hereby, the first term on the right hand side is a diffusion term with a manifestly positive nonlinear diffusion coefficient, while the second term is interpreted as an advection term with a fluid velocity that depends itself on $\partial_\varphi M$ as well as the other prefactors.
		The terms proportional to the derivatives of $H$ enter as source terms in the evolution equation of $M$ but have in general the same structure as the bosonic advection and diffusion terms.
		Overall, the clear advantage of this version is that none of the coefficients in front of derivatives of $M$ and $H$ contain denominators that depend on spatial derivatives of $M$ or $H$.
		In particular, this is advantageous when the propagators have to be evaluated close to their poles or velocities on the level of the \gls{pde} have to be calculated.
		
		\item	A similar interpretation can be given to the flow equation for the Yukawa coupling $H ( t, \varphi )$, see \cref{eq:flow_H_1}.
		There is also a diffusion term with manifestly positive nonlinear diffusion coefficient, the second term, which is the tadpole contribution in \gls{frg} language.
		The first and third terms, which correspond to the fermion self-energies, are interpreted as advective contributions, since they depend on the first derivative of $H$.
		Again, the velocity field is proportional to $\partial_\varphi H$ itself.
		However, the flow equation for the Yukawa coupling -- to the best of our knowledge -- cannot be rewritten in a manifestly conservative form.
		A similar observation was made in Ref.~\cite{Grossi:2021ksl}, where the authors rewrote the flow equation of a similar system into a conservative part and nonconservative fluxes, which are then treated separately in the numerics.

		Playing the same trick as for the flow equation of the potential and defining $h ( t, \varphi ) \equiv \partial_\varphi H ( t, \varphi )$, such that
			\begin{align}
				\partial_t h = \, &
				\tfrac{\dd}{\dd \varphi} \Big( 2 \, \big( \tfrac{1}{2} \, \partial_t r_\boson \big) \, \tfrac{1}{( r_\boson + \partial_\varphi^2 U )^2} \, \tfrac{1}{r_\fermion + H} \, h^2 +	\Vdistance	\label{eq:flow_h_conservative}
				\\
				& \quad - \big( \tfrac{1}{2} \, \partial_t r_\boson \big) \, \tfrac{1}{( r_\boson + \partial_\varphi^2 U )^2} \, \partial_\varphi h +	\Vdistance	\nonumber
				\\
				& \quad + 2 \, \big( \tfrac{1}{2} \, \partial_t r_\fermion \big) \, \tfrac{1}{r_\boson + \partial_\varphi^2 U} \, \tfrac{1}{( r_\fermion + H )^2} \, h^2 \Big) \, ,	\Vdistance	\nonumber
			\end{align}
		does indeed not lead to the desired conservative form.
		As one can see, the problem here is that the flux on the \gls{rhs} as well as the fluxes in the first lines of \cref{eq:flow_du,eq:flow_m2} still depend on $H$, which is no longer the variable that is evolved in the equation, but has to be reconstructed from $h$ via integration over the spatial domain or a constraint equation.
		Hence, the problem formally turned into an integro-differential equation, which prevents us from directly applying techniques developed in previous works and is in general an involved numerical problem.
	\end{enumerate}

	In our previous works and the works of colleagues, where \gls{rg} flow equations were already analyzed using (computational) fluid dynamics, the respective \glspl{pde} were studied in their conservative form (or forms that are close to conservative forms).
	This drastically simplifies the search for suitable numerical methods as well as their applicability, because lots of numerical schemes are designed for (hyperbolic/parabolic) conservation laws, \eg{}, finite volume methods, discontinuous Galerkin methods, \etc.~\cite{LeVeque:1992,LeVeque:2002,RezzollaZanotti:2013,Hesthaven2007}.
	As already mentioned, for flow equations involving field-dependent Yukawa couplings~\cite{Pawlowski:2014zaa,Fejos:2020lli,Grossi:2021ksl,Ihssen:2023xlp}, this is not always possible and similar issues arise for flow equations of field-dependent wave-function renormalizations in higher-dimensional systems~\cite{Batini:2023nan,Ihssen:2022xkr,Rennecke:2022ohx,Johnson:2022cqv,Ihssen:2023nqd,Ihssen:2024ihp,Bonanno:2025mon}.
	Consequently, we deemed it necessary to develop a new understanding/interpretation of the respective \gls{rg} flow equation paired with adequate numerical methods based on the resulting structural insights.
	In \cref{subsec:wetterich_ii} we already established the interpretation of the flow \cref{eq:wetterich_equation_Gii} as a functional (infinite-dimensional), viscous \gls{hjb} \cref{eq:wetterich_equation_Fii}, which is central for our practical developments in the following. 
	In the next \cref{subsec:frgHJB0d} we will elaborate on the practical implications of this structural insight and introduce a suitable numerical scheme based on it.

\subsection{FRG flow equations as coupled viscous Hamilton-Jacobi systems}%
\label{subsec:frgHJB0d}

	In order to explicitly introduce our new ideas for the (numerical) solution of the \gls{frg} flow equations, let us slightly rearrange \cref{eq:flow_H_1,eq:flow_m2} in their primitive form,
	\begin{widetext}
		\begin{align}
			\partial_t M & - \tfrac{\partial_t r_\boson}{( r_\boson + M )^3} \, ( \partial_\varphi M )^2 + \tfrac{2 \, \partial_t r_\fermion}{( r_\fermion + H )^3} \, ( \partial_\varphi H )^2 = - \tfrac{\frac{1}{2} \, \partial_t r_\boson}{( r_\boson + M )^2} \, \partial_\varphi^2 M  %
			+ \tfrac{\partial_t r_\fermion}{( r_\fermion + H )^2} \, \partial_\varphi^2 H  \, ,	\Vdistance	\label{eq:flow_M_hj}
			\\
			\partial_t H & - \tfrac{1}{r_\boson + M} \, \tfrac{1}{r_\fermion + H} \, \Big( \tfrac{\partial_t r_\boson}{r_\boson + M} + \tfrac{\partial_t r_\fermion}{r_\fermion + H} \Big) \, ( \partial_\varphi H )^2 = - \tfrac{\frac{1}{2} \, \partial_t r_\boson}{( r_\boson + M )^2} \, \partial_\varphi^2 H \, ,	\Vdistance	\label{eq:flow_H_hj}
		\end{align}
	\end{widetext}
	and compare these equations to the standard form of (nonautonomous) (viscous) \gls{hj}-(Bellman)-type problems introduced in \cref{sec:wetterich},
		\begin{align}
			\partial_t u + \mathcal{H} \big( t, \vec{x}, u, \vec{\nabla}_{x} u \big) = \varepsilon \, \Delta u \, .	\label{eq:hamilton_jacobi_diffusion}
		\end{align}
	Clearly, the \gls{rg} flow equations are each of rather similar form, where the terms on the \gls{lhs} play the role of the respective Hamiltonians and the terms on the \gls{rhs} are the diffusion terms with more general diffusion coefficients.\footnote{The rightmost diffusion-type-term $\propto \partial_\varphi^2 H$ in the \gls{pde}~\eqref{eq:flow_M_hj} for $M$ has negative coefficient.
	This is not a problem, because it factually acts like a source in the evolution of $M$ and does not lead to instabilities.
	The same applies to the last term $\propto ( \partial_\varphi H )^2$ on the \gls{lhs} of \cref{eq:flow_M_hj}.}
	\gls{hj}-(Bellman)-type problems appear in many areas of physics, but also in the context of finance, optimal control/transport, and many other fields.
	A first systematic concept of their solution (originally in the absence of the diffusion term) was developed by Crandall and Lions~\cite{Crandal1983} in terms of so-called viscosity solutions.
	This was generalized by Jensen~\cite{Jensen1988} to the static second-order nonlinear degenerate elliptic \glspl{pde} of the form\footnote{In fact, \gls{frg} flow equations in higher spacetime dimensions in the fixed-point limit are of this type.}
		\begin{align}
			F ( x, u, \vec{\nabla} u, \Delta u ) = 0 \, .
		\end{align}
	According to Barles~\cite{Barles2013} also parabolic/first-order evolution equations like the viscous \gls{hj} equation \labelcref{eq:hamilton_jacobi_diffusion} and our \gls{rg} flow equations can be considered as degenerate elliptic for $\varepsilon \geq 0$, which allows to find viscosity solutions.
	Without going too much into the mathematical details, which is beyond the scope of this work, let us briefly try to sketch the notion of viscosity solutions and their connection to weak and entropy solutions of conservation laws.
		
	First, we recapitulate the difference between a strong and a weak solution of a \gls{pde}.
	A strong solution is a function that satisfies the \gls{pde} pointwise, while a weak solution is a function that satisfies the \gls{pde} in a distributional sense.
	A weak solution is a solution of the integral form of the equation and is often obtained by multiplying the \gls{pde} with a test function and integrating over the domain.
	This allows to relax the requirements on the differentiability of the solution and to consider a larger class of functions, such as functions with nonanalytic structures.
	Since nonlinear \glspl{pde} in general tend to form nonanalyticities in the solution, weak solutions are often the only practically relevant solutions.

	Second, weak solutions are not unique and one has to impose additional conditions to obtain a unique solution.
	One way to do this is to consider entropy solutions, which are weak solutions that satisfy the entropy condition.
	It ensures that entropy -- ideally defined from physics considerations -- is non-decreasing.
	This implies that the solution is stable under small perturbations.
	It further singles out the correct physical solution for shocks via the Rankine-Hugoniot condition.
	For conservation laws, the entropy condition is often formulated in terms of the entropy flux and the entropy production, which have to satisfy certain inequalities~\cite{LeVeque:1992,LeVeque:2002}.
	Though, usually it is hard or not clear, how to find a suitable entropy for a given conservation law and the entropy condition is not always applicable.
	Still, decent numerical schemes for conservation laws converge to the entropy solution, even if the entropy is not known, because they usually introduce a small amount of numerical diffusion.
	Convergence, however, is only achieved for numerical schemes that are based on the conservative form of the \gls{pde}.

	Third, a viscosity solution is a concept that was developed to find a unique and stable solution to degenerate elliptic \glspl{pde} and \gls{hj}-(Bellman)-type problems that are in nonconservative form \cite{Crandal1983,Crandall:1992,Barles2013,Jensen1988}.
	They are defined in terms of the comparison principle, which states that the solution is the largest subsolution and the smallest supersolution of the \gls{pde}.
	In general, viscosity solutions do not have to be smooth and can have nonanalyticities and singularities and can therefore be regarded as a generalization of weak solutions.

	The connection between these concepts is that both, entropy and viscosity solutions, emerge as the unique and physically meaningful solutions.
	Oftentimes this is practically achieved with vanishing viscosity methods (especially when an entropy is not known for conservation laws):
	One adds a small diffusion term to the \gls{pde} by hand or via numerical diffusion of the discretization scheme and then takes the limit of the diffusion coefficient to zero and $\Delta x \to 0$.
	For conservation laws, this is the Lax-Friedrichs method, which is used to stabilize and single out the correct numeric solution also for shocks.
	For \gls{hj}-type problems, this is the vanishing viscosity method of Crandall and Lions.
	In both cases, the underlying idea is that the diffusion term regularizes the \gls{pde} and turns it into a parabolic problem with a unique smoothened solution.
	Then, in one or the other way, the limit of the diffusion coefficient to zero is taken, which defines the unique solution of the original \gls{pde}.
	For some \glspl{pde}, such as Burgers equation, there exists a formulation in terms of a conservation law as well as a formulation in terms of a \gls{hj}-type problem and entropy and viscosity solutions are equivalent~\cite{Caselles:1992}.
	The same seems to be the case for the flow equation for the effective potential alone.
	Maybe, recent developments, see Ref.~\cite{Floerchinger:2023qpw} and Refs.\ therein, might even help to identify a suitable entropy for \gls{frg} problems.

	However, it is important to note that for \glspl{pde} that cannot be converted into a conservative form, such as general \gls{hj}-type problems, the viscosity solution is only unique in the sense that it is unique for a particular choice of artificial diffusion.
	This implies that the numeric solution of a \gls{hj}-problem that is dominated by the advection terms and has nonanalyticities in the solution depends on the choice of the numeric scheme, \eg{}, the choice of limiter functions.
	For the present \gls{frg} problems, however, there are always diffusive terms -- the tadpole contributions, which gives hope that the solution is unique and stable.
	This is also expected, because the solution is in general also obtainable from the path integral, which, in the presence of sufficient regularization, should mean that it exists and is unique.

	Let us discuss this in more detail for the full \gls{frg} problem.
	Here, of course, things are more involved, because we have a dynamical system with time-dependent coefficients.
	However, what the \gls{frg} flow equations actually seem to do is very similar to the vanishing viscosity method -- but automatically and in a self-consistent way.
	As long as and wherever the diffusion coefficients are positive, the flow equations are parabolic and the solution should be unique.
	However, for late times, where $\partial_t r_\mathrm{b/f}$ and $r_\mathrm{b/f}$ both tend to zero, we have to distinguish between two different scenarios:
	\begin{enumerate}
		\item Wherever the denominator of the diffusion coefficient does not approach its pole, the diffusion term vanishes and the flow approaches its viscosity solution.
		The reason is that also the coefficients in front of the advective terms (first-derivative-square terms in the Hamiltonian) approach zero and the dynamic overall stops.

		\item If the denominators in the diffusion coefficient as well as in the prefactors of the advective terms approach their poles, there is in general a lot of dynamics and velocities are high.
		Even in the presence of diffusion we expect shocks to form, because the number of propagators in advective terms is larger than in the diffusion terms.
		However, the presence of diffusion should ensure that the solution is unique and stable and converges against the viscosity solution.
	\end{enumerate}
	The interesting points in the solution of the \gls{frg} flow equations are those points, where the behavior changes, \ie{}, where $M$ and $H$ change their sign as a function of $\varphi$ while the regulator $r_\mathrm{b/f}$ approaches zero.
	Here, we expect the solution to form nonanalyticities and singularities.
	Another region, where we expect nonanalytic behavior is, where the diffusion coefficient in \cref{eq:flow_H_hj} vanishes for late times (positive $M$), while $r_\mathrm{f} + H$ approaches zero (negative $H$).
	However, also here, we expect a self-regulation of the system towards the correct solution.

	We close the present discussion with three remarks:
	\begin{enumerate}
		\item In general, nonanalytic behavior is also physically totally expected, if one considers for example the flow equation of the effective potential of a $\mathbb{Z}_2$-model in its symmetry broken regime in nonzero spacetime dimensions.
		Here, the \gls{ir} potential consists of a flat plateau and a steep slope starting at the field-expectation value.
		For $M = \partial_\varphi^2 U$ this implies that it takes negative values for small $| \varphi |$ and positive values for large $| \varphi |$ during the flow and approaches the region with $M = 0$ (the plateau of the potential) from below.
		
		Another scenario, where nonanalytic behavior is totally expected is in the presence of a chemical potential that introduces an external shock into the system.

		\item The attentive reader might have noticed that our analysis of the \gls{frg} flow equations in terms of coupled \gls{hj}-type problems is based on the flow equations for the field-dependent two-point functions.
		One might therefore ask, if this structure and analogy is more general.
		Indeed, as already mentioned in \cref{subsec:wetterich_ii}, for full field-dependent \gls{frg} flow equations that are extracted from the two-point vertex function (\eg{}, the field-dependent wave-function renormalization), the same structure can be found.
		The reason is rather simple:
		Diagrammatically one always has contributions from tadpole-type diagrams, which are proportional to second derivatives of vertices and are therefore diffusion type operators, and contributions from self-energy diagrams, which are proportional to two first derivatives of vertices and are therefore advective operators.
		We come back to this point in \cref{sec:higher_dim_exp} in the context of applications to higher-dimensional systems.

		\item Before we turn to the explicit numerical scheme let us return to the remark that the flow equation for the effective potential alone can also be rewritten as a conservation law.
		Indeed, we experienced that it is numerically advantageous in terms of convergence and stability to treat the purely bosonic contributions in the flow equation for the field-dependent bosonic mass conservatively, while the fermionic source-type contribution are handled within the \gls{hj} framework that is also used for the Yukawa coupling.
	\end{enumerate}

\section{Numerical scheme}%
\label{sec:numerical_scheme}

	Having reformulated the \gls{pde} system of the \gls{frg} flow equations as a system of coupled viscous \gls{hj}-type problems, we can now think about suitable numerical methods to solve these.
	Since we are not experts on the construction of numerical methods, we follow our own repeatedly stated advise and search for and adapt well-tested methods from the literature.

	Hence, we suggest to use a high-resolution semi-discrete central scheme for (viscous) \gls{hj} equations that was developed by Kurganov and Tadmor~\cite{KT2000:HamiltonJacobi}.
	This scheme shares some similarities with the semi-discrete \gls{muscl} that we previously used for the solution of \gls{lpa} flow equations in their conservative formulation~\cite{Koenigstein:2021syz,Koenigstein:2021rxj,Steil:2021cbu,Stoll:2021ori,Murgana:2023xrq,Koenigstein:2023wso,Steil:2023sfd,Zorbach:2024rre} as well the scheme presented in Appendix B of Ref. \cite{Capellino:2023cxe}.
	The diffusive part of the former scheme is still used for the conservative contributions to the flow equation of the field-dependent boson mass in this work.

	However, before we briefly present the scheme and all necessary expressions for an implementation, we introduce the following notation for partial derivatives:
		\begin{align}
			u_t \equiv \, & \partial_t u ( t, x ) \, ,	\vdistance
			\\
			u_x \equiv \, & \partial_x u ( t, x ) \, ,	\vdistance
			\\
			u_{xx} \equiv \, & \partial_x^2 u ( t, x ) \, .	\vdistance
		\end{align}

	In their publication, Kurganov and Tadmor consider problems of the form
		\begin{align}
			u_t + \mathcal{H} ( u_x ) = \varepsilon \, u_{xx} \, ,	\label{eq:hamilton_jacobi_diffusion_original}
		\end{align}
	with additional initial and boundary conditions.
	As already discussed at the end of the previous section, our flow equations are more general, which calls for several modifications of the original scheme.
	However, let us first present the original scheme and then discuss the necessary modifications afterwards.

\subsection{Original semi-discrete scheme}

	Most of the finite-volume and finite-element schemes for \glspl{pde} are based on a multi-step procedure:
	The spatial domain is subdivided into cells and the solution within each cell is approximated by a low-order polynomial.
	This approximated solution is what is actually evolved by the time stepper.
	Then, for each time step the fluxes at the cell interfaces have to be computed.
	To this end, it is necessary to reconstruct the solution at the cell interfaces or some staggered grid.
	Usually, this reconstruction involves a limiting procedure of the gradients of the solution to avoid spurious oscillations.
	Another limiting procedure is applied to the fluxes, which involves an estimation of the local wave speeds.
	Finally, the fluxes are computed and the solution is updated.
	The entire procedure is, however, strongly dependent on the conservative form of the \gls{pde}.

	Nevertheless, also the high-resolution semi-discrete central scheme of Kurganov and Tadmor -- in the following \gls{kthj}-scheme -- is based on the same concepts.
	First, one discretizes the solution on an equidistant grid,
		\begin{align}
			x_j = x_0 + j \, \Delta x \, ,
		\end{align}
	with $j = 0, \ldots, N$, $x_\mathrm{min} = x_0$, $x_\mathrm{max} = x_0 + N \, \Delta x$, and $\Delta x = \tfrac{x_\mathrm{max} - x_\mathrm{min}}{N}$,
		\begin{align}
			u_j  = u ( x_j, t ) \, .
		\end{align}
	Here, $x_\mathrm{min}$ is the left boundary of the spatial domain and $x_\mathrm{max}$ the right boundary and in total there are $N + 1$ grid points.
	Additionally, one defines midpoints between the grid points -- the staggered grid,
		\begin{align}
			x_{j + \frac{1}{2}} = x_j + \tfrac{1}{2} \, \Delta x \, ,
		\end{align}
	which is the analogue to the cell interfaces in the finite-volume method.
	Instead of reconstructing the solution itself on $x_{j + \frac{1}{2}}$ using slope-limiters for $u_x$, the \gls{kthj}-scheme is based on the use of limiters on the level of the second derivatives of the solution, $u_{xx}$.
	Hence, using the definition of the first-order forward and backward differences%\cite[Eq.~(2.1) f]{KT2000:HamiltonJacobi}
		\begin{align}
			\frac{( \Delta u )_{j + \frac{1}{2}}}{\Delta x} \equiv \frac{u_{j + 1} - u_j}{\Delta x} \, ,	\vdistance
		\end{align}
	one can estimate the forward and backward reconstruction of the first derivatives by%\cite[Eq.~(4.10)]{KT2000:HamiltonJacobi}
		\begin{align}
			u_{x,j}^\pm = \frac{( \Delta u )_{j \pm \frac{1}{2}}}{\Delta x} \mp \frac{( \Delta u )^\prime_{j \pm \frac{1}{2}}}{2 \Delta x} \, ,
		\end{align}
	Here, $( \Delta u )^\prime_{j \pm \frac{1}{2}}$ is the limited second derivative of the solution on the midpoints.%\cite[Eq.~(2.2)]{KT2000:HamiltonJacobi}
	\begin{widetext}
		\begin{align}
			&	( \Delta u )^\prime_{j + \frac{1}{2}} = \mathrm{minmod} \Big( \theta \, \big( \Delta u_{j + \frac{3}{2}} - \Delta u_{j + \frac{1}{2}} \big), \tfrac{1}{2} \, \big( \Delta u_{j + \frac{3}{2}} - \Delta u_{j - \frac{1}{2}} \big), \theta \, \big( \Delta u_{j + \frac{1}{2}} - \Delta u_{j - \frac{1}{2}} \big) \Big) \, ,	&&	\theta \in [ 1, 2 ] \, .\label{eq:hj22}
		\end{align}
	\end{widetext}
	Smaller values of $\theta$ lead to more dissipative schemes, while larger values of $\theta$ lead to less dissipative schemes.
	In particular, we use the minmod limiter with $\theta = 1$ for this work as in the original publication, \ie{},%\cite[Eq.~(2.3)]{KT2000:HamiltonJacobi}
		\begin{align}
			& \mathrm{minmod} ( x_1, x_2, \ldots ) =	\Vdistance
			\\
			= \, &
			\begin{cases}
				\min_j \{ x_j \} \, ,	& \text{if } \forall j: x_j > 0 \, ,	\vdistance
				\\
				\max_j \{ x_j \} \, ,	& \text{if } \forall j: x_j < 0 \, ,	\vdistance
				\\
				0 \, ,	& \text{otherwise.}	\vdistance
			\end{cases}	\nonumber
		\end{align}
	Additionally, we test the superbee and MUSCL\footnote{The MUSCL limiter in the present context is sometimes also called monotonized central (MC) limiter.} limiters~\cite{RezzollaZanotti:2013,WikiFluxLimiter2025}, for which \cref{eq:hj22} may be expressed As
		\begin{align}
			& ( \Delta u )^\prime_{j + \frac{1}{2}} =	\Vdistance
			\\
			= \, & \big( \Delta u_{j + \frac{3}{2}} - \Delta u_{j + \frac{1}{2}} \big) \, \phi \bigg( \frac{\Delta u_{j + \frac{1}{2}} - \Delta u_{j - \frac{1}{2}}}{\Delta u_{j + \frac{3}{2}} - \Delta u_{j + \frac{1}{2}} }  \bigg) \, ,\Vdistance	\nonumber
		\end{align}
	with
		\begin{align}
			\phi_\mathrm{superbee}(r) & \equiv \max (0,\min (1,2\mkern2mu r),\min (2,r)) \, ,\vdistance
			\\
			\phi_\mathrm{MUSCL}(r) & \equiv \max (0,\min (2,2\mkern2mu r,(1+r)/2)) \, .\vdistance
		\end{align}
	While the minmod limiter is the most dissipative one, the superbee limiter is the least dissipative second-order accurate limiter that is total variation diminishing.
	The MUSCL limiter is a compromise in between the two, which has the advantage of limiting symmetrically in and opposed to the velocity direction.
	As an additional ingredient one needs a local estimate of the wave speeds.
	This is given by
		\begin{align}
			a_j = \max \big( | \mathcal{H}^\prime ( u_{x,j}^+ ) |, | \mathcal{H}^\prime ( u_{x,j}^- ) | \big) \, ,
		\end{align}
	where $\mathcal{H}^\prime$ is the derivative of the Hamiltonian with respect to the first derivative of the solution -- the canonical momentum.
	This is rather intuitive to understand, because the coefficient in front of a $u_x$ term can be considered as the velocity and the respective term as an advection term.
	Lastly, one needs a discretization of the diffusion/Laplace operator.
	Since the remaining scheme is of second order in $\Delta x$ it is natural to use
		\begin{align}
			u_{xx, j} = \frac{u_{j + 1} - 2 \, u_j + u_{j - 1}}{\Delta x^2} \, .
		\end{align}
	The full semi-discrete scheme is then given by the following equation,%\cite[Eq.~(4.9)]{KT2000:HamiltonJacobi}
		\begin{align}
			u_{t,j} = \, & - \frac{\mathcal{H} ( u_{x,j}^+ ) + \mathcal{H} ( u_{x,j}^- ) - a_j \, ( u_{x,j}^+ - u_{x,j}^- )}{2} +	\vdistance	\label{eq:KT_HJ_scheme_original}
			\\
			& + \varepsilon \, \frac{( u_{j + 1} - 2 \, u_j + u_{j - 1} )}{\Delta x^2} \, .	\vdistance	\nonumber
		\end{align}
	Here, however, we abstain from further details of its derivation and refer to the original publication~\cite{KT2000:HamiltonJacobi} for any details.
	Before discussing our modifications let us remark that there exists also a first-order accurate version of the \gls{kthj}-scheme, which is also presented in the original publication.
	Furthermore, note that due to its semi-discrete nature, the scheme can be combined with any time-stepping method.
	For this work, we use the implicit time steppers in terms of the \texttt{IDA}\footnote{
		We use as options for \texttt{IDA} a \texttt{MaxDifferenceOrder} of $5$ and \texttt{Newton} as an \texttt{ImplicitSolver} with \texttt{LinearSolverMethod} \texttt{Band} and total \texttt{BandWidth} of $11$ for the two component systems considered in this work.
		To use such a sparse/band solver it is imperative to implement the discretized \gls{pde} systems in \textit{'point-major'} order (${[\ldots,u_{j,1},u_{j,2},\ldots]}$) as opposed to \textit{'component-major'} order (${[u_{0,1},\ldots,u_{N,1},u_{0,2},\ldots,u_{N,2}]}$), \cf{} \cref{subsec:modKTHJ}. 
		The use of a sparse/banded linear solver over a dense one significantly increases the performance/decreases the runtime especially for systems with a large number $N$ of cells. 
	} method in \texttt{NDSolve}~\cite{MathematicaNDSolveIDA} from \textit{Mathematica 14.2}~\cite{Mathematica:14.2} with \texttt{PrecisionGoal} and \texttt{AccuracyGoal} of $\geq 10$ and $\leq 14$.
	Time-stepping is not a focus of this work and we refer the interested reader to the excellent Ref.~\cite{Ihssen:2023qaq} discussing the issue in the context of \gls{frg} in detail. 
	Let us also remark at this point that we crosschecked our implementation with the sample cases from the original publication and found good agreement.

\subsection{Modifications of the KT-HJ scheme for FRG flow equations}%
\label{subsec:modKTHJ}

	For this work, we consider the following modifications of the original \gls{kthj}-scheme for the solution of the \gls{frg} flow equations.
	\begin{enumerate}
		\item We allow for an additional dependence of the Hamiltonian on the time $t$ and $u$ itself,
			\begin{align}
				& \mathcal{H} ( u_x ) \quad \Rightarrow \quad \mathcal{H} ( t, u, u_x ) \, ,
			\end{align}
		such that
			\begin{align}
				&	\mathcal{H} ( u_{x,j}^\pm )	\quad 	\Rightarrow	\quad 	\mathcal{H} ( t, u_j, u_{x,j}^\pm ) \, ,
			\end{align}
		and the new estimate of the wave speeds is given by
			\begin{align}
				& \mkern6mu\qquad  a_j = \max \big( | \mathcal{H}^\prime ( t, u_j, u_{x,j}^+ ) |, | \mathcal{H}^\prime ( t, u_j, u_{x,j}^- ) | \big) \, .
			\end{align}
		The reason, why we believe that this modification is still valid, is that $t$ and $u$ solely appear in prefactors of $u_x^2$-terms in the Hamiltonian and there is no other $u_x$-dependence.

		\item The same applies to the coefficient of the diffusion terms, $\varepsilon$,
			\begin{align}
				& \varepsilon \quad  \Rightarrow \quad  \varepsilon ( t, u ) \, ,
			\end{align}
		or on a discrete level
			\begin{align}
				& \mkern2mu\qquad \varepsilon \quad \Rightarrow \quad  \varepsilon_j = \varepsilon ( t, u_j ) \, .
			\end{align}
		
		\item In addition, we consider systems of $f \in \mathbb{N}$ coupled Hamilton-Jacobi-type \glspl{pde}, such that \cref{eq:hamilton_jacobi_diffusion_original} is replaced by
			\begin{align}
				\vec{u}_t + \vec{\mathcal{H}} ( t, \vec{u}, \vec{u}_x ) = \mathcal{E} ( t, \vec{u} \, ) \cdot \vec{u}_{xx} \, .	\label{eq:hamilton_jacobi_diffusion_vector}
			\end{align}
		Hereby,
			\begin{align}
				\vec{u} = ( u_1, \ldots, u_f )^T \, ,
			\end{align}
		and
			\begin{align}
				\vec{\mathcal{H}} ( t, \vec{u}, \vec{u}_x ) =
				\begin{pmatrix}
					\mathcal{H}_1 ( t, \vec{u}, \vec{u}_x )
					\\
					\vdots
					\\
					\mathcal{H}_f ( t, \vec{u}, \vec{u}_x )
				\end{pmatrix} \, ,	\label{eq:hamiltonian_vector}
			\end{align}
		and $\mathcal{E} ( t, \vec{u} \, )$ is matrix valued.
		On the discretized level the diffusion term on the \gls{rhs}\ generalizes trivially.
		A good -- the best possible -- estimate for the local wave speed is very important for convergence and numerical performance.
		We tested various approaches and have come to the conclusion that the most promising approach is to estimate the local velocities separately for each component of the system by only studying the respective Hamiltonian and its dependence on the first derivatives,
			\begin{align}
				& \mkern6mu\qquad a_{j,m,n} = \max_{\vec{u}_{x,j} \in \{ \vec{u}_{x,j}^{\, +}, \vec{u}_{x,j}^{\, -} \}}  \bigg|  \frac{\partial  \mathcal{H}_m ( t, \vec{u}_j, \vec{u}_{x,j} )}{\partial u_{x,j,n}} \bigg| \, ,	\label{eq:KT_HJ_aj_vector}
			\end{align}
		where $j$ labels the spatial grid point, $m$ the component of the Hamiltonian \labelcref{eq:hamiltonian_vector}, and $n$ the component of the first derivative $\vec{u}_x$.
		(Another possible approach can be found in Ref.~\cite{Capellino:2023cxe}.)
		Hence, the advective term in \cref{eq:KT_HJ_scheme_original} for the $m$-th field generalizes to
		\begin{widetext}
			\begin{align}
				u_{t,j,m} = \, & - \frac{\mathcal{H}_m ( t, \vec{u}_j, \vec{u}_{x,j}^{\, +} ) + \mathcal{H}_m ( t, \vec{u}_j, \vec{u}_{x,j}^{\, -} ) - a_{j,m,n} \, ( u_{x,j,n}^+ - u_{x,j,n}^- )}{2} + \ldots \, ,	\label{eq:KT_HJ_scheme_vector}
			\end{align}
		\end{widetext}
		where summation over $n$ is implied.
			
		Combining this adapted \gls{kthj} with a conservative treatment for the bosonic contributions to the field-dependent boson mass using the diffusion term of the \gls{kt} scheme has proven to be the most promising numerical implementation.
		To be explicit, for \cref{eq:flow_M_hj,eq:flow_H_hj} we have
			\begin{align}
				\mathcal{H}_1 = \, & \tfrac{2 \, \partial_t r_\fermion}{( r_\fermion + H )^3} \, ( \partial_\varphi H )^2 \, ,	\Vdistance
				\\
				\mathcal{H}_2 = \, & - \tfrac{1}{r_\boson + M} \, \tfrac{1}{r_\fermion + H} \, \Big( \tfrac{\partial_t r_\boson}{r_\boson + M} + \tfrac{\partial_t r_\fermion}{r_\fermion + H} \Big) \, ( \partial_\varphi H )^2 \, ,	\Vdistance
			\end{align}
		such that the velocities are
			\begin{align}
				a_{1,1} = \, & 0 \, ,	\Vdistance
				\\
				a_{1,2} = \, & 4 \, \tfrac{\partial_t r_\fermion}{( r_\fermion + H )^3} \, ( \partial_\varphi H ) \, ,	\Vdistance
				\\
				a_{2,1} = \, & 0 \, ,	\Vdistance
				\\
				a_{2,2} = \, & - 2 \, \tfrac{1}{r_\boson + M} \, \tfrac{1}{r_\fermion + H} \, \Big( \tfrac{\partial_t r_\boson}{r_\boson + M} + \tfrac{\partial_t r_\fermion}{r_\fermion + H} \Big) \, \partial_\varphi H \, ,	\Vdistance
			\end{align}
		because the \gls{pde} \labelcref{eq:flow_M_hj} for $M$ reads in semi-conservative form
			\begin{align}
				& \partial_t M + \tfrac{2 \, \partial_t r_\fermion}{( r_\fermion + H )^3} \, ( \partial_\varphi H )^2 =	\Vdistance
				\\
				= \, & \tfrac{\partial_t r_\fermion}{( r_\fermion + H )^2} \, \partial_\varphi^2 H + \tfrac{\partial}{\partial \varphi} \Big( - \tfrac{\frac{1}{2} \, \partial_t r_\boson}{( r_\boson + M )^2} \, \partial_\varphi M \Big) \, .	\Vdistance	\nonumber
			\end{align}
		For the conservative part, we use the discretization of the diffusion term as in \gls{kt} scheme~\cite{KTO2-0}, where
			\begin{align}
				u_t = \cdots + \partial_x Q ( t, u, u_x )
			\end{align}
		is discretized as follows
			\begin{align}
				u_{t,j} = \, & \cdots + \tfrac{1}{\Delta x} \, \big( P_{j + \frac{1}{2}} - P_{j - \frac{1}{2}} \big) \, ,	\Vdistance	\label{eq:KT_conservative_term}
				\\
				P_{j + \frac{1}{2}} = \, & \tfrac{1}{2} \, \Big( Q \big( t, u_j, \tfrac{u_{j + 1} - u_j}{\Delta x} \big) + Q \big( t, u_{j + 1}, \tfrac{u_{j + 1} - u_j}{\Delta x} \big) \Big) \, .	\Vdistance	\nonumber
			\end{align}
		Note, however, that also treating the entire coupled \gls{pde} system as a pure diffusive \gls{hj} problem and using solely the here presented \gls{kthj} scheme with the spectral radius as an estimate for the $a_j$ lead to satisfactory results for most problems.		
	\end{enumerate}
		\begin{figure*}%Test 0: Flows - Fig. 1.
			\subfloat[\label{fig:test_0_U_I_RG_flow} For \gls{uv} potential \labelcref{eq:test_0_U_I}.]{%
				\centering
				\includegraphics{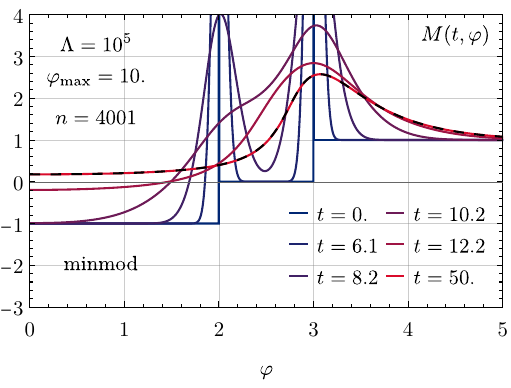}
			}\hfill
			\subfloat[\label{fig:test_0_U_II_RG_flow} For \gls{uv} potential \labelcref{eq:test_0_U_II}.]{%
				\centering
				\includegraphics{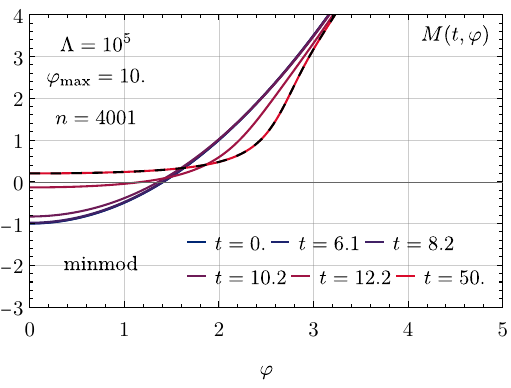}
			}
			
			\subfloat[\label{fig:test_0_U_III_RG_flow} For \gls{uv} potential \labelcref{eq:test_0_U_III}.]{%
				\centering
				\includegraphics{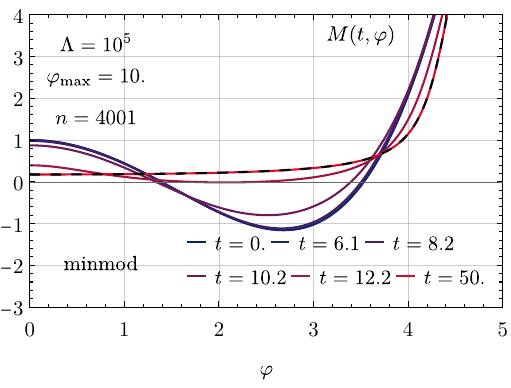}
			}\hfill
			\subfloat[\label{fig:test_0_U_IV_RG_flow} For \gls{uv} potential \labelcref{eq:test_0_U_IV}.]{%
				\centering
				\includegraphics{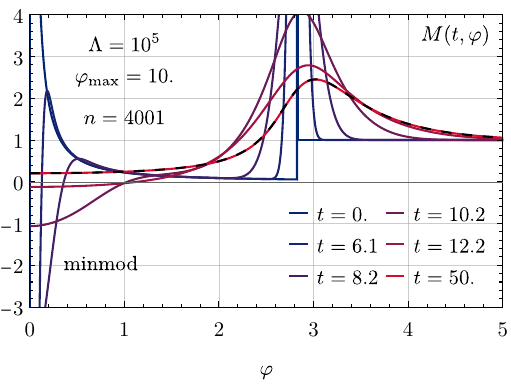}
			}
			\caption{Test 0:~\gls{rg} flow of $M ( t, \varphi )$ (colored) and the exact \gls{ir} reference solution (dashed, black).}%
			\label{fig:test_0_U_RG_flow}
		\end{figure*}
\subsection{Boundary conditions}

	Lastly, we have to provide boundary conditions for the flow equations as well as their discretization.
	Otherwise a \gls{pde} problem is ill posed.
	Within the numerical schemes under consideration, spatial boundary conditions are usually implemented via so-called ghost cells/points, which are located outside the computational domain.
	The number of ghost cells/points depends on the order of the spatial discretization scheme and is two for the \gls{kthj} scheme, because it is a second-order accurate scheme.
	For a detailed discussion of this topic, we refer to Ref.~\cite{Koenigstein:2021syz}.
	Here, we only provide a short version that is used in the following.
	
	In fact, our \gls{pde} problem is defined on $\mathbb{R}$, such that there are actually no spatial boundaries and the initial condition suffices to have a well-posed problem.
	However, to numerically solve the \glspl{pde}, we have to discretize the spatial domain and therefore have to use a finite computational domain.
	Instead of mapping the infinite domain to a finite one, we truncate the domain at some large $\pm \varphi_\mathrm{max}$, where we can assume that the solution does not change anymore or the change is almost trivial due to the shape of the \glspl{pde} and initial conditions.
	At this artificial boundary, we simply use linear extrapolation\footnote{%
		Extrapolation using the asymptotics of the \gls{uv} initial condition can be a suitable improvement over linear extrapolation for certain problems.
	} of the solution to ghost cells $\varphi_{N}$ and $\varphi_{N + 1}$ outside the computational domain, \eg{},
		\begin{align}
			u_{N + 1} = \, & 2 \, u_{N} - u_{N - 1} \, ,	\vdistance	\label{eq:ghost_cells}
			\\
			u_{N + 2} = \, & 3 \, u_{N} - 2 u_{N - 1} \, ,	\vdistance
		\end{align}
	where $N$ is the last grid point of the computational domain, because the first computational grid point is denoted as $\varphi_0$.
	This boundary condition is used for both fields $M$ and $H$.

	However, due to the $\varphi \to - \varphi$ symmetry of the entire problem, we can restrict our computational domain to $\varphi \in [ \varphi_\mathrm{min} = 0, \varphi_\mathrm{max} = \varphi_{N} ]$.
	In consequence, we have to provide another boundary condition at $\varphi = 0$, which is
		\begin{align}
			&	u_{-1} = u_{1} \, ,	&&	u_{-2} = u_{2} \, ,
		\end{align}
	without constraints on $u_0$ for both functions $M$ and $H$.

\section{Tests and discussion in zero dimensions}%
\label{sec:zero_dim_exp}

	Next, we turn to some explicit tests of the newly introduced interpretation of the \gls{frg} problem as a \gls{hj}-type system and the numerical \gls{kthj} scheme.
	In particular, we consider different test cases (initial conditions), which challenge different aspects of the numerical scheme or simulate various physical situations.
	Besides the explicit tests, we also discuss the modeled situations on a qualitative level.

	As a quantitative measure of the quality of the numerical scheme we mainly use the $L^1$ and $L^\infty$ norm/error between the numerical solution from the modified \gls{kthj} scheme and the exact solution generated via \cref{eq:gamma2_of_phi,eq:gamma_theta_theta_of_phi}.
	(Note, however, that also the ``exact'' reference solution involves comparatively small numerical error, which we tried to minimize but could not completely avoid due to a necessary interpolation of the discrete $J$-$\varphi$-data and the numerical integrations.)
	To be explicit, the $L^1$ and $L^\infty$ norm/error is defined as
		\begin{align}
			L_1 ( u ) = \, & \tfrac{1}{J} \sum_{j = 0}^{J} | u_{\text{KT-HJ},j} - u_{\text{exact},j}  | \, ,	\Vdistance	\label{eq:L1_norm}
			\\
			L_\infty ( u ) = \, & \max_{j \in \{ 0, \ldots, J \}} | u_{\text{KT-HJ},j} - u_{\text{exact},j} | \, ,	\Vdistance	\label{eq:Linf_norm}
		\end{align}
	where $u \in \{ M, H \}$ and $u_{\text{KT-HJ},j}$ and $u_{\text{exact},j}$ are the numerical and exact solution at grid point $j$, respectively, and $J$ is the number of grid points included.
	We choose $J = \lfloor N/2 \rfloor$, where $N$ is the number of grid points in the computational domain $[ \varphi_\mathrm{min} = 0, \varphi_\mathrm{max} ]$.
	The latter is much larger than the region with nontrivial dynamics.
	The reason that we do not use all grid points for benchmarking is that we do not want to include possible numerical errors from the artificial boundary conditions at $\varphi_\mathrm{max}$ and focus on the $\varphi$-values in the ``physical'' region.
	For initial conditions for $M$ and $H$ that are nonlinear for large $| \varphi |$ one finds some nonnegligible boundary effects close to $\varphi_\mathrm{max}$.
	However, if $\varphi_\mathrm{max}$ is sufficiently large, these errors do not propagate at all into the physical region.
	This is shown in plots that present the relative error of the numerical solution to the exact solution as a function of $\varphi$ at some resolution $N$,
		\begin{align}
			R_N ( u_j ) = \bigg| \frac{u_{\text{KT-HJ},j} - u_{\text{exact},j}}{u_{\text{exact},j}} \bigg| \, .	\label{eq:relative_error}
		\end{align}
	Furthermore, for some of the tests, it turns out that the $L_1$ and $L_\infty$ norms are unsuited as error measures:
	If the solution develops moving shocks/singularities a slight offset in the position of the shock leads to a huge error in the $L_1$ and $L_\infty$ norm.
	Hence, it is more appropriate to directly compare the numeric and the exact solution by visual inspection.

\subsection{Test 0: Decoupling}
	\begin{figure}%Test 0: Error in phi - Fig. 2.
		\centering
		\includegraphics{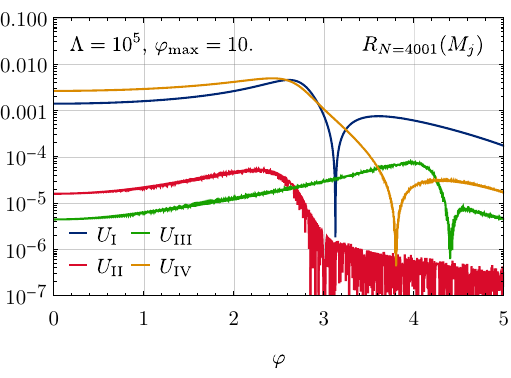}
		\caption{Test 0:~Relative error~\eqref{eq:relative_error} of the numerical solution to the exact solution as a function of $\varphi$ for the \gls{uv} potentials~\eqref{eq:test_0_U_I}-\eqref{eq:test_0_U_IV} at a resolution of $N = 4001$ points for $\varphi_\mathrm{max} = 10$.}%
		\label{fig:test_0_U_relative_error}
	\end{figure}	
	Let us turn to our first minimal test case, which serves as a crosscheck with our previous works that where based on the conservative form of the flow equations~\cite{Koenigstein:2021syz,Koenigstein:2021rxj,Steil:2021cbu,Zorbach:2024rre}.
	To this end, we consider the purely bosonic $\mathbb{Z}_2$-symmetric system.
	This can be achieved in our model by setting the Yukawa coupling to a constant, \gls{wlog}\ $H ( \phi ) = 1 \Rightarrow H ( t, \varphi ) = 1$, and solely considering the potential for the bosonic field as being \gls{rg}-time dependent.
	In this test case we treat the flow equation for $M$ with the \gls{hj} scheme \cref{eq:KT_HJ_scheme_original}, instead of using the \gls{kt} conservative formulation of \cref{eq:KT_conservative_term}.
	We tested the latter extensively in previous works~\cite{Koenigstein:2021syz,Koenigstein:2021rxj,Stoll:2021ori}.
		\begin{figure*}%Test 0: Error scaling -- Fig. 3.
			\subfloat[\label{fig:test_0_U_I_LNorm} For \gls{uv} potential \labelcref{eq:test_0_U_I}.]{%
				\centering
				\includegraphics{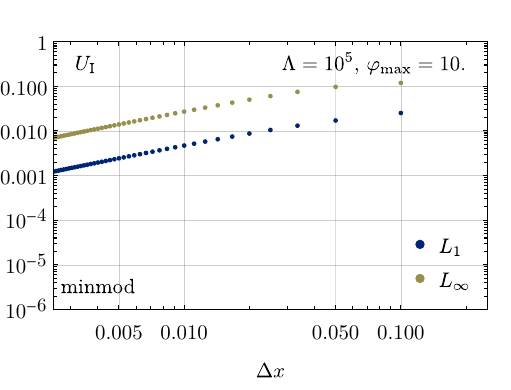}
			}\hfill
			\subfloat[\label{fig:test_0_U_II_LNorm} For \gls{uv} potential \labelcref{eq:test_0_U_II}.]{%
				\centering
				\includegraphics{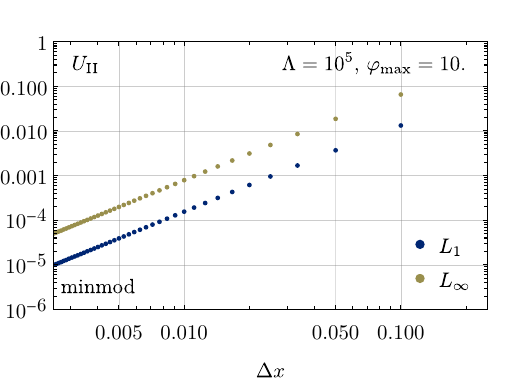}
			}
			
			\subfloat[\label{fig:test_0_U_III_LNorm} For \gls{uv} potential \labelcref{eq:test_0_U_III}.]{%
				\centering
				\includegraphics{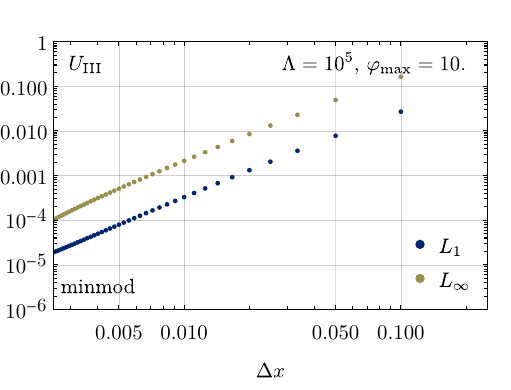}
			}\hfill
			\subfloat[\label{fig:test_0_U_IV_LNorm} For \gls{uv} potential \labelcref{eq:test_0_U_IV}.]{%
				\centering
				\includegraphics{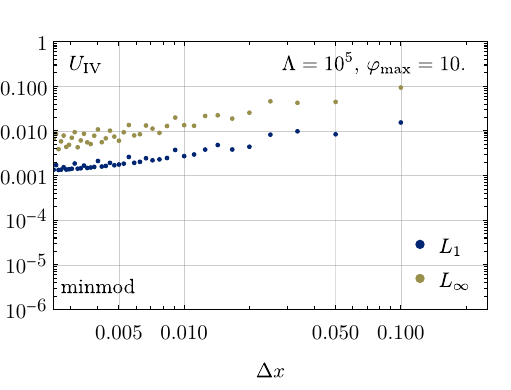}
			}
			\caption{Test 0:~$L^1$ and $L^\infty$ norm, see \cref{eq:L1_norm,eq:Linf_norm}, of the deviation between the numerical and the exact solution on the computational domain as a function of the spatial resolution for the \gls{uv} potentials~\eqref{eq:test_0_U_I}-\eqref{eq:test_0_U_IV}.}%
			\label{fig:test_0_U_LNormScaling}
		\end{figure*}
	As explicit initial conditions of our test, we choose the test potentials put forward in our previous works~\cite{Koenigstein:2021syz,Koenigstein:2021rxj,Steil:2021cbu,Zorbach:2024rre}, which where developed as challenging test cases for the numerical solution of the \gls{rg} flow equations.
	In particular, these are the following four \gls{uv} initial potentials:
		\begin{align}
			U_\mathrm{I} ( \phi ) = \, &
			\begin{cases}
				- \tfrac{1}{2} \, \phi^2 \, ,	& \text{if } | \phi | \leq 2 \, ,	\vdistance
				\\
				- 2 \, ,	& \text{if } 2 < | \phi | \leq 3 \, ,	\vdistance
				\\
				+ \tfrac{1}{2} \, ( \phi^2 - 13 ) \, ,	& \text{if } 3 < | \phi | \, ,	\vdistance	\label{eq:test_0_U_I}
			\end{cases}
			\\
			U_\mathrm{II} ( \phi ) = \, & - \tfrac{1}{2} \, \phi^2 + \tfrac{1}{4!} \, \phi^4 \, ,	\vdistance	\label{eq:test_0_U_II}
			\\
			U_\mathrm{III} ( \phi ) = \, & \tfrac{1}{2} \, \phi^2 - \tfrac{1}{20} \, \phi^4 + \tfrac{1}{6!} \, \phi^6 \, ,	\vdistance	\label{eq:test_0_U_III}
			\\
			U_\mathrm{IV} ( \phi ) = \, &
			\begin{cases}
				- ( \phi^2 )^\frac{1}{3} \, ,	& \text{if } | \phi | \leq \sqrt{8} \, ,	\vdistance
				\\
				+ \tfrac{1}{2} \, \phi^2 - 6 \, ,	& \text{if } \sqrt{8} < | \phi | \, .	\vdistance	\label{eq:test_0_U_IV}
			\end{cases}
		\end{align}
	For a detailed discussion on their choice and the simulated dynamics and challenges, we refer to Ref.~\cite{Koenigstein:2021syz}.
	Here, let us simply present the corresponding \gls{rg} flows of $M ( t, \varphi )$ in \cref{fig:test_0_U_RG_flow}.
	Already on a qualitative level, we observe from these plots that the numerical solution from the \gls{kthj} scheme, shown in red, is in very good agreement with the exact \gls{ir} reference solution, depicted as dashed black lines.
	In particular, we stress that despite the nontrivial shape of the initial conditions, the numerical scheme was able to handle all test cases without any problems like spurious oscillations or instabilities.
	The final \gls{ir} times $t = 50$ corresponds to an \gls{ir} cutoff of $r(t = 50) \simeq 10^{-17}$, which is sufficiently small to ensure that all dynamics are frozen out, which is also confirmed by the match with the exact \gls{ir} reference solution.
	To get a better quantitative understanding of the quality of the numerical solution, we present the relative error of the numerical solution to the exact solution as a function of $\varphi$ in \cref{fig:test_0_U_relative_error}.
	One observes that the relative errors are always below 1\% and in most cases even significantly smaller at a spatial resolution of $\Delta x = 0.0025$.
	The $\varphi$-region, where the deviations of the numerical from the exact solution are largest are always close to those points, where most of the dynamics takes place, \eg{}, where shocks are smeared out \etc.
	Even more interesting than the pointwise relative error at fixed resolution is the scaling of the error with the resolution.
	This is shown in \cref{fig:test_0_U_LNormScaling}, where we present the $L^1$ and $L^\infty$ norm of the deviation between the numerical and the exact solution as a function of the spatial resolution.
	For all test cases one finds that the error decreases with increasing resolution as should be the case.
	Furthermore, the $L_\infty$ error is always larger than the $L_1$ error, which is expected, because the $L_\infty$ error is the maximum error over all grid points, while the $L_1$ error is the average error over all grid points.
	Hence, there are single grid points which have significantly larger error than the average error, which is already seen from \cref{fig:test_0_U_RG_flow,fig:test_0_U_relative_error}.
	Note that the error scaling is always polynomial.
	However, it never perfectly matches the expected/ideal $\Delta x^2$ scaling.
	First, it is seldomly the case that one finds perfect matching of the scaling exponent with the expected one, which explains \cref{fig:test_0_U_II_RG_flow,fig:test_0_U_III_RG_flow}, where the exponent is close to 2 but not exactly 2.
	For the other two test cases, \cref{fig:test_0_U_I_RG_flow,fig:test_0_U_IV_RG_flow}, the scaling is actually closer to $\Delta x^1$.
	Also this is rather natural, because we already imprinted nonanalytic structures in the initial condition.
	In the vicinity of nonanalyticities it is rather natural that the error scaling reduces to the one of a first-order accurate scheme.
	Concerning \cref{fig:test_0_U_IV_RG_flow}, we also observe minimal `oscillations' in the scaling. 
	Here we are not sure about the reason for this behavior.
	However, we suspect that it is not caused by the numerical scheme itself but by the approximation/discretization of the pole in the initial condition.
	Furthermore, let us mention that exchanging the slope limiters in the \gls{kthj} has no effect on the results and the results are almost identical for the superbee and MUSCL limiters.
	Hence, we conclude that the \gls{kthj} scheme can be used instead of conservative schemes suggested in Refs.~\cite{Koenigstein:2021syz} to handle field-dependent flow equations for the second derivative of the effective potential that come as a single \gls{pde}.

\FloatBarrier%
\subsection{Test 1: Minimal test}
		\begin{figure}% Test 1: Flows -- Fig. 4.
			\subfloat[\label{fig:test_1_RG_flow_u}$\partial_\varphi U ( t, \phi )$]{%
				\centering
				\includegraphics{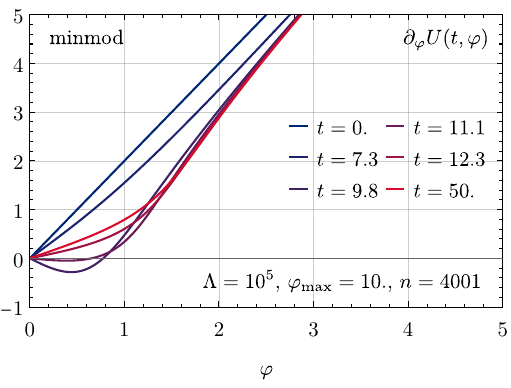}
			}

			\subfloat[\label{fig:test_1_RG_flow_M}$M ( t, \phi )$]{%
				\centering
				\includegraphics{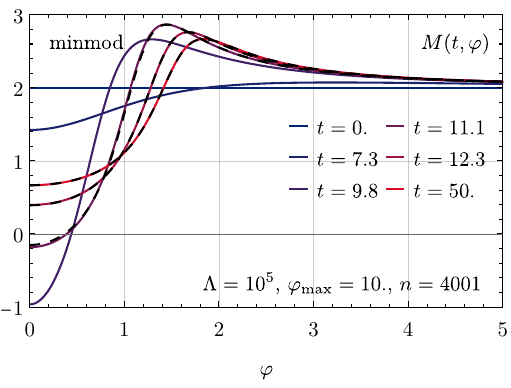}
			}

			\subfloat[\label{fig:test_1_RG_flow_H}$H ( t, \phi )$]{%
				\centering
				\includegraphics{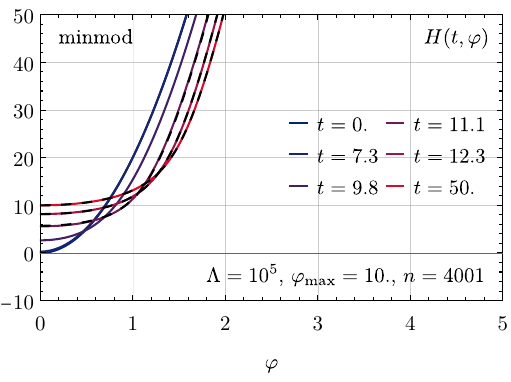}
			}
			\caption{Test 1:~\gls{rg} flow for the initial conditions~\eqref{eq:test_U_minimal} and~\eqref{eq:test_H_minimal}.
			The exact reference solution for the last three times is shown black dashed.}%
			\label{fig:test_1_RG_flow}
		\end{figure}
		\begin{figure}% Test 1: Errors in x -- Fig. 5.
			\centering
			\subfloat[For the minmod limiter.]{%
				\centering
				\includegraphics{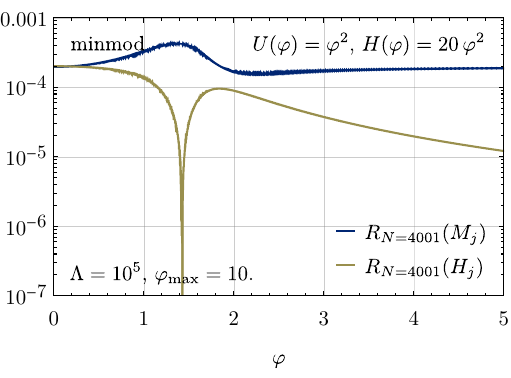}
			}

			\subfloat[For the MUSCL limiter.]{%
				\centering
				\includegraphics{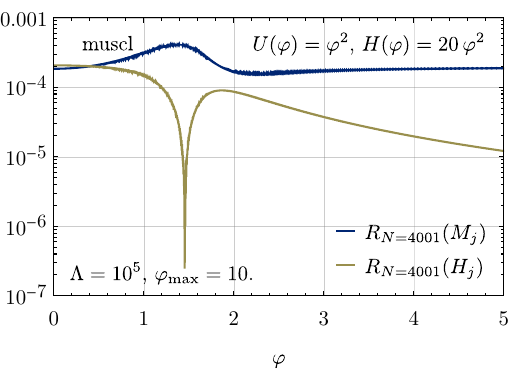}
			}

			\subfloat[For the Superbee limiter.]{%
				\centering
				\includegraphics{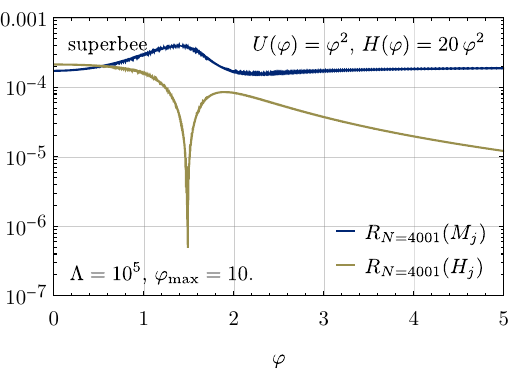}
			}
			\caption{Test 1:~Relative error~\eqref{eq:relative_error} of the numerical solution to the exact solution as a function of $\varphi$ for the initial conditions~\eqref{eq:test_U_minimal} and~\eqref{eq:test_H_minimal} for different limiters at a resolution of $N = 4001$ points with $\varphi_\mathrm{max} = 10$.}%
			\label{fig:test_1_relative_error}
		\end{figure}
	So far, we have only considered the \gls{rg} flow of the effective potential $U(t,\varphi)$, respectively the bosonic mass $M(t,\varphi) = \partial_\varphi^2 U(t,\varphi)$ in a purely bosonic $\mathbb{Z}_2$ system.
	Consequently, let us come to our first test case, where we also include the fermionic degrees of freedom.
	Hence, let use study the full \gls{pde} system \cref{eq:flow_M_hj,eq:flow_H_hj} for the effective potential $U(t,\varphi)$ and the Yukawa coupling $H(t,\varphi)$.
	This time and from now on, we deploy \cref{eq:KT_conservative_term} for the conservative part of the \gls{pde} for the bosonic mass $M(t,\varphi)$, while all other terms as well as the \gls{pde} for $H(t,\varphi)$ are treated with the modified \gls{kthj} scheme.
	We start with a rather simple test, which we call the ``minimal test'' with the following initial conditions:
		\begin{align}
			U ( \phi ) = \, & \phi^2 \, ,	\vdistance	\label{eq:test_U_minimal}
			\\
			H ( \phi ) = \, & 20 \, \phi^2 \, .	\vdistance	\label{eq:test_H_minimal}
		\end{align}
	Note, that all expectation values from \cref{subsec:exp_c} in fact boil down to the calculation of Gaussian moments, which is easily seen from \cref{eq:normalization,eq:expvalf,eq:expvalg}.
	In terms of an \gls{rg} flow, the problem is still nontrivial and corresponds to solving a complicated \gls{pde} system.
	The actual \gls{rg} flows of $M(t,\varphi)$, $H(t,\varphi)$, and also $\partial_\varphi U(t,\varphi)$, which are obtained via the \gls{kthj} scheme, are shown in \cref{fig:test_1_RG_flow}.

\subsubsection{Qualitative discussion}

	Qualitatively, one first observes perfect agreement of the numerical solution with the exact reference solutions at late \gls{rg} times, which are depicted as dashed black lines and obtained via the procedure outlined in \cref{sec:field-dependent-vertex-functions}.\footnote{The reason why we solely plot the exact reference solution of the latest three times is that reference solutions for earlier \gls{rg} times require an excessively large domain in $J$-space for the Legendre transformation and the formulae from \cref{sec:field-dependent-vertex-functions} because of the large \gls{uv} cutoff and there is little to learn from these curves.}
	However, let us also inspect the actual dynamics taking place during the \gls{rg} flow:
	We find that caused by the fermion-boson interaction the effective potential $U(t,\varphi)$ develops a nontrivial minimum, which is directly seen from the zero-crossing in its derivative $\partial_\varphi U ( t, \varphi )$, see \cref{fig:test_1_RG_flow_u} at $t=9.8$.
	At late times, this $\mathbb{Z}_2$-symmetry breaking vanishes again -- as expected in zero spacetime dimensions -- and the potential turns convex, which is easily seen from the positivity of $M(t,\varphi) = \partial_\varphi^2 U(t,\varphi)$ in the \gls{ir}.
	Note that both phase transitions during the flow are of the second order.
	The behavior of the Yukawa coupling $H(t,\varphi)$ during the \gls{rg} flow is rather subdued in comparison.
	The shape changes, while it is ultimately always positive, which is expected.
	Still, let us stress that $r ( t ) + M ( t, \varphi )$ as well as $r ( t ) + H ( t, \varphi )$ always remain positive during the entire flow, which is a necessary condition for the well-posedness of the scale-dependent version of the effective action $\Gamma ( t, \phi, \tilde{\vartheta}, \vartheta )$ in \cref{eq:effective_action} being the Legendre transform of the scale-dependent version of the Schwinger functional $W ( t, J, \tilde{\eta}, \eta )$ in \cref{eq:schwinger_functional}.

	Maybe, the most interesting aspect is that the fermions dynamically acquire some mass during the flow, which manifests as a constant offset in $H(t,\varphi)$.
	Note, that in contrast to nonzero spacetime dimensions a nonzero fermion mass is not in contradiction with the $\mathbb{Z}_2$-symmetry restoration in the \gls{ir}, because here $H(t,\varphi)$ is in general an even function of $\varphi$.
	Before we turn to the quantitative analysis, let us mention that the strength of the symmetry breaking during the flow is controlled by the prefactor in the initial condition for $H(t,\varphi)$ and for too small values there is no symmetry breaking and restoration at all, hence our choice of $20$ in \cref{eq:test_H_minimal} to facilitate a more interesting discussion.
	Note, however, that the \gls{ir} result is always in the symmetric phase.

\subsubsection{Quantitative discussion}

	Again, let us first turn to the relative error of the numerical solution to the exact solution as a function of $\varphi$ in \cref{fig:test_1_relative_error} at constant $\Delta x = 0.0025$ to get a first impression of the quality of the numerical solution as well as the position-dependence of the error.
	Overall we find that due to the rather smooth dynamics of the \gls{rg} flow the relative error is very small for all $\varphi$-values, below $0.1\%$ in fact.
	There are small spatial modulations in the relative error and the largest error is again found in the vicinity of the region in field space with the strongest dynamic during \gls{rg} flow.
	Note, that the relative error for $H(t,\varphi)$ is always smaller than the one of $M(t,\varphi)$.
	We speculate that this is due to the fact that the \gls{pde} of $H(t,\varphi)$ does not depend on derivatives of $M(t,\varphi)$ such that the velocity estimates~\eqref{eq:KT_HJ_aj_vector} are slightly more accurate.
		\begin{figure}% Test 1: Scaling -- Fig. 6.
			\centering
			\subfloat[For the minmod limiter.]{%
				\centering
				\includegraphics{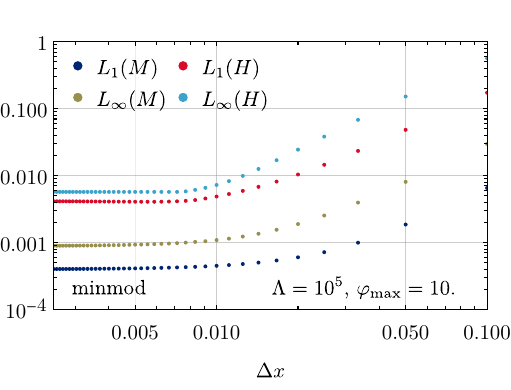}
			}

			\subfloat[For the MUSCL limiter.]{%
				\centering
				\includegraphics{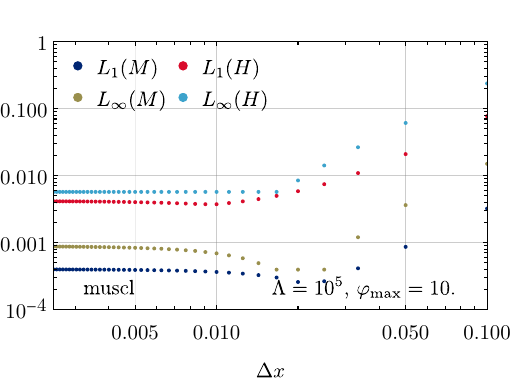}
			}

			\subfloat[For the Superbee limiter.]{%
				\centering
				\includegraphics{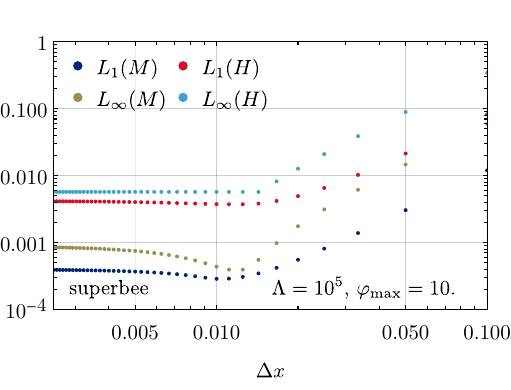}
			}
			\caption{Test 1:~$L^1$ and $L^\infty$ norm between the numerical and the exact solution on the computational domain as a function of the spatial resolution for initial conditions~\eqref{eq:test_U_minimal} and~\eqref{eq:test_H_minimal} for different limiters.}%
			\label{fig:test_1_LNormScaling}
		\end{figure}
	Next, we turn to the scaling of the error with the resolution in \cref{fig:test_1_LNormScaling}.
	For both $L$ norms and for both fields $M(t,\varphi)$ and $H(t,\varphi)$ we find that the error decreases with increasing resolution and the scaling is indeed almost $\Delta x^2$ for all limiters and rather large $\Delta x$.
	However, we observe that there seems to be some saturation of the error for $\Delta x \lesssim 0.01$.
	We were not able to identify the cause/reason for this issue/limitation.

	Still, we conclude that the numerical \gls{kthj} scheme seems to be suited for treating our system of coupled \gls{rg} flow equations -- at least for the rather smooth dynamics of this minimal test case.

\subsection{Test 2: First order phase transition and shock development}
		\begin{figure*}%Test 2: Flows -- Fig. 7.
			\subfloat[\label{fig:test_2_RG_flow_U}$U ( t, \phi )$]{%
				\centering
				\includegraphics{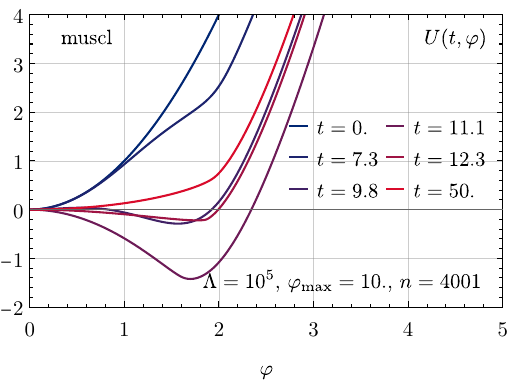}
			}\hfill
			\subfloat[\label{fig:test_2_RG_flow_u}$\partial_\varphi U ( t, \phi )$]{%
				\centering
				\includegraphics{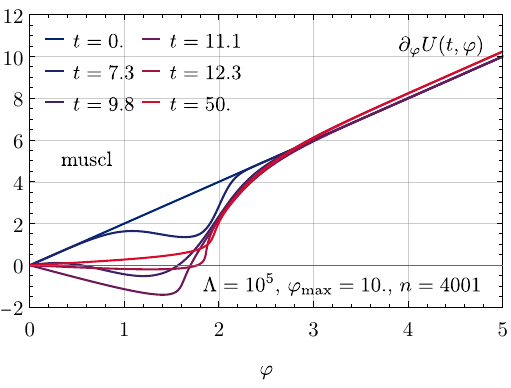}
			}

			\subfloat[\label{fig:test_2_RG_flow_M}$M ( t, \phi ) = \partial_\varphi^2 U ( t, \varphi)$]{%
				\centering
				\includegraphics{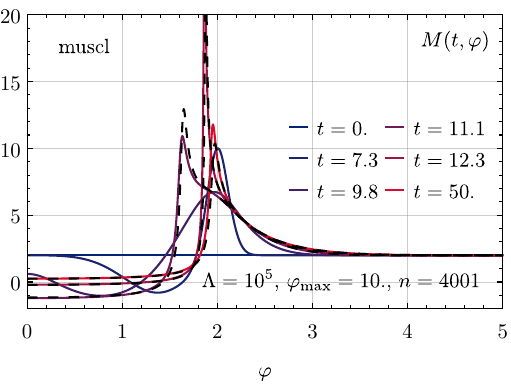}
			}\hfill
			\subfloat[\label{fig:test_2_RG_flow_H}$H ( t, \phi )$]{%
				\centering
				\includegraphics{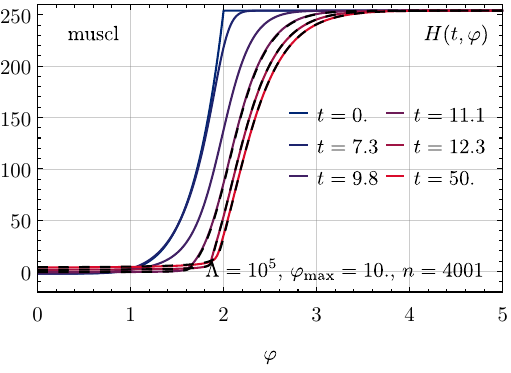}
			}
			\caption{Test 2:~\gls{rg} flow for the initial conditions~\eqref{eq:test_U_first_order} and~\eqref{eq:test_H_first_order}.
			The exact reference solution for the last three times is shown black dashed.}
		\end{figure*}
	In order to challenge the numerical scheme further, we now consider another test case, where the $\mathbb{Z}_2$-symmetry of the ground state is (spontaneously) broken and restored dynamically during the \gls{rg} flow.
	Here, however, we trigger a first-order-like phase transition by choosing the following initial conditions
		\begin{align}
			U ( \phi ) = \, & \phi^2 \, ,	\vdistance	\label{eq:test_U_first_order}
			\\
			H ( \phi ) = \, &
			\begin{cases}
				- 2 + 4 \, \phi^6 \, ,	&	\text{if } | \phi | \leq 2 \, ,	\vdistance
				\\
				254 \, ,	&	\text{if } | \phi | > 2 \, .	\vdistance	\label{eq:test_H_first_order}
			\end{cases}
		\end{align}
	Again, the initial condition for the bosons is of Gaussian type and the nontrivial dynamics is caused by the boson-fermion-interactions.
	The reason we define $H(\phi)$ in \cref{eq:test_H_first_order} as a piecewise function with some cusp at $\varphi = 2$ is mainly due to the large-$\varphi$ behavior of the entire system:
	In general, we argued that the only restriction for $H(\phi)$ for large $\phi$ from the perspective of the (path) integral~\eqref{eq:partition_function} is that it should not grow faster than polynomial.
	Also the \gls{pde} system, which is an initial value problem on $\varphi \in \mathbb{R}$, is still well-defined for such functions.
	However, inspecting \cref{eq:flow_H_hj} one finds that for $H(t,\varphi)$ which asymptotically grows faster than $\varphi^2$ the first term on the \gls{lhs} is not necessarily suppressed for large $\varphi$ -- even at $t = 0$, such that initial conditions like $\lim_{\phi \to \infty} H(\phi) \sim \phi^6$ would at least lead to large errors from the artificial boundary conditions at $\varphi_\mathrm{max}$.
	Since we are not focused on the issue of boundary conditions and correctly handling the asymptotic, but want to benchmark the numeric scheme, we decided to choose a piecewise function with some nonanalyticity as the initial condition.

\subsubsection{Qualitative discussion}

	First, we turn to a qualitative discussion of the \gls{rg} flows of $M(t,\varphi)$ and $H(t,\varphi)$ as well as the integrated quantities, $\partial_\varphi U (t,\varphi)$ and the potential $U(t,\varphi)$ itself.
	Overall, especially from \cref{fig:test_2_RG_flow_U,fig:test_2_RG_flow_u}, we observe that the system exhibits a first-order-like phase transition during the \gls{rg} flow, while ultimately in the \gls{ir} the symmetry of the ground state is restored.
	However, it is much more interesting to inspect the dynamics during the flow on the level of $M(t,\varphi)$ and $H(t,\varphi)$, the quantities that are actually evolved by the \glspl{pde}, see \cref{fig:test_2_RG_flow_M,fig:test_2_RG_flow_H}.
	Here, we find that the contribution of $H(t,\varphi)$ causes the development of a shock in the bosonic mass $M(t,\varphi)$, which is directly seen from the spike that forms at $\varphi \approx 2$ and slightly moves to the right for increasing $t$.
	Ultimately, this shock is smeared out a bit and freezes, but the remnant is clearly visible in the first derivative of the effective potential $\partial_\varphi U(t,\varphi)$, see \cref{fig:test_2_RG_flow_u}, as well as in the potential itself, see \cref{fig:test_2_RG_flow_U}.
	Also note that the Yukawa coupling $H(t,\varphi)$ as well as $M(t,\varphi)$ both get shifted to values larger than zero in the \gls{ir}.
	This is consistent with the fact that they are both two-point functions, which should be positive or zero in order to obtain a convex \gls{ir} \gls{ea} and well-defined/non-singular propagators.
	Again, we remark that $r(t) + M(t, \varphi)$ as well as $r(t) + H(t, \varphi)$ are always positive during the entire flow.
	Otherwise one would overshoot the pole of the propagators in the flow equations \labelcref{eq:flow_H_1,eq:flow_m2} and $\Gamma ( t, \varphi, \vartheta, \tilde{\vartheta} )$ would not be well-defined in terms of a Legendre transform of $W ( t, J, \eta, \bar{\eta} )$.

	Furthermore, we can compare our solutions to the exact reference solution at late \gls{rg} times, which are depicted as dashed black lines.\footnote{Again, we only show the last three exact curves, because little new information is gained from also comparing the earlier times, while the computation of the exact reference solutions turns rather costly.}
	Here, we find overall convincing agreement between the numerical and the exact solution.
	Only very close to the spike in $M(t,\varphi)$ there are deviations, which can, however, be reduced by increasing the spatial resolution.

\subsubsection{Quantitative discussion}

	This can be further analyzed on a more quantitative level.
		\begin{figure}%Test 2: Error x -- Fig. 8.
			\centering
			\subfloat[For the minmod limiter.]{%
				\centering
				\includegraphics{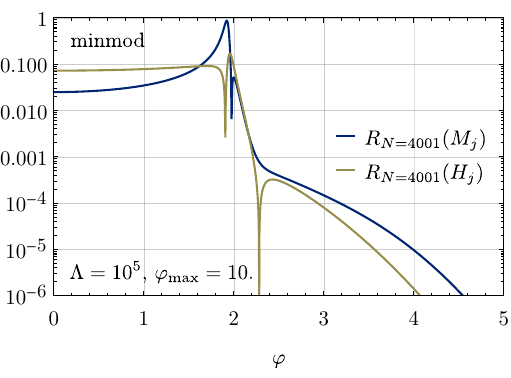}
			}

			\subfloat[For the MUSCL limiter.]{%
				\centering
				\includegraphics{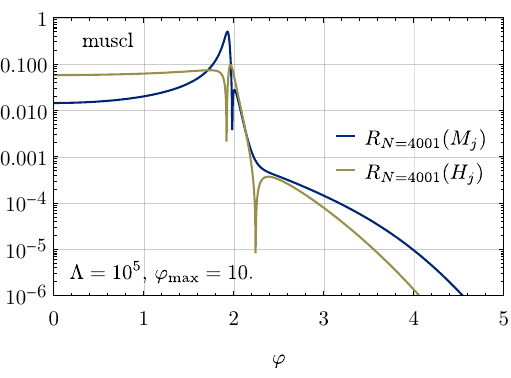}
			}

			\subfloat[For the superbee limiter.]{%
				\centering
				\includegraphics{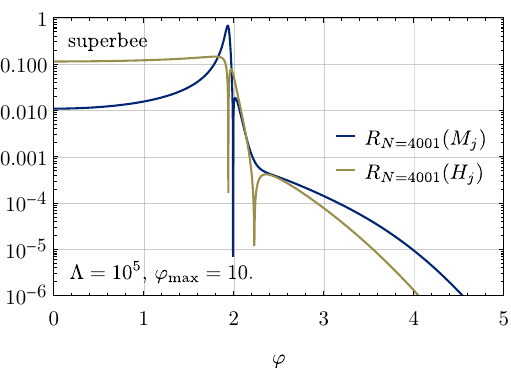}
			}
			\caption{Test 2:~Relative error~\eqref{eq:relative_error} of the numerical solution to the exact solution as a function of $\varphi$ for the initial conditions~\eqref{eq:test_U_first_order} and~\eqref{eq:test_H_first_order} for different limiters at a resolution of $N = 4001$ points with $\varphi_\mathrm{max} = 10$.}%
			\label{fig:test_2_relative_error}
		\end{figure}
	First, we turn to the relative error of the numerical solution to the exact solution as a function of $\varphi$ in \cref{fig:test_2_relative_error} at constant $\Delta x = 0.0025$.
	We find that the large-$\varphi$ numerical error is very small and the numerical solution is comparable to the solution from the numerical evaluation of the expectation values.
	At small field values, $|\varphi| < 2$, in between the smeared shocks at $\varphi \approx \pm 2$, we find rather good agreement with the exact solution and relative errors of approximately $1\%$.
	Only close to the shock, the relative error increases up to almost $100\%$.
	This might sound worrying, but one has to keep in mind that the shock is a nonanalytic structure, which is hard to resolve with a numerical scheme.
	Additionally, these seemingly large errors can be traced back to a minimal offset and difference in height of the shock position in the numeric solution compared to the exact solution, as can be seen in \cref{fig:test_2_RG_flow_M}.
	However, for very singular structures, a minimal offset in the position can lead to large relative errors, even though the numeric solution is very close to the exact solution and satisfying for practical purposes.

\begin{figure}%Test 2: Error scaling -- Fig. 9.
			\centering
			\subfloat[For the minmod limiter.]{%
				\includegraphics{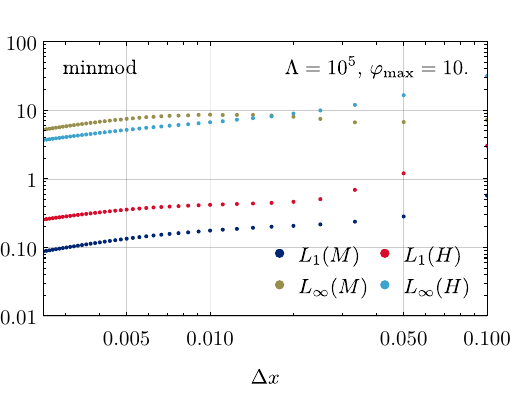}
			}

			\subfloat[For the MUSCL limiter.]{%
				\includegraphics{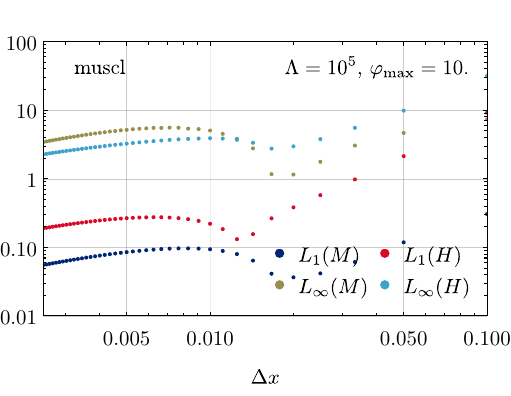}
			}

			\subfloat[For the superbee limiter.]{%
				\includegraphics{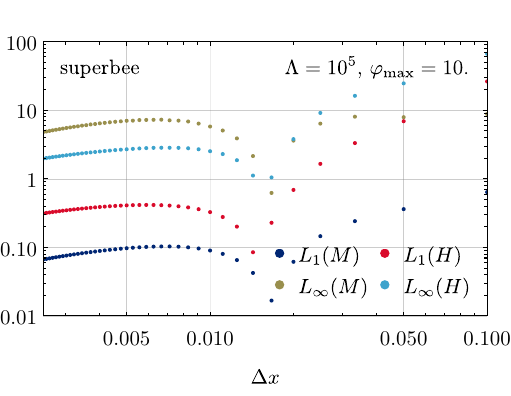}
			}
			\caption{Test 2:~$L^1$ and $L^\infty$ norm between the numerical and the exact solution on the computational domain as a function of the spatial resolution for the initial conditions~\eqref{eq:test_U_first_order} and~\eqref{eq:test_H_first_order} for different limiters.}%
			\label{fig:test_2_LNormScaling}
		\end{figure}

	Finally, we turn to the scaling of the error with the resolution in \cref{fig:test_2_LNormScaling}.
	Here, we find that for small resolution, thus large $\Delta x$, the error scaling is again polynomial.
	However, the explicit errors are rather large for all limiters -- especially the $L_\infty$ error.
	Still, this is expected, since it measures the maximum error over all grid points and for low resolutions, the shock is not resolved and its position is off.
	However, even for small $\Delta x$ we have already mentioned that the slight offset of the shock position can still lead to rather large relative errors.
	For the $L_1$ norm, we find that the error is overall smaller, because it is the absolute deviation averaged over all grid points.
	In general, we find for both error measures that for rather large $\Delta x$ they scale approximately like $\Delta x^2$.
	However, for small $\Delta x$ one even finds scalings that are worse than $\Delta x$, which seems to be related to the resolution of the shock structure.

	In general, we are satisfied with the performance of the nonconservative scheme for such a challenging test, which motivates us to further investigate the capabilities of the \gls{kthj} scheme.

\subsection{Test 3: A ``mild'' sign problem and ``shockingly'' hard dynamics}
	
	As indicated in the introduction, this work is not only about testing new numerical schemes for \gls{frg} flow equations but also about conceptual aspects of the Wetterich equation.
	Therefore, we now turn to a test case, which is not only extremely challenging for the numerics but also an interesting realization of a sign problem, nonanalyticities, phase transitions, and a prime example of Yang-Lee zeros on the level of \gls{rg} flows.
	To study this interesting dynamics and ``physical'' situation we propose the initial conditions  
		% \begin{align}
		% 	U ( \phi ) = \, & \phi^2 \, ,	\vdistance	\label{eq:test_initial_U_nonanalytic}
		% 	\\
		% 	H ( \phi ) = \, & 1 - \frac{5}{\cosh^2 ( \phi^2 - 4 )} \, .	\vdistance	\label{eq:test_initial_H_nonanalytic}
		% \end{align}
		\begin{align}
			U ( \phi ) = \, & \phi^2 \, ,	\vdistance	\label{eq:test5_initial_U_nonanalytic}
			\\
			H ( \phi ) = \, &
			\begin{cases}
				\tfrac{1}{2} \, \phi^2 - \tfrac{1}{18} \, \phi^4 + \tfrac{1}{720} \, \phi^6 \, ,	&	\text{if } | \phi | \leq 8 \, ,	\vdistance
				\\
				\frac{2528}{15} \, ,	&	\text{if } 8 \leq | \phi | \, .	\vdistance	\label{eq:test5_initial_H_nonanalytic}
			\end{cases}
		\end{align}
	Hence, $U(\phi)$ is again of the Gaussian type, while $H(\phi)$ is of polynomial type with two dips at $\phi \approx \pm 4.5$, where $H(\phi)$ even turns negative, see also the blue curves in \cref{fig:RGFlowSignFlow} (right column).
	The reason, why we choose $H(\phi)$ as a piecewise function with a cusp at $\phi = \pm 8$ is the same as explained below \cref{eq:test_H_first_order} and does only minimally affect the actual dynamics in the interesting region $|\phi| < 7$.
	It simply avoids dealing with the large-$\varphi$ boundary conditions in a more sophisticated way.
	
\subsubsection{Discussion from the path-integral perspective}
\label{sec:maxwell-construction}

	Before we turn to the discussion of this problem with the \gls{rg} flow equations, let us first analyze the exact reference solution on the level of the path integral, the (connected) correlation functions and the vertex functions in the presence of sources $J$, respectively.

\begin{figure*}
			\subfloat[\label{fig:test5mildsignW}$\mathcal{W} ( J )$.]{%
				\centering
				\includegraphics{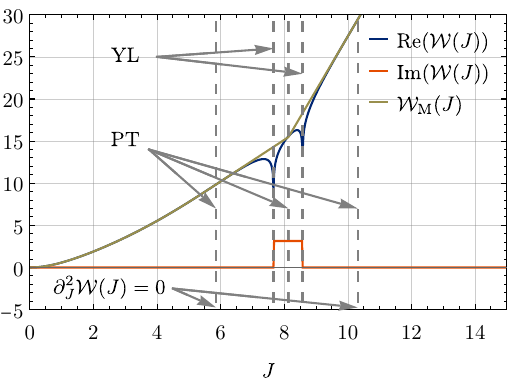}
			}\hfil
			\subfloat[\label{fig:test5mildsigndWdJ}$\partial_J \mathcal{W} ( J )$.]{%
				\centering
				\includegraphics{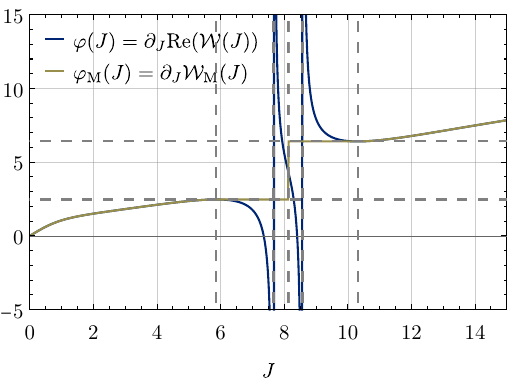}
			}

			\subfloat[\label{fig:test5mildsignddWddJ}$\partial_J^2 \mathcal{W} ( J )$.]{%
				\centering
				\includegraphics{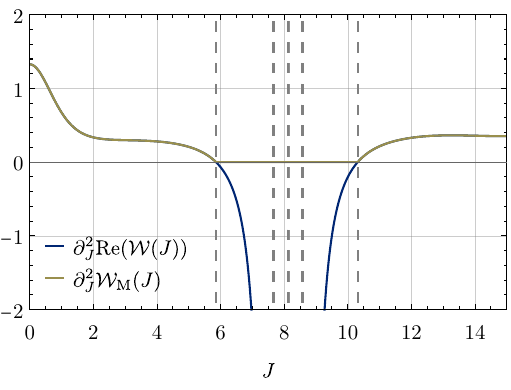}
			}\hfil
			\subfloat[\label{fig:test5mildsignObservables}$u ( \varphi ) = \partial_\varphi U ( \varphi )$, $M ( \varphi ) = \partial_\varphi^2 U ( \varphi )$, $H ( \varphi )$.]{%
				\centering
				\includegraphics{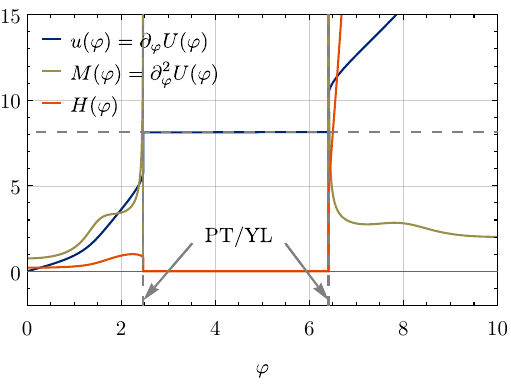}
			}
			\caption{Test 3:~Exact (connected) field-dependent correlation functions and vertex functions in the presence of sources $J$ for \cref{eq:test5_initial_U_nonanalytic,eq:test5_initial_H_nonanalytic}.}%
			\label{fig:test5mildsign}
		\end{figure*}

	To this end, we directly compute the Schwinger functional $\mathcal{W} ( J ) = \mathcal{W} ( J, 0, 0)$ and its first and second derivatives \gls{wrt}\ $J$ from \cref{eq:schwinger_functional,eq:partition_function} and plot the results in \cref{fig:test5mildsignW,fig:test5mildsigndWdJ,fig:test5mildsignddWddJ} in blue (and red).
	We immediately observe that the Schwinger functional $\mathcal{W}(J)$ is smooth and convex for small and large $|J|$, while there is a region between $J = \pm 5.8(6)$ and $J = \pm 10.3(1)$ where $\mathcal{W}(J)$ is nonconvex, which is also directly seen from $\partial_J \mathcal{W}(J)$ and $\partial_J^2 \mathcal{W}(J)$.
	In this region $\mathcal{W}(J)$ even has poles and gains an imaginary part.
	From a statistical physics perspective we identify these poles as Yang-Lee ($\mathrm{YL}$) zeros in the partition function \labelcref{eq:partition_function}, which are associated with phase transitions, see Refs.~\cite{Connelly:2020gwa,Rennecke:2022ohx,Johnson:2022cqv,Ihssen:2022xjv} for recent works in the \gls{frg} context.
	Here, these phase transitions are triggered by cranking up the external source $J$ and should realize as kinks and jumps in the bosonic expectation values.
	Hence, at this point we already conclude that the fermionic degrees of freedom can indeed cause nonanalytic structures and phase transitions even in zero spacetime dimensions, which is not possible in purely bosonic systems.

	Next, we have to ask the question of how these structures in $\mathcal{W}(J)$ are reflected in $\Gamma ( \varphi, \tilde{\vartheta}, \vartheta )$, the vertex functions, and consequently the \gls{rg} flow of $M(t,\varphi)$ and $H(t,\varphi)$.
	We know, that the result  of the Legendre transformation~\eqref{eq:effective_action} is a convex function, which after Legendre-backtransformation again yields a convex $\mathcal{W}$, see Refs.~\cite{Fujimoto:1982tc,Wipf:2013vp} for details.
	Let us therefore find the convex $\mathcal{W}(J)$ directly via a Maxwell construction.
	The corresponding results are depicted in ochre/gold in \cref{fig:test5mildsignW,fig:test5mildsigndWdJ,fig:test5mildsignddWddJ}.
	We start with the second derivative, see \cref{fig:test5mildsignddWddJ}, and argue that it should never turn negative in order to obtain a convex $\mathcal{W}(J)$.
	Hence, the region between the zeros in $\partial_J^2 \mathcal{W}(J)$ is replaced by $\partial_J^2 W_\mathrm{M} ( J ) = 0$.
	On the level of the slope, $\partial_J \mathcal{W} ( J )$, see \cref{fig:test5mildsigndWdJ}, this corresponds to a constant slope, where we simply take the value of $\partial_J \mathcal{W} ( J )$ at $J = \pm 5.8(6)$ and $J = \pm 10.3(1)$ and extrapolate (integrate $\partial_J^2 \mathcal{W}_\mathrm{M} ( J )$) to the left/right.
	The points $J = \pm 8.1(3)$, where $\partial_J \mathcal{W}_\mathrm{M} ( J )$ jumps are obtained via $\mathcal{W}_\mathrm{M} (J)$ itself:
	It is simply the point, where the two straight lines intersect.
	In summary, we find the Maxwell construction $\mathcal{W}_\mathrm{M} ( J )$, which contains (for positive $J$) three kinks and two straight lines.
	The latter are the phase-coexistence regions and we denote the kinks as phase transitions ($\mathrm{PT}$) points.

	From $\mathcal{W}_\mathrm{M} (J)$ it is now straight forward to calculate the Legendre transformation and to extract $U(\varphi)$, $u ( \varphi ) = \partial_\varphi U ( \varphi )$, $M(\varphi) = \partial_\varphi^2 U (\varphi)$ and $H(\varphi)$ as explained in \cref{sec:field-dependent-vertex-functions}.
	These are the \gls{ir} reference solutions for the \gls{rg} flows with \cref{eq:test5_initial_U_nonanalytic,eq:test5_initial_H_nonanalytic} as initial conditions.
	Inspecting \cref{fig:test5mildsignObservables}, where we plot $u(\varphi)$, $M(\varphi)$, and $H(\varphi)$, we immediately recognize that this is prospective \gls{ir} solution will present a very challenging problem on the level of the \gls{frg} and \gls{pde} numerics:
	The \gls{ir} solution has poles as well as a flat region in between, which is extremely hard to resolve and to flow into.\footnote{Similar highly challenging problems are used as tests for fluid-dynamic numerics.
	For example, the Sod's shock tube problems~\cite{Sod1978Apr} with suitable initial conditions can produce rarefaction waves and approach vacuum solutions.}
	As we discuss in the next paragraphs, we find that on the level of the \glspl{pde} this is caused by the $H(t,\varphi)$ being negative for some $\varphi$.

	In fact, negative $H(\phi)$ is exactly what we denoted as a sign problem in \cref{sec:the_model}.
	On the level of the (path) integral~\eqref{eq:partition_function}, we have argued that a (partially) negative $H(\phi)$ leads to (in parts) negative or oscillating probability distributions, \cf{} \cref{eq:expvalf}.
	Hence, already before solving the actual \gls{rg} flows (numerically), we claim that some sign problems in \glspl{qft} might show up as challenging nonanalyticities on the level of the \gls{frg} flow equations.
	Similar arguments were already put forward in Refs.~\cite{Stoll:2021ori,Grossi:2021ksl,Ihssen:2023xlp,Ihssen:2022xkr,Ihssen:2022xjv}, where it was seen that a chemical potential causes external shocks in \gls{rg} flows.

\subsubsection{Discussion from the FRG perspective}

\begin{figure*}
			\subfloat[\label{fig:RGFlowSignFlowMminmod}%
			$M ( t, \phi)$ for the minmod limiter.]{%
				\centering
				\includegraphics[width=\columnwidth]{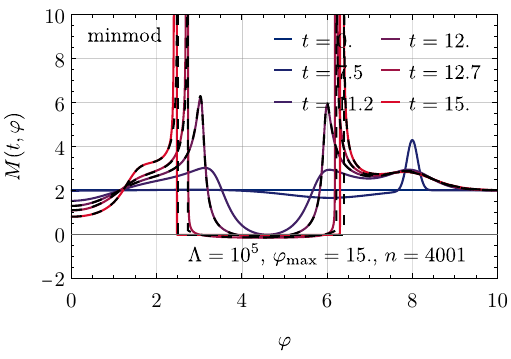}
			}\hfill
			\subfloat[\label{fig:RGFlowSignFlowHminmod}%
			$H ( t, \phi)$ for the minmod limiter.]{%
				\centering
				\includegraphics[width=\columnwidth]{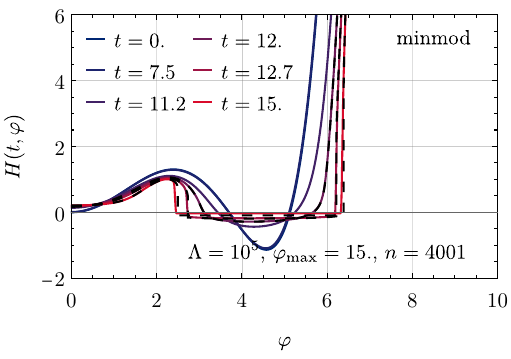}
			}

			\subfloat[\label{fig:RGFlowSignFlowMmuscl}%
			$M ( t, \phi)$ for the MUSCL limiter.]{%
				\centering
				\includegraphics[width=\columnwidth]{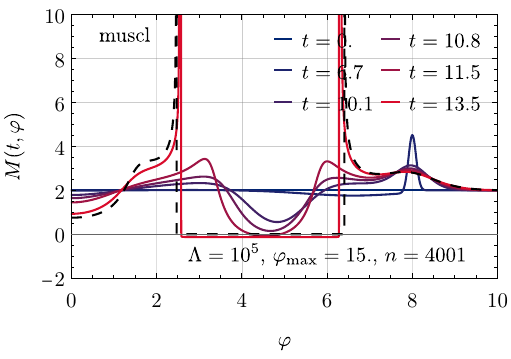}
			}\hfill
			\subfloat[\label{fig:RGFlowSignFlowHmuscl}%
			$H ( t, \phi)$ for the MUSCL limiter.]{%
				\centering
				\includegraphics[width=\columnwidth]{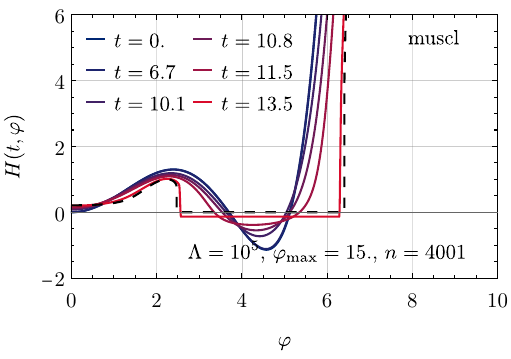}
			}

			\subfloat[\label{fig:RGFlowSignFlowMsuperbee}%
			$M ( t, \phi)$ for the Superbee limiter.]{%
				\centering
				\includegraphics[width=\columnwidth]{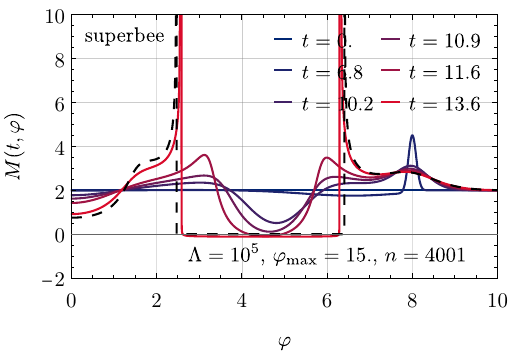}
			}\hfill
			\subfloat[\label{fig:RGFlowSignFlowHsuperbee}%
			$H ( t, \phi)$ for the Superbee limiter.]{%
				\centering
				\includegraphics[width=\columnwidth]{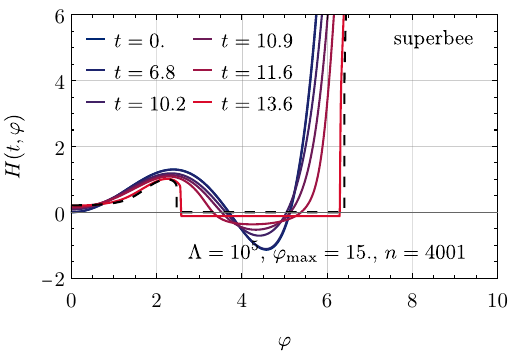}
			}
			\caption{\label{fig:RGFlowSignFlow}
				Test 3:~\gls{rg} flow with the initial conditions~\eqref{eq:test5_initial_U_nonanalytic} and~\eqref{eq:test5_initial_H_nonanalytic} for different limiters.
				The analytic solution for the last three \gls{rg} times in \cref{fig:RGFlowSignFlowMminmod,fig:RGFlowSignFlowHminmod} as well as the \gls{ir} ($t \to \infty$) reference solution in \cref{fig:RGFlowSignFlowMmuscl,fig:RGFlowSignFlowHmuscl,fig:RGFlowSignFlowMsuperbee,fig:RGFlowSignFlowHsuperbee} are shown as black dashed curves.}
		\end{figure*}

	Next, let us analyze the same problem within the framework of the \gls{frg} and the \gls{rg} flow equations~\eqref{eq:flow_M_hj} and~\eqref{eq:flow_H_hj}.
	The corresponding \gls{rg} flows for $M(t,\varphi)$ and $H(t,\varphi)$ are depicted in \cref{fig:RGFlowSignFlow} for the three different limiters.
	However, for the qualitative part of the discussion we simply focus on the plots \cref{fig:RGFlowSignFlowMmuscl,fig:RGFlowSignFlowHmuscl}.
	Since the \gls{uv} potential $U ( t, \varphi )$ is Gaussian, the \gls{rg} flow for $M ( t, \varphi )$ starts with a constant function $M ( t = 0, \varphi ) = 2$, while the flow of $H ( t, \varphi )$ starts as a polynomial with two dips below zero (blue curves).
	During the flow, we observe the following:
	First, the cusp at $\varphi = \pm 8$ in $H ( t, \varphi)$ causes a small bump in $M ( t, \varphi )$ at the same position.
	Both, the cusp and this bump are still visible in the final solution even though they are smeared out and do not significantly influence the other dynamics.
	The main dynamics takes place at $\varphi$-values around the region, where $H ( t, \varphi )$ used to be negative in the \gls{uv}.
	Here, the propagator $1/(r + H)$ is closer to its pole than in other regions, which causes a strong response in the flow of both, $M ( t, \varphi )$ and $H ( t, \varphi )$.
	In fact, $M(t, \varphi)$ is lowered in this region and even turns negative for some $\varphi$.
	In turn, this also causes the propagator $1/(r + M)$ to be closer to its pole, fostering even faster dynamics.
	In particular, we observe some strong advective dynamics, which pushes $M ( t, \varphi )$ and $H ( t, \varphi )$ to the left and right of the negative region, while the negative region itself flattens and approaches zero from below.
	Thereby, $M ( t, \varphi )$ develops two shocks/poles and $H ( t, \varphi )$ develops two cusps, which we immediately identify as the phase transition points from the path-integral perspective, compare with the black-dashed curves and \cref{fig:test5mildsignObservables}.
	Before we turn to some comments on the quantitative performance, let us mention the following interesting aspects:
	As becomes clear from the final \gls{rg} times, which correspond to regulator values of $r \approx 0.1$, we were not able to flow arbitrarily deep into the \gls{ir}.
	The reason for this is that the propagators $1/(r + M)$ and $1/(r + H)$ approach their poles for $r \to 0$ in the flat region and the \gls{pde} system turns into an extremely stiff problem.
	Even with the advanced implicit time-stepping algorithm, which is still stable, the number of time-steps simply explodes and the computation time becomes unfeasible, which is why we stopped after 600k time steps.
	The reason, why it was possible to flow deeper to the \gls{ir} with the minmod limiter, see \cref{fig:RGFlowSignFlowMminmod,fig:RGFlowSignFlowHminmod}, is that it leads to more diffusive flows.
	By smearing out the poles, $M(t,\varphi)$ and $H(t,\varphi)$ do turn slightly positive in the flat region instead of correctly approaching zero from below, as it is the case for the MUSCL and Superbee limiters.
	Still, as can be seen from the black dashed curves in \cref{fig:RGFlowSignFlowMminmod,fig:RGFlowSignFlowHminmod}, which correspond to the exact reference solutions at the respective latest three \gls{rg} times obtained from the formulae form \cref{sec:field-dependent-vertex-functions}, the \gls{frg} results almost perfectly agree with the exact ones.\footnote{We could have presented the same curves for the MUSCL and Superbee limiters as well as for earlier \gls{rg} times.
	However, for earlier \gls{rg} times we again face the problem of excessivily large computational domains in the source field $J$ to obtain $\varphi ( J )$ in the interesting region.
	Additionally, for late \gls{rg} times, after formation of the pole structures, one has to perform the Maxwell constructuion from \cref{sec:maxwell-construction} for each \gls{rg} step.
	Since new insights are limited, we abstained from both endeavour.}

		\begin{figure*}
			\subfloat[$M ( t, \phi)$ for the minmod limiter.]{%
				\centering
				\includegraphics[width=\columnwidth]{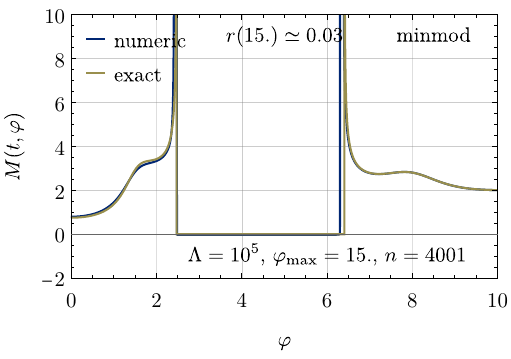}
			}\hfill
			\subfloat[$H ( t, \phi)$ for the minmod limiter.]{%
				\centering
				\includegraphics[width=\columnwidth]{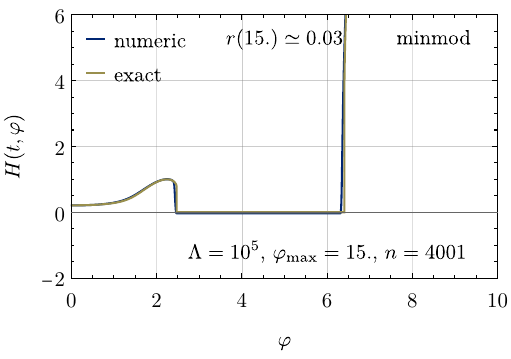}
			}

			\subfloat[$M ( t, \phi)$ for the MUSCL limiter.]{%
				\centering
				\includegraphics[width=\columnwidth]{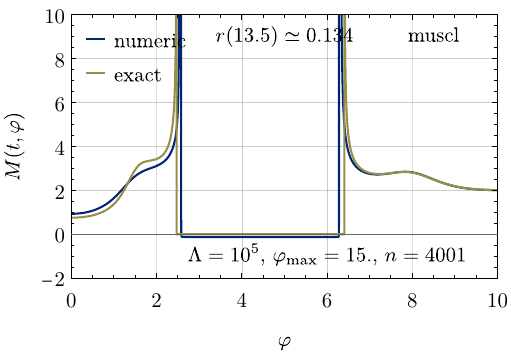}
			}\hfill
			\subfloat[$H ( t, \phi)$ for the MUSCL limiter.]{%
				\centering
				\includegraphics[width=\columnwidth]{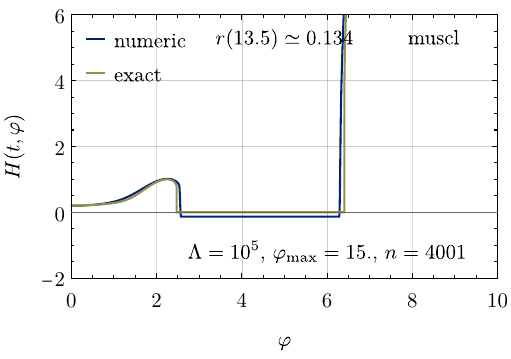}
			}

			\subfloat[$M ( t, \phi)$ for the Superbee limiter.]{%
				\centering
				\includegraphics[width=\columnwidth]{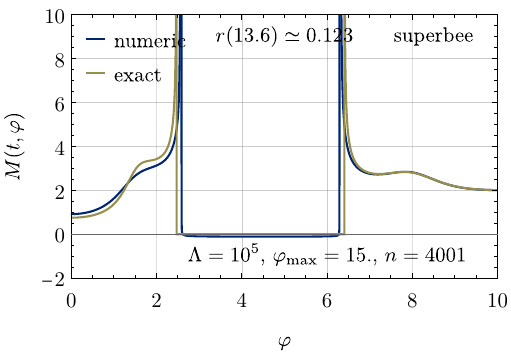}
			}\hfill
			\subfloat[$H ( t, \phi)$ for the Superbee limiter.]{%
				\centering
				\includegraphics[width=\columnwidth]{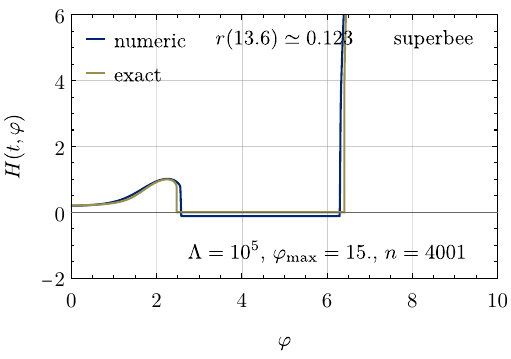}
			}
			\caption{\label{fig:RGFlowSignIR}
				Test 3:~Comparison of the exact \gls{ir} reference solution ($t \to \infty$) from \cref{fig:test5mildsignObservables} (gold) to the numerical solution at finite $t$ (blue) from the \gls{rg} flow with the initial conditions~\eqref{eq:test5_initial_U_nonanalytic} and~\eqref{eq:test5_initial_H_nonanalytic} for different limiters.}
		\end{figure*}
	
	In fact, this general behavior of $M(t,\varphi)$ approaching zero from below is well-known from higher-dimensional \gls{rg} problems in their symmetry broken regimes.
	In the present case, however, we can directly see that this is not only related to condensation, but also to the presence of negative $H ( t, \varphi )$ and the associated sign problem.
	We therefore conclude that sign problems on the level of the path integral may indeed show up in \gls{rg} flows as nonanalytic structures and may turn into stiff \gls{pde} problems.
	This definitely questions the common lore that \gls{frg} approaches are not affected by sign problems and calls for further investigations.
	On a conceptual level sign problems might be less apparent in the \gls{frg} but on a practical/numerical level they might present severe challenges.

	Finally, we comment on the quantitative performance of the \gls{kthj} scheme for this challenging problem.
	As mentioned earlier and as it is visible in \cref{fig:RGFlowSignFlow} it does not really make sense to show relative or $L_1$ and $L_\infty$ errors.
	Even though the scheme is in general able to resolve and propagate the shocks and cusps without spurious oscillations, there are still small offsets in their positions.
	In addition, $M(t,\varphi)$ and $H(t,\varphi)$ will never exactly reach the final solution, since we cannot flow all the way to $r = 0$.
	Especially in the flat region, this would cause large deviations.
	Hence, we abstain from a deeper quantitative analysis and are satisfied with the qualitative performance of the scheme for this extremely challenging test case.
	To this end, we solely compare the solutions at the latest \gls{rg} time to the exact reference solution in \cref{fig:RGFlowSignIR} for different limiters.

	Equipped with the insights from our zero-dimensional toy model, we next turn to some higher-dimensional examples.

\section{Applications to higher dimensional models}%
\label{sec:higher_dim_exp}

	In this section, we turn to some explicit applications of our developments to higher-dimensional systems.
	Note that within this work we are not aiming at deeper insights into the physics of these systems, but rather at demonstrating the performance of the \gls{kthj} scheme and the new formulation of the flow equations for more realistic systems.
	We consider the following two setups:
	\begin{enumerate}
		\item First, we discuss a $\mathbb{Z}_2$-symmetric system of a single real scalar field in three Euclidean dimensions.
		In particular, we study the \gls{frg} flow of the effective potential and the wave-function renormalization with full field dependence.
		Thereby, we use two different types of regulators.

		\item Second, we consider another $\mathbb{Z}_2$-symmetric system.
		Here, however, we analyze the \gls{rg} flow of the effective potential and field-dependent Yukawa coupling in the Gross-Neveu-Yukawa model in $1+1$ dimensions at nonzero temperature and chemical potential.
	\end{enumerate}

\subsection{Field-dependent wave-function renormalization in the \texorpdfstring{$\mathbb{Z}_2$}{Z2}-symmetric model}%
\label{sec:field-dependent-wave-function-renormalization}

	Let us turn to our first example -- the $\mathbb{Z}_2$-symmetric \gls{frg} flow of the effective potential and the wave-function renormalization in three Euclidean dimensions.
	In terms of a truncation, the \gls{eaa} for the model is given by
		\begin{align}
			\bar{\Gamma}_t [ \varphi ]= \int \dd^d x \, \big[ U_t [ \varphi ] + Z_t [ \varphi ] \, \tfrac{1}{2} \, ( \partial_\mu \varphi )^2 + \mathcal{O} ( \partial^4 ) \big] \, ,	\label{eq:effective_average_action_z2_three_dim}
		\end{align}
	with the $\mathbb{Z}_2$-symmetric scale-dependent effective potential and wave-function renormalization,
		\begin{align}
			&	 U_t [ \varphi ] = U_t [ - \varphi ] \, ,	&&	Z_t [ \varphi ] = Z_t [ - \varphi ] \, .
		\end{align}
	We ignore terms, which are of higher order in the derivatives of $\varphi (x)$, hence nonquadratic momentum dependences.
	Furthermore, we define the \gls{rg} time
		\begin{align}
			&	t = - \ln \big( \tfrac{k}{\Lambda} \big) \, ,	&	t \in [ 0, \infty ) \, ,
		\end{align}
	where $\Lambda$ is the ultraviolet cutoff and $k$ is the renormalization group scale, which are both of dimension of energy.
	For the moment, we keep $d \geq 1$ arbitrary and only specify it later as $d = 3$.

\subsubsection{Flow equations}
	
	The corresponding \gls{rg} flow equations for the second derivative of the effective potential $M ( t, \sigma ) = \partial_\varphi^2 U_t [ \varphi ] \big|_{\varphi ( x ) = \sigma}$ and $Z ( t, \sigma ) = Z_t [ \sigma ]$, where $\sigma$ is some constant background field configuration are both directly extracted from the flow equation~\eqref{eq:flow_2point_z2_three_dim} of the full momentum-dependent bosonic two-point function.
	The latter is directly derived from the Wetterich equation \labelcref{eq:wetterich_equation} by taking two functional derivatives with respect to the field $\varphi$, \cf{} \cref{eq:wetterich_equation_Gii}.
	In momentum space, we again recover a \gls{hj}-type flow equation, \cf{} \cref{sec:wetterich},
	\begin{widetext}
			\begin{align}
				& \partial_t \bar{\Gamma}^{(2)}_{t} ( q_2, q_1 ) =	\Vdistance	\label{eq:flow_2point_z2_three_dim}
				\\
				= \, & - \int_{p_1, \ldots, p_4} \big( \tfrac{1}{2} \, \partial_t R_t ( p_1, p_2 ) \big) \, G_t ( p_2, p_3 ) \, \bar{\Gamma}_{t}^{(4)} ( p_3, p_4, q_1, q_2 ) \, G_t ( p_4, p_1 ) +	\Vdistance	\nonumber
				\\
				& + 2 \int_{p_1, \ldots, p_6} \big( \tfrac{1}{2} \, \partial_t R_t ( p_1, p_2 ) \big) \, G_t ( p_2, p_3 ) \, \bar{\Gamma}_{t}^{(3)} ( p_3, p_4, q_1 ) \, G_t ( p_4, p_5 ) \, \bar{\Gamma}_{t}^{(3)} ( p_5, p_6, q_2 ) \, G_t ( p_6, p_1 ) \, ,	\Vdistance	\nonumber
			\end{align}
	\end{widetext}
	Here, we use the abbreviation
		\begin{align}
			\int_{p_1, \ldots, p_n} = \int_{- \infty}^{\infty} \frac{\dd^d p_1}{( 2 \uppi )^d} \ldots  \frac{\dd^d p_n}{( 2 \uppi )^d}
		\end{align}
	and introduced the regulator
		\begin{align}
			R_t ( p_1, p_2 ) = \, & ( 2 \uppi )^d \, \delta^{(d)} ( p_1 + p_2 ) \, R ( t, p_1^2 ) \, ,	\vdistance
			\\
			R ( t, p^2 ) = \, & p^2 \, r_t ( p^2 ) \, ,	\vdistance
		\end{align}
	where $r_t ( p^2 )$ is a suitable regulator shape function.
	In particular, we use the Callan-Symanzik type regulator
		\begin{align}
			r_t ( p^2 ) = \tfrac{k^2}{p^2} \, ,	\label{eq:callan_symanzik_regulator}
		\end{align}
	and the Litim/flat regulator,
		\begin{align}
			r_t ( p^2 ) = \big( \tfrac{k^2}{p^2} - 1 \big) \, \Theta \big( \tfrac{k^2}{p^2} - 1 \big) \, ,	\label{eq:litim_regulator}
		\end{align}
	both without including any dependence on the field, the couplings, or the wave-function renormalization.
	Let us also note that due to momentum conservation,
		\begin{align}
			\bar{\Gamma}^{(2)}_{t} ( q_2, q_1 ) = ( 2 \uppi )^d \, \delta^{(d)} ( q_2 + q_1 ) \, \bar{\Gamma}^{(2)} ( t, q_1 ) \, .
		\end{align}
	Inserting the ansatz \labelcref{eq:effective_average_action_z2_three_dim} in \cref{eq:flow_2point_z2_three_dim}, evaluating the equation on a constant background field configuration $\varphi ( x ) = \sigma$, as well as evaluating the integrals, we find the flow equation for $\bar{\Gamma}^{(2)} ( t, q_1 )$.
	By either evaluating this at $q_1 = 0$ or taking a derivative with respect to $q_1^2$ and only afterwards setting $q_1 = 0$, we find the flow equations for $M ( t, \sigma )$ and $Z ( t, \sigma )$, respectively.
	The symbolic computation is a nontrivial task with some subtle details, which we elaborate on in Appendix~\ref{app:flow_equations_z2_three_dimensions}.
	The final expressions for the flow equations using the regulator shape functions \labelcref{eq:callan_symanzik_regulator,eq:litim_regulator} are rather lengthy and are presented explicitly in Appendix~\ref{app:flow_equations_z2_three_dimensions}, see also Refs.~\cite{Connelly:2020gwa,Rennecke:2022ohx,Johnson:2022cqv}.	
	However, note, that overall, we find the following structure of the flow equations for both regulators,
		\begin{align}
			& \partial_t M + \#_1 \, ( M^\prime )^2 + \#_2 \, M^\prime \, Z^\prime + \#_3 \, ( Z^\prime )^2 =	\vdistance
			\\
			= \, & \#_4 \, M^{\prime \prime} + \#_5 \, Z^{\prime \prime} \, ,	\vdistance	\nonumber
		\end{align}
	and
		\begin{align}
			& \partial_t Z + \tilde{\#}_1 \, ( M^\prime )^2 +  \tilde{\#}_2 \, M^\prime \, Z^\prime +  \tilde{\#}_3 \, ( Z^\prime )^2 =	\vdistance
			\\
			= \, & \tilde{\#}_5 \, Z^{\prime \prime} \, ,	\vdistance	\nonumber
		\end{align}
	where $\#_i$ and $\tilde{\#}_i$ are coefficients, which depend on $t$, $M$, and $Z$.
	Consequently, the structure is -- as already anticipated -- similar to the one of \cref{eq:flow_M_hj,eq:flow_H_hj} and the \gls{kthj} scheme is applicable.
	However, analogously to the zero-dimensional case, we solely solve the flow equation for $Z ( t, \sigma )$ via the \gls{hj} approach, while we treat the entire flow equation for $M ( t, \sigma )$ as a conservation law, see \cref{eq:flow_M_callan_symanzik_conservative,eq:flow_M_litim_conservative}, and use the conservative \gls{kt} scheme \cref{eq:KT_conservative_term}.

\subsubsection{Discussion}

	Next, let us turn to the discussion of the \gls{rg} flows and \gls{ir} results for $M ( t, \sigma )$ and $Z ( t, \sigma )$.
	To this end, we always fix the scales of the problem by setting $\Lambda = 1$ and consider the \gls{uv} initial conditions of the type
		\begin{align}
			U ( t = 0, \sigma ) = \, & \tfrac{1}{4} \, \Big( \tfrac{\sigma^2}{2} - \tfrac{\sigma_0^2}{2} \Big)^2 \, ,
		\end{align}
	where we choose different $\sigma_0$ in order to either end the \gls{rg} flow in the symmetric or symmetry broken phase.
	Furthermore, we set $Z ( t = 0, \sigma ) = 1$.
	For each regulator -- Callan-Symanzik and Litim -- we do the same analysis:
	We take two different values for $\sigma_0$, one, which ends the flow in the symmetric phase and one, which ends the flow in the symmetry broken phase, close to the phase transition.
	Then we compute the \gls{rg} flows of $M ( t, \sigma )$ and $Z ( t, \sigma )$ within two different truncations:
	First, we use the \gls{lpa} truncation, where we keep $Z ( t, \sigma ) = 1$ fixed on the \gls{rhs} of the flow equations.
	Still, we compute the flow of $Z ( t, \sigma )$ evaluated on the \gls{lpa} solution for $M ( t, \sigma )$ without feeding it back on the \gls{rhs}\ of the flow equations.
	Second, we use the \gls{lpap} truncation, where we take into account the full flow of $Z ( t, \sigma )$ on the \gls{rhs} of the flow equation for $M ( t, \sigma )$ and $Z ( t, \sigma )$ itself.
	This allows us to study the impact of a dynamic field-dependent wave-function renormalization.
	
\begin{figure}
			\subfloat[$u ( t, \sigma ) = \partial_\sigma U ( t, \sigma )$]{%
				\centering
				\includegraphics[width=\columnwidth]{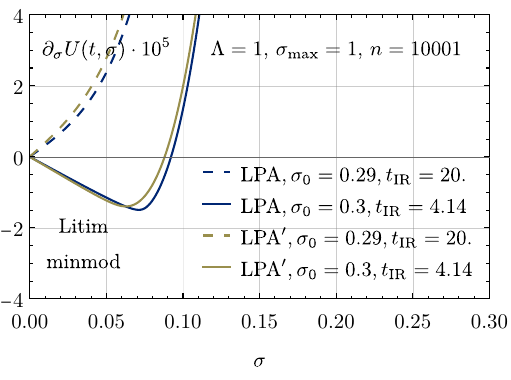}
			}

			\subfloat[$M ( t, \sigma ) = \partial_\sigma^2 U ( t, \sigma )$]{%
				\centering
				\includegraphics[width=\columnwidth]{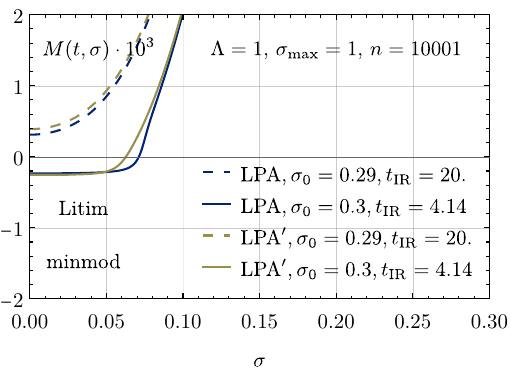}
			}

			\subfloat[$Z ( t, \sigma )$]{%
				\centering
				\includegraphics[width=\columnwidth]{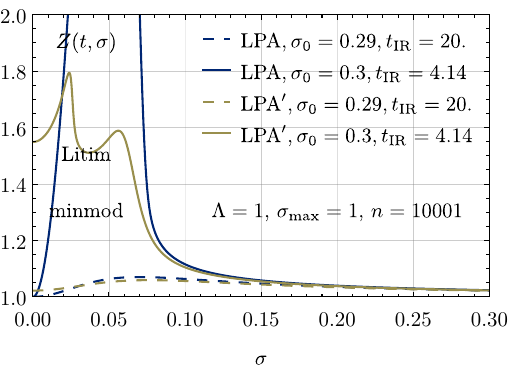}
			}
			\caption{\label{fig:RGFlowLitim}%
				\gls{ir} results of \gls{rg} flows of $\partial_\sigma U(t, \sigma)$, $M(t, \sigma) = \partial_\sigma^2 U(t, \sigma)$, and $Z(t, \sigma)$ with the Litim regulator into the symmetric and symmetry-broken regime in \gls{lpa} and \gls{lpap}.}
		\end{figure}

\begin{figure*}
			\subfloat[$M ( t, \sigma ) = \partial_\sigma^2 U ( t, \sigma )$]{%
				\centering
				\includegraphics[width=\columnwidth]{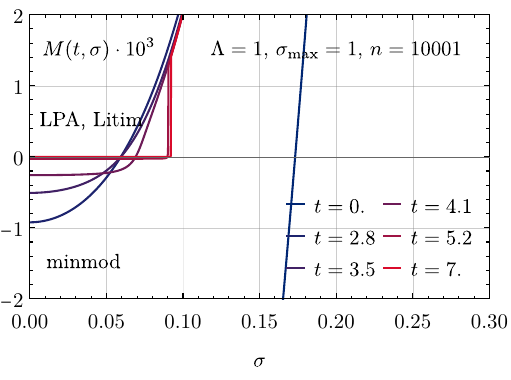}
			}\hfill
			\subfloat[$M ( t, \sigma ) = \partial_\sigma^2 U ( t, \sigma )$]{%
				\centering
				\includegraphics[width=\columnwidth]{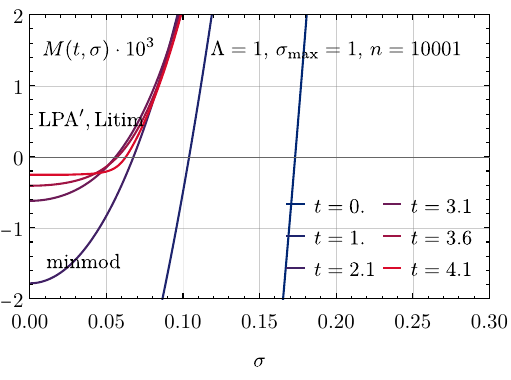}
			}

			\subfloat[$Z ( t, \sigma )$]{%
				\centering
				\includegraphics[width=\columnwidth]{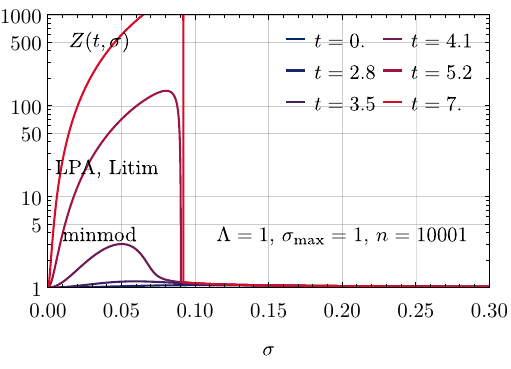}
			}\hfill
			\subfloat[$Z ( t, \sigma )$]{%
				\centering
				\includegraphics[width=\columnwidth]{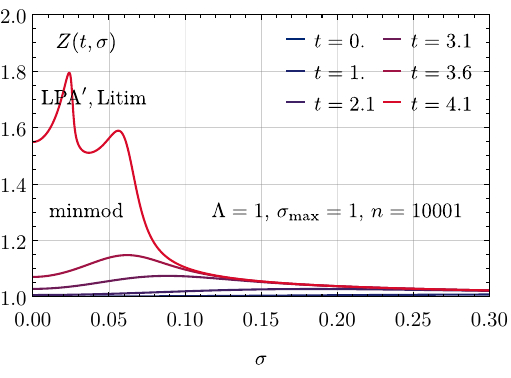}
			}
			\caption{\label{fig:RGFlowLitimFlows}%
				\gls{rg} flows of $M(t, \sigma) = \partial_\sigma^2 U(t, \sigma)$ and $Z(t, \sigma)$ with the Litim regulator in the symmetry-broken regime in \gls{lpa} and \gls{lpap}.}
		\end{figure*}

	Let us begin the discussion with the results for the Litim regulator, which are shown in \cref{fig:RGFlowLitim,fig:RGFlowLitimFlows}.
	First, we discuss the \gls{ir} results shown in \cref{fig:RGFlowLitim}.
	If the flow ends in the symmetric phase, one can flow arbitrarily close to $k = 0$.
	We simply stop the flow at $t = 20$, since no significant changes have occurred for some time beforehand.
	In this case, we find a smooth convex effective potential with a single minimum at $\sigma = 0$ and a positive mass function $M ( t, \sigma )$.
	The wave-function renormalization $Z ( t, \sigma )$ is close to one and only shows a mild field dependence.
	Deviations between \gls{lpa} and \gls{lpap} are only marginal.
	Second, we consider the flow into the symmetry-broken phase.
	Here, the system turns stiff at $t = 4.14$ in \gls{lpap} and while it is still possible to flow further to the \gls{ir} in \gls{lpa}, see \cref{fig:RGFlowLitim}, where we stopped the flow at $t = 7$.
	However, comparing the \gls{ir} results of both truncations in the broken phase at the same $t_\mathrm{IR} = 4.14$, we find significant deviations.
	In particular, the position of the \gls{ir} minimum of the effective potential is moved to slightly smaller field values.
	Even more important is the deviation in the profile of the wave-function renormalization.
	In \gls{lpa}, $Z ( t, \sigma )$ sticks to $Z ( t, 0 ) = 1$ but shows a strong peak in the flat region of the potential, while it falls off very fast at field-values larger than the minimum of the potential.
	In contrast, in \gls{lpap} it develops a nontrivial shape around $\sigma = 0$, which we expect to turn into a fully-fledged peak in the \gls{ir} limit in even better truncations.
	The peak close to the minimum is still visible but much smaller than in \gls{lpa}.
	Still, it is interesting, that the large-field behavior is almost the same in both truncations and especially at the physical point -- the minimum itself -- the effect of a dynamic wave-function renormalization is only mild.
	
	In order to get a better understanding of the actual dynamics, we show the full \gls{rg} flows of $M ( t, \sigma )$ and $Z ( t, \sigma )$ in \cref{fig:RGFlowLitimFlows} in the symmetry-broken regime.
	Here, it should also become clear that the reason for stopping the \gls{rg} flow at some finite $t$ is not related to the numeric scheme but rather to the stiffness of the \gls{pde} problem itself.
	It is even unclear to the authors whether the flow can be extended to arbitrarily large $t$ in principle within a truncation or if the $t \to \infty$ limit is only accessible for the untruncated Wetterich equation.
	
\begin{figure}
			\subfloat[$u ( t, \sigma ) = \partial_\sigma U ( t, \sigma )$]{%
				\centering
				\includegraphics[width=\columnwidth]{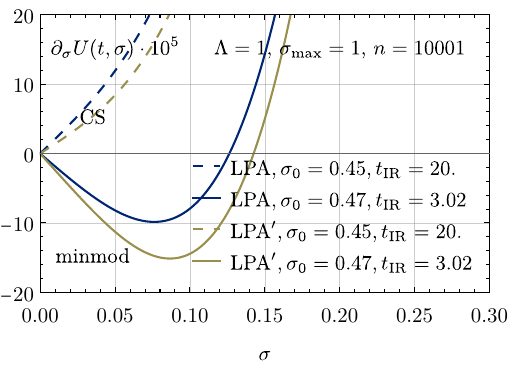}
			}

			\subfloat[$M ( t, \sigma ) = \partial_\sigma^2 U ( t, \sigma )$]{%
				\centering
				\includegraphics[width=\columnwidth]{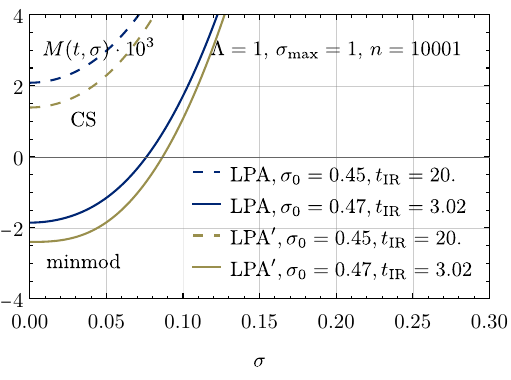}
			}

			\subfloat[$Z ( t, \sigma )$]{%
				\centering
				\includegraphics[width=\columnwidth]{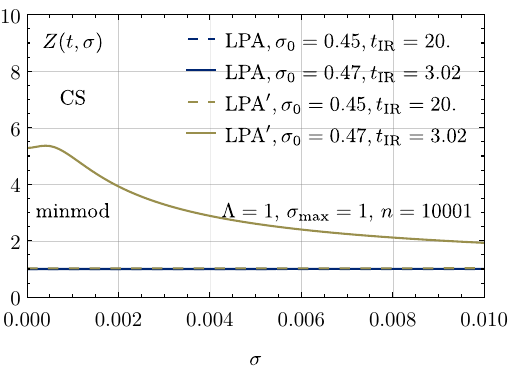}
			}
			\caption{\label{fig:RGFlowCS}%
				\gls{ir} results of \gls{rg} flows of $\partial_\sigma U(t, \sigma)$, $M(t, \sigma) = \partial_\sigma^2 U(t, \sigma)$, and $Z(t, \sigma)$ with the Callan-Symanzik regulator into the symmetric and symmetry-broken regime in \gls{lpa} and \gls{lpap}.}
		\end{figure}

\begin{figure*}
			\subfloat[$M ( t, \sigma ) = \partial_\sigma^2 U ( t, \sigma )$]{%
				\centering
				\includegraphics[width=\columnwidth]{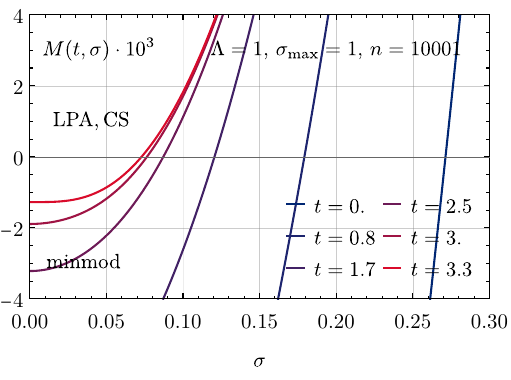}
			}\hfill
			\subfloat[$M ( t, \sigma ) = \partial_\sigma^2 U ( t, \sigma )$]{%
				\centering
				\includegraphics[width=\columnwidth]{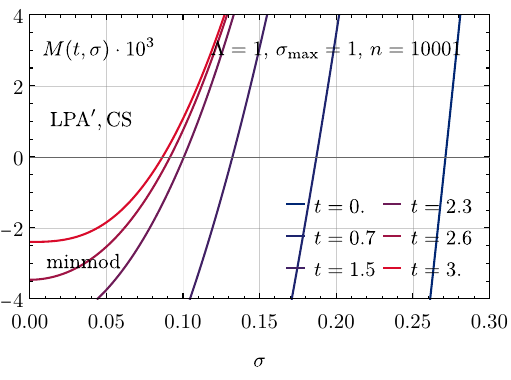}
			}

			\subfloat[$Z ( t, \sigma )$]{%
				\centering
				\includegraphics[width=\columnwidth]{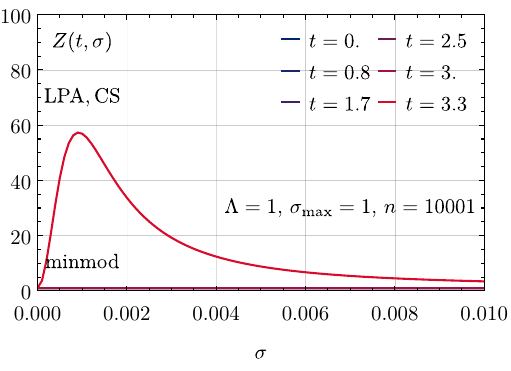}
			}\hfill
			\subfloat[$Z ( t, \sigma )$]{%
				\centering
				\includegraphics[width=\columnwidth]{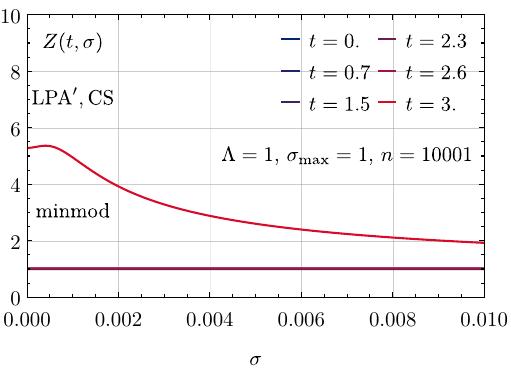}
			}
			\caption{\label{fig:RGFlowCSFlows}%
				\gls{rg} flows of $M(t, \sigma) = \partial_\sigma^2 U(t, \sigma)$ and $Z(t, \sigma)$ with the Callan-Symanzik regulator in the symmetry-broken regime in \gls{lpa} and \gls{lpap}.}
		\end{figure*}

	Next, let us briefly turn to the results for the Callan-Symanzik regulator, which are shown in \cref{fig:RGFlowCS,fig:RGFlowCSFlows}.
	First, note that the explicit values for $\sigma_0$ are different than for the Litim regulator, which implies that the critical point is regulator dependent.
	Next, instead of repeating the entire discussion, let us only highlight the differences.
	For the Callan-Symanzik regulator we observe that the system is in general much stiffer than for the Litim regulator and we cannot flow deep into the \gls{ir} in the symmetry-broken phase, independent of the truncation.
	Apart from this, the results are qualitatively similar to the ones for the Litim regulator and we again find huge differences for the wave-function renormalization in the flat region of the potential between \gls{lpa} and \gls{lpap}, while the effect of the wave-function renormalization at the physical point and beyond as well as in the symmetric phase is mild.
	
	Let us conclude this section with the remark that we have demonstrated the general applicability and stability of our new numerical scheme to solve \gls{rg} flows involving field-dependent wave-function renormalizations.
	We hope that this will pave the way for more advanced studies of \gls{frg} flows including nontrivial momentum dependencies in the future that go far beyond the limited application and discussion presented here.
	
\subsection{Gross-Neveu-Yukawa model}%
\label{subsec:gny_model}

	Within the next paragraphs, let us turn to our last example of this work.
	We consider the \gls{gny} model in $1 + 1$ dimensions at nonzero temperature $T$ and fermion chemical potential $\mu$.
	The \gls{gny} model is a well-known prototype \gls{qft}, which is used as a toy model for various phenomena in particle physics, condensed matter physics, and statistical physics, \eg{}, dimensional transmutation, asymptotic freedom, dynamical (chiral) symmetry breaking/restoration, critical phenomena, thermal and quantum phase transitions, phases of spatially inhomogeneous condensation and so on.
	It is the bosonized version of the \gls{gn} model, which describes the interactions of $N$ species of relativistic fermions via local scattering in the scalar channel.
	Therefore, it is the perfect testing ground to apply our new numerical scheme and to demonstrate the performance for systems including field-dependent Yukawa couplings.
	For details on the \gls{gn} and \gls{gny} model, we refer to the nonexhaustive list of Refs.~\cite{Gross:1974jv,Rosenstein:1990nm,ZinnJustin:2002ru,Quinto:2021lqn,Chodos:1993mf,Takayama:1980zz,Affleck:1981bn,Shankar:1985zc,Harrington:1974te,Harrington:1974tf,Jacobs:1974ys,Dashen:1974xz,Dashen:1975xh,Wolff:1985av,Treml:1989,Pausch:1991ee,Karbstein:2006er,Thies:2003br,Thies:2003kk,Schnetz:2004vr,deForcrand:2006zz,Braun:2014fga,Pannullo:2019bfn,Lenz:2020bxk,Lenz:2020bxk,Lenz:2020cuv,Stoll:2021ori}.

	Let us start by providing the ansatz for the \gls{eaa} of the \gls{gny} model within a \gls{sclpa},
		\begin{align}
			& \bar{\Gamma}_t [ \bar{\psi}, \psi, \varphi ] =	\Vdistance
			\\
			= \, & \int_{- \infty}^{\infty} \dd x \int_{0}^{\frac{1}{T}} \dd \tau \, \big[ \bar{\psi} \, \big( \gamma^\nu \partial_\nu - \mu \, \gamma^0 + m_t [ \varphi ] \big) \, \psi +	\Vdistance	\nonumber
			\\
			& \quad + \tfrac{N}{2} \, ( \partial_\mu \varphi )^2 + N \, U_t [ \varphi ] \big] \, .	\Vdistance	\nonumber
		\end{align}
	Hereby, $\varphi ( \tau, x )$ describes the bosonic scalar field that is generated from the four-fermion interaction via bosonization.
	The fermionic fields $\psi ( \tau, x )$ and $\bar{\psi} ( \tau, x )$ are two-component spinors, which are defined in the Euclidean spacetime and come in $N$ species (sometimes called flavors).
	Hence, the gamma matrices $\gamma^\nu$ are $2 \times 2$ matrices with $\nu \in \{ 0, 1 \}$.
	The bosons couple to the fermions via the field- and scale-dependent fermion mass $m_t [ \varphi ]$.
	In the \gls{uv} limit, this term is simply given by a Yukawa interaction $h_\Lambda \, \varphi$ that can be generated from the four-fermion interaction via a Hubbard-Stratonovich transformation.
	In general, we work with zero bare fermion mass, hence in the chiral limit.
	For the sake of the simplicity, we choose $h_\Lambda = 1$ to set the scales of the model (the Yukawa coupling has dimension energy).
	Additionally, we have a kinetic term for the bosonic field as well as a scale-dependent effective potential $U_t [ \varphi ]$, which are both multiplied by the number of fermions, $N$, because we are working in $\frac{1}{N}$-rescaled quantities to better match the mean-field (infinite-$N$) limit.
	Self-evidently, within a better truncation, one would also include the (field-dependent) wave-function renormalization for all fields or higher-order derivative interactions, which is however, beyond the scope of this work.
	The bosonic potential is initialized in the \gls{uv} with the mass term $U_\Lambda [ \varphi ] = \tfrac{1}{2} \, m_\Lambda^2 \, \varphi^2$ that is generated from the Hubbard-Stratonovich transformation.
	From the mean-field-renormalization procedure, we obtain an initial mass~\cite{Stoll:2021ori}
		\begin{align}
					& m^2_\Lambda =	\Vdistance
					\\
					= \, & \tfrac{d_\gamma}{4 \uppi} \, h^2 \bigg[ \mathrm{arcoth} \bigg( \sqrt{1 + \big( \tfrac{\Sigma_0 h}{\Lambda} \big)^2} \bigg) - \Big( 1 + \big( \tfrac{\Sigma_0 h}{\Lambda} \big)^2 \Big)^{-\frac{1}{2}} \bigg] \, ,	\Vdistance	\nonumber
				\end{align}
	which yields the typical $m_\Lambda^2 \propto \frac{1}{\lambda_\Lambda} \sim \ln ( \Lambda )$ behavior of the asymptotically free four-fermion interaction.
	Here, $d_\gamma$ is the dimensionality of the gamma matrices, which is $2$ in our case.
	Indeed, in the limit $N \to \infty$, where only fermionic quantum fluctuations are taken into account, this reproduces a constant vacuum chiral condensate $\Sigma_0$ and the limit $\Lambda \to \infty$ can be taken to remove the cutoff.
	Consequently, in the $N \to \infty$ limit the renormalized model has a unique phase diagram, whose dimensions are set by $h \, \Sigma_0$.
	In the following all dimensionful quantites --  \ie{} temperature $T$, chemical potential $\mu$, and chiral condensate $\sigma$ -- are to be understood in units of $h \, \Sigma_0$.
	Also the position of the condensate $\Sigma_0$ in dimensionless field space is conveniently chosen to be $\Sigma_0 = 1$.
	
	Since we have already mentioned chiral symmetry breaking/restoration and the vacuum (chiral) condensate, let us briefly mention that the \gls{uv} theory has a $\mathbb{Z}_2$ discrete chiral symmetry,
		\begin{align}
			& \psi \mapsto \gamma_\mathrm{ch} \, \psi \, ,	&&	\bar{\psi} \mapsto - \bar{\psi} \, \gamma_\mathrm{ch} \, ,	&&	\varphi \mapsto - \varphi \, ,	\label{eq:chiral_symmetry}
		\end{align}
	and
		\begin{align}
			U_t [ \varphi ] \mapsto U_t [ - \varphi ] \, ,	&&	m_t [ \varphi ] \mapsto - m_t [ - \varphi ] \, ,	\label{eq:chiral_symmetry_potential}
		\end{align}
	where $\gamma_\mathrm{ch}$ is the two-dimensional analog of the chiral gamma matrix $\gamma^5$.
	This chiral symmetry can be spontaneously broken by the attractive fermionic quantum fluctuations and at least in the $N \to \infty$ limit one finds a nonzero chiral condensate\footnote{Within this work we ignore the possibility of spatially inhomogeneous chiral condensation and refer to Refs.~\cite{Carignano:2012yli,Buballa:2014tba,Koenigstein:2021llr,Motta:2023pks,Steil:2023sfd} for further discussion and references.} $\sigma_0 ( \mu, T ) \neq 0$ for small values of the chemical potential $\mu$ and temperature $T$ -- $\Sigma_0 = \sigma_0 ( 0, 0 ) = 1$.

	A long standing question is whether this condensate is an artifact of the $N \to \infty$ limit, because in $1+1$ dimensions one would not expect condensation at nonzero temperatures, if bosonic fluctuations are taken into account.
	In preceding works~\cite{Stoll:2021ori,Koenigstein:2023wso,Steil:2023sfd}, we indeed showed within the \gls{frg} framework in the \gls{lpa} and also with the inclusion of field-independent wave-function renormalization and Yukawa coupling that condensation is an artifact of the $N \to \infty$ limit and is only possible at $T = 0$ in one spatial dimension.
	However, we also revealed that the imprint of the fermions and the chemical potential on the bosonic effective potential are still tremendous, especially for small temperatures, and necessitate state of the art numerics that can deal with nonanalyticities like shock waves within the \gls{frg} flow equations.

	Within this work, we are interested in extending the truncation to the inclusion of a field-dependent Yukawa coupling/fermion mass and to inspect their impact on the dynamics in field-space during the \gls{rg} flow.
	Hereby, our strategy is as follows:

	First, we recapitulate the \gls{rg} flow equations for our truncation without explicit derivation.
	Nevertheless, we briefly discuss their general structure \gls{wrt}\ to our new approach and compare it to the zero-dimensional fermion-boson model from \cref{sec:the_model}.
	Second, we turn to some explicit \gls{rg} flows.
	Here, we will solve the flow equations for a specific point in the $\mu$-$T$-phase diagram and specific choice of $N$ and compare the results for two different approximations -- \gls{lpa} (constant Yukawa coupling) and \gls{sclpa} (field-dependent Yukawa coupling).

\subsubsection{Flow equations}
	
	Let us start with the \gls{rg} flow equations for the \gls{gny} model.
	All details of their derivation as well as the corresponding results for the Matsubara sums can be found in Refs.~\cite{Koenigstein:2023wso} and are not repeated here.

	For their derivation we have used dimensionally reduced Litim regulators for the bosons and fermions that solely regulate the spatial momenta and keep the Matsubara frequencies untouched.
	This ensures Silver-Blaze symmetry and the correct pole structure of the propagators, but introduces deviations for the $T \to 0$ limit.
	The flow equation for the scale-dependent effective potential, evaluated for a constant bosonic background field configuration $\varphi ( x ) = \sigma$, reads
		\begin{align}
			\partial_t U ( t, \sigma ) = \, & - A_d \, \tfrac{1}{N} \, k^{d + 2} \, T \sum_{n = - \infty}^{\infty} \frac{1}{\omega_n^2 + E_{\mathrm{b}}^2} +	\Vdistance
			\\
			& + A_d \, d_\gamma \, k^{d + 2} T \sum_{n = - \infty}^{\infty} \frac{1}{( \nu_n + \ii \mu )^2 + E_{\mathrm{f}}^2} \, ,\Vdistance	\nonumber
		\end{align}
	where $d = 1$ is the number of spatial dimensions,
		\begin{align}
			&	A_d = \frac{\Omega_d}{d \, ( 2 \uppi )^d} \, ,	&&	\Omega_d = \frac{2 \uppi^{\frac{d}{2}}}{\Gamma ( \frac{d}{2} )} \, ,
		\end{align}
	and
		\begin{align}
			& \omega_n = 2 \uppi T n \, ,	&&	\nu_n = 2 \uppi T \big( n + \tfrac{1}{2} \big) \, ,
		\end{align}
	are the Matsubara frequencies for bosons and fermions, respectively.
	The bosonic and fermionic energies are given by	
		\begin{align}
			&	E_{\mathrm{b}}^2 = k^2 + \partial_\sigma^2 U ( t, \sigma ) \, ,	&&	E_{\mathrm{f}}^2 = k^2 + m^2 ( t, \sigma ) \, .	\Vdistance
		\end{align}
	From this, it is straightforward to derive the flow equation for the second derivative of the bosonic potential $M ( t, \sigma ) = \partial_\sigma^2 U ( t, \sigma )$,
	\begin{align}
		& \partial_t M =	\Vdistance	\label{eq:flow_M_gny}
		\\
		= \, & \frac{\partial}{\partial \sigma} \bigg( A_d \, \tfrac{1}{N} \, k^{d + 2} \, ( \partial_\sigma M ) \, T \sum_{n = - \infty}^{\infty} \frac{1}{[ \omega_n^2 + E_{\mathrm{b}}^2 ]^2} \bigg) +	\Vdistance	\nonumber
		%\\
		%& - 2 \, A_d \, \tfrac{1}{N} \, k^{d + 2} \, ( \partial_\sigma M )^2 \, T \sum_{n = - \infty}^{\infty} \frac{1}{[ \omega_n^2 + E_{\mathrm{b}}^2 ]^3} +	\Vdistance	\nonumber
		\\
		& - A_d \, d_\gamma \, k^{d + 2} \, 2 \, [ m \, \partial_\sigma^2 m + ( \partial_\sigma m )^2 ] \times	\Vdistance	\nonumber
		\\
		& \quad \times T \sum_{n = - \infty}^{\infty} \frac{1}{[ ( \nu_n + \ii \mu )^2 + E_{\mathrm{f}}^2 ]^2} +	\Vdistance	\nonumber
		\\
		& + A_d \, d_\gamma \, k^{d + 2} \, 8 \, m^2 \, ( \partial_\sigma m )^2 \, T \sum_{n = - \infty}^{\infty} \frac{1}{[ ( \nu_n + \ii \mu )^2 + E_{\mathrm{f}}^2 ]^3} \, .	\Vdistance	\nonumber
	\end{align}
	Similarly, one can derive the flow equation for the field-dependent fermion mass~\cite[Eq.~(E.110) -- the sign of the boson tadpole contribution is wrong in the reference]{Koenigstein:2023wso},
	\begin{align}
		& \partial_t m ( t, \sigma ) =	\Vdistance	\label{eq:flow_m_gny}
		\\
		= \, & A_d \, \tfrac{1}{N} \, ( \partial_\sigma^2 m ) \, k^{d + 2} \, T \sum_{n = - \infty}^{\infty} \frac{1}{[ \omega_n^2 + E_{\mathrm{b}}^2 ]^2} +	\Vdistance	\nonumber
		\\
		& - 2 \, A_d \, \tfrac{1}{N} \, m \, ( \partial_\sigma m )^2 \, k^{d + 2} \times	\Vdistance	\nonumber
		\\
		& \times \mathrm{Re} \bigg( T \sum_{n = - \infty}^{\infty} \frac{1}{[ \omega_n^2 + E_{\mathrm{b}}^2 ]^2} \frac{1}{( \nu_n + \ii \mu )^2 + E_{\mathrm{f}}^2} +	\Vdistance	\nonumber
		\\
		& \quad + T \sum_{n = - \infty}^{\infty} \frac{1}{\omega_n^2 + E_{\mathrm{b}}^2} \frac{1}{[ ( \nu_n + \ii \mu )^2 + E_{\mathrm{f}}^2 ]^2} \bigg) \, ,	\Vdistance	\nonumber
	\end{align}
	where we did not indicate the $t$ and $\sigma$ dependence of the fermion mass on the \gls{rhs}.

	Note that structurally as well as on the level of the Feynman graphs, the \gls{rg} flow equation of this example is identical to \cref{eq:flow_u,eq:flow_H_1} -- apart from the symmetry properties of the fermion mass.
	Again, we find that the tadpole contributions in both equations correspond to diffusion type operators on the \gls{pde} level with complicated diffusion coefficients, while the self-energy diagrams turn out to be of the advective type and can be interpreted as Hamiltonians and therefore be treated with the \gls{kthj} scheme.
	However, as before, we experienced slightly better numerical stability and convergence when treating the bosonic contribution on the \gls{rhs}\ of \cref{eq:flow_M_gny} with the conservative \gls{kt} scheme \cref{eq:KT_conservative_term} and all other parts of the \gls{pde} system with the \gls{kthj} scheme.

\subsubsection{RG flows}

	Let us now turn to the \gls{rg} flows of the \gls{gny} model.
	Here, we shall focus on a single point in the $\mu$-$T$-phase diagram, which is located at $\mu = 0.6$ and $T = 0.00625$.
	This perfectly suffices to demonstrate the performance of our numerical scheme and a detailed reanalysis of the entire phase diagram with the present numerics is postponed to future work.
	Furthermore, \gls{wlog}\ we set the number of fermionic flavors to $N = 16$.
	These parameters turn out to be a challenging test case for our numerical scheme due to the following reasons:
	We are working at rather large chemical potential and small temperature.
	Still, the chemical potential is not too large, such that in the $N \to \infty$ limit, the \gls{gny} model is in the phase of spontaneous discrete chiral symmetry breaking.
	For $N = 16$, however, we expect the condensate to vaporize at nonzero temperature, while the imprint of the fermions on the dynamics of the bosonic effective potential (and field-dependent fermion mass) is still significant.
	Furthermore, exactly for these values there are reference solutions available from previous works~\cite{Stoll:2021ori,Koenigstein:2023wso,Steil:2023sfd}, which we can use for comparison.
	These solutions where obtained in \gls{lpa} with the \gls{kt} scheme applied to $u ( t, \sigma ) = \partial_\sigma U ( t, \sigma )$.

\begin{figure}
			\subfloat[$M ( t, \sigma ) = \partial_\sigma^2 U ( t, \sigma )$ at $\mu = 0.6$, $T = 0.015$]{%
				\centering
				\includegraphics[width=\columnwidth]{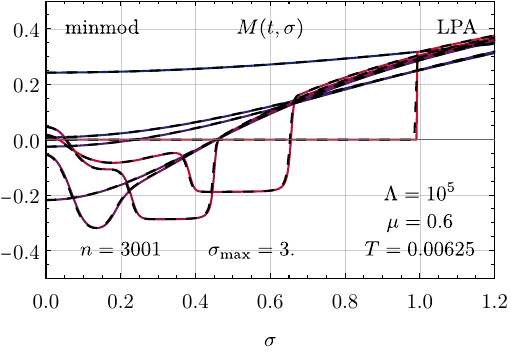}
			}

			\subfloat[$u ( t, \sigma ) = \partial_\sigma U ( t, \sigma )$ at $\mu = 0.6$, $T = 0.0625$]{%
				\centering
				\includegraphics[width=\columnwidth]{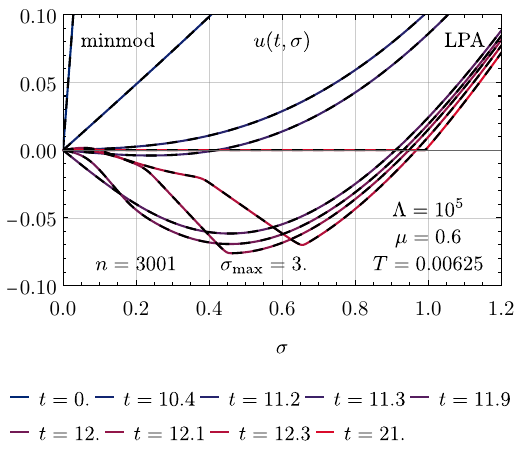}
			}

			\caption{\label{fig:RGFlowGNYminmod}%
				\gls{rg} flow in the \gls{lpa} of derivative of the effective potential $u ( t, \sigma ) = \partial_\sigma U ( t, \sigma )$ and the boson mass $M ( t, \sigma ) = \partial_\sigma^2 U ( t, \sigma )$ of the \gls{gny} model with the \gls{kt} (black dashed) and \gls{kthj} (blue to red) scheme.
				For the \gls{kthj} $u ( t, \sigma )$ is generated by numerical integration of $M ( t, \sigma )$ in $\sigma$-direction.}
		\end{figure}

\begin{figure}
			\subfloat[$M ( t, \sigma ) = \partial_\sigma^2 U ( t, \sigma )$ at $\mu = 0.6$, $T = 0.015$]{%
				\centering
				\includegraphics[width=\columnwidth]{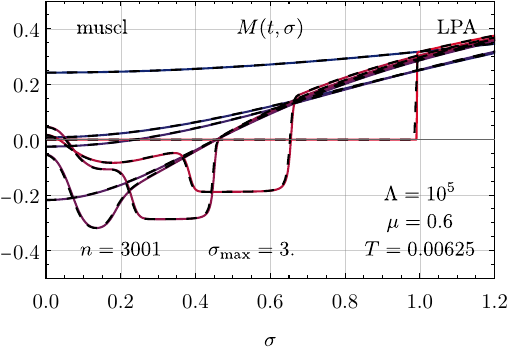}
			}

			\subfloat[$u ( t, \sigma ) = \partial_\sigma U ( t, \sigma )$ at $\mu = 0.6$, $T = 0.0625$]{%
				\centering
				\includegraphics[width=\columnwidth]{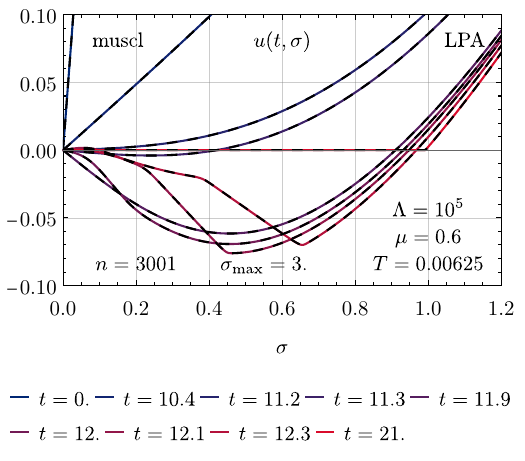}
			}

			\caption{\label{fig:RGFlowGNYmuscl}%
				\gls{rg} flow in the \gls{lpa} of derivative of the effective potential $u ( t, \sigma ) = \partial_\sigma U ( t, \sigma )$ and the boson mass $M ( t, \sigma ) = \partial_\sigma^2 U ( t, \sigma )$ of the \gls{gny} model with the \gls{kt} (black dashed) and \gls{kthj} (blue to red) scheme.
				For the \gls{kthj} $u ( t, \sigma )$ is generated by numerical integration of $M ( t, \sigma )$ in $\sigma$-direction.}
		\end{figure}

\begin{figure}
			\subfloat[$M ( t, \sigma ) = \partial_\sigma^2 U ( t, \sigma )$ at $\mu = 0.6$, $T = 0.015$]{%
				\centering
				\includegraphics[width=\columnwidth]{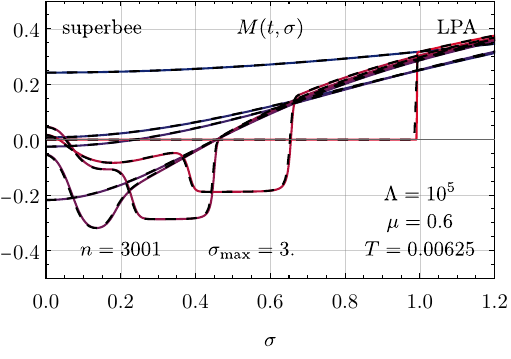}
			}

			\subfloat[$u ( t, \sigma ) = \partial_\sigma U ( t, \sigma )$ at $\mu = 0.6$, $T = 0.0625$]{%
				\centering
				\includegraphics[width=\columnwidth]{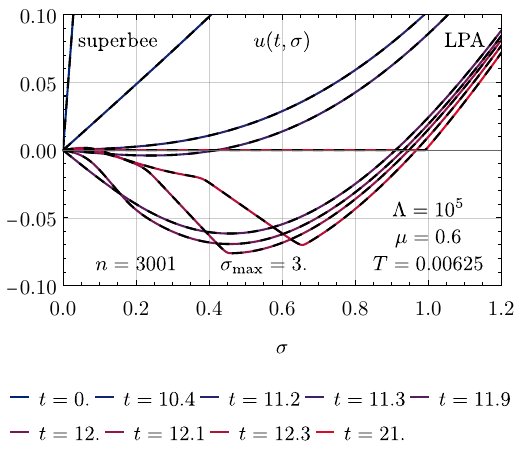}
			}

			\caption{\label{fig:RGFlowGNYsuperbee}%
				\gls{rg} flow in the \gls{lpa} of derivative of the effective potential $u ( t, \sigma ) = \partial_\sigma U ( t, \sigma )$ and the boson mass $M ( t, \sigma ) = \partial_\sigma^2 U ( t, \sigma )$ of the \gls{gny} model with the \gls{kt} (black dashed) and \gls{kthj} (blue to red) scheme.
				For the \gls{kthj} $u ( t, \sigma )$ is generated by numerical integration of $M ( t, \sigma )$ in $\sigma$-direction.}
		\end{figure}

	Hence, let us start with the first test of our new setup.
	For the start, we work in the \gls{lpa} and with a constant Yukawa coupling, \ie{}, $m ( t, \sigma ) = \sigma$.
	However, in contrast to our previous work and reference solution, we use the flow equation for $M ( t, \sigma ) = \partial_\sigma^2 U ( t, \sigma )$, namely \cref{eq:flow_M_gny}, and apply the \gls{kt} scheme to the conservative part, while the fermionic contributions are treated with the \gls{kthj} scheme.
	The results for the \gls{rg} flow of $M ( t, \sigma )$ and $u ( t, \sigma ) = \partial_\sigma U ( t, \sigma )$ are shown in \cref{fig:RGFlowGNYminmod,fig:RGFlowGNYmuscl,fig:RGFlowGNYsuperbee} for three different flux limiters -- minmod, muscl, and superbee, which all show identical results.

	We find perfect agreement of our new numerical scheme with the reference solution, which is indicated by the black dashed lines in the plots and which has been computed with the \gls{kt} scheme for $u$.
	This is a strong cross-check and proves that the conservative \gls{kt} scheme can also be applied on the level of the field-dependent mass $M ( t, \sigma ) = \partial_\sigma^2 U ( t, \sigma )$ instead of $u ( t, \sigma ) = \partial_\sigma U ( t, \sigma )$.
	The independence of the results from the flux limiter is expected, since the limiters only enter the Hamilton part of the \gls{pde} system with the fermion mass, which does not evolve in \gls{lpa}.
	Let us also briefly comment on the dynamics during the \gls{rg} flow:
	One starts with a Gaussian potential in the \gls{uv}, while at some intermediate \gls{rg} time the fermionic fluctuations induce a negative curvature around the origin in field space, which signals spontaneous chiral symmetry breaking and a nontrivial minimum of the potential.
	This is also reflected in the zero-crossing of $u ( t, \sigma )$.
	At later \gls{rg} times, where $k \approx \mu$, the chemical potential enters from the origin in field space and induces a shock wave in $M ( t, \sigma )$ that travels outwards to larger field values.
	Meanwhile, the bosonic fluctuations work against the fermionic ones and push to restore chiral symmetry.
	Finally, at the end of the flow, chiral symmetry is restored and the field-dependent mass $M ( t \to \infty, \sigma )$ is positive for all field values.
	The latter is not visible in the plots due to the chosen scales, but was checked in the data.
	Very similar dynamics was already observed in Ref.~\cite{Ihssen:2023xlp} for \gls{left} for \gls{qcd}.

\begin{figure}
			\subfloat[$m ( t, \sigma )$ at $\mu = 0.6$, $T = 0.0625$]{%
				\centering
				\includegraphics[width=\columnwidth]{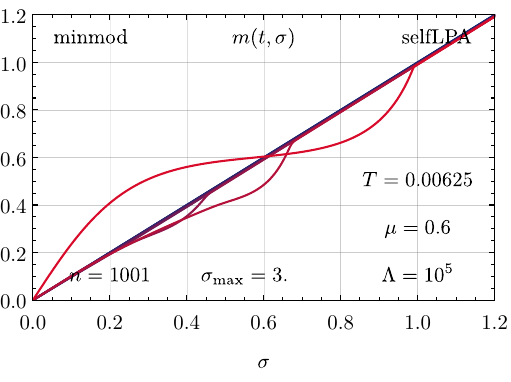}
			}

			\subfloat[$M ( t, \sigma ) = \partial_\sigma^2 U ( t, \sigma )$ at $\mu = 0.6$, $T = 0.0625$]{%
				\centering
				\includegraphics[width=\columnwidth]{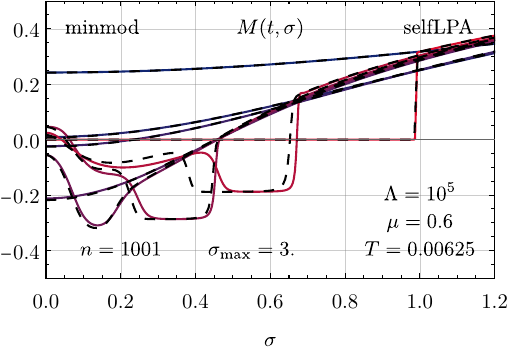}
			}

			\subfloat[$u ( t, \sigma ) = \partial_\sigma U ( t, \sigma )$ at $\mu = 0.6$, $T = 0.0625$]{%
				\centering
				\includegraphics[width=\columnwidth]{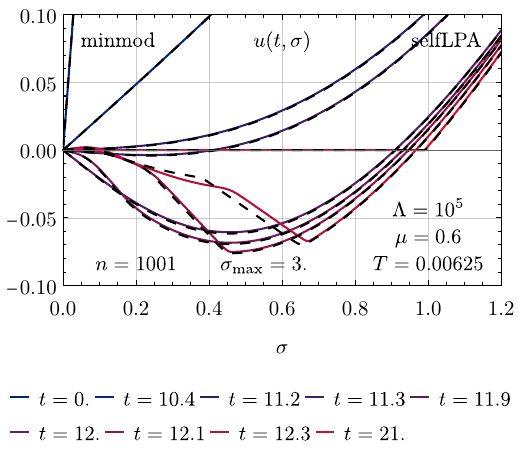}
			}

			\caption{\label{fig:RGFlowGNYminmodSelf}%
				\gls{rg} flow in the \gls{sclpa} of derivative of the effective potential $u ( t, \sigma ) = \partial_\sigma U ( t, \sigma )$, the boson mass $M ( t, \sigma ) = \partial_\sigma^2 U ( t, \sigma )$, and the fermion mass $m ( t, \sigma )$ of the \gls{gny} model with the \gls{kthj} (blue to red) scheme compared to the \gls{lpa} result from the \gls{kt} scheme (black dashed).
				For the \gls{kthj} $u ( t, \sigma )$ is generated by numerical integration of $M ( t, \sigma )$ in $\sigma$-direction.}
		\end{figure}

\begin{figure}
			\subfloat[$m ( t, \sigma )$ at $\mu = 0.6$, $T = 0.0625$]{%
				\centering
				\includegraphics[width=\columnwidth]{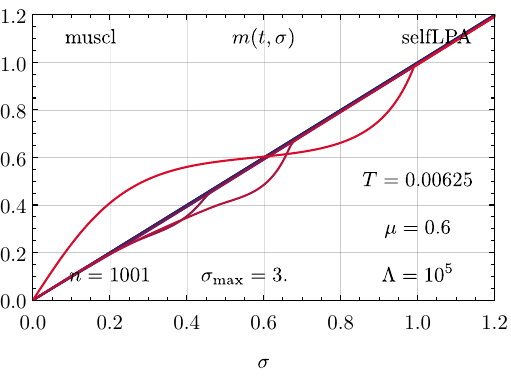}
			}

			\subfloat[$M ( t, \sigma ) = \partial_\sigma^2 U ( t, \sigma )$ at $\mu = 0.6$, $T = 0.0625$]{%
				\centering
				\includegraphics[width=\columnwidth]{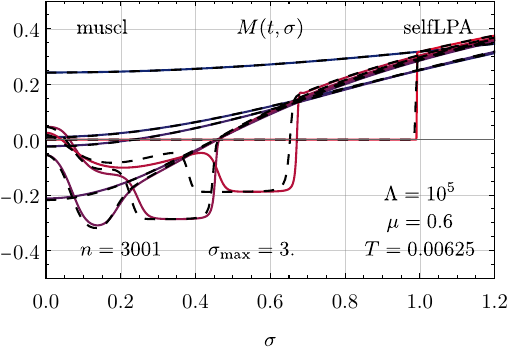}
			}

			\subfloat[$u ( t, \sigma ) = \partial_\sigma U ( t, \sigma )$ at $\mu = 0.6$, $T = 0.0625$]{%
				\centering
				\includegraphics[width=\columnwidth]{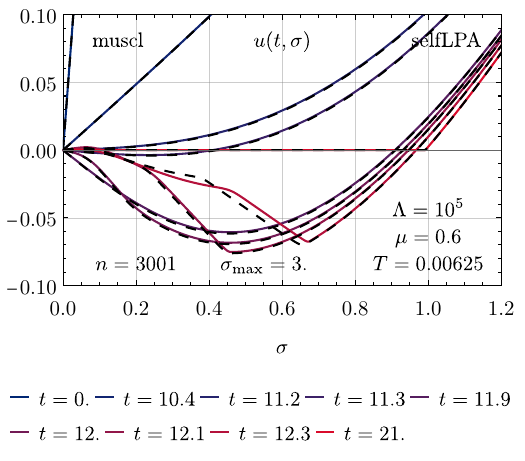}
			}

			\caption{\label{fig:RGFlowGNYmusclSelf}%
				\gls{rg} flow in the \gls{sclpa} of derivative of the effective potential $u ( t, \sigma ) = \partial_\sigma U ( t, \sigma )$, the boson mass $M ( t, \sigma ) = \partial_\sigma^2 U ( t, \sigma )$, and the fermion mass $m ( t, \sigma )$ of the \gls{gny} model with the \gls{kthj} (blue to red) scheme compared to the \gls{lpa} result from the \gls{kt} scheme (black dashed).
				For the \gls{kthj} $u ( t, \sigma )$ is generated by numerical integration of $M ( t, \sigma )$ in $\sigma$-direction.}
		\end{figure}

\begin{figure}
			\subfloat[$m ( t, \sigma )$ at $\mu = 0.6$, $T = 0.0625$]{%
				\centering
				\includegraphics[width=\columnwidth]{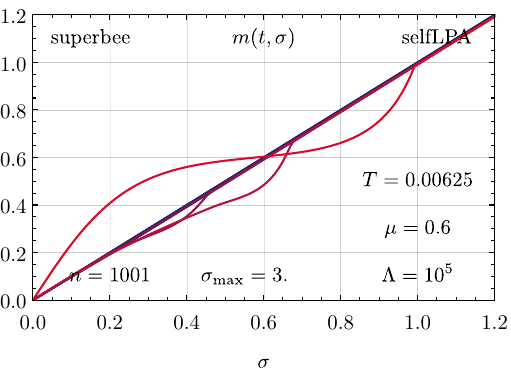}
			}

			\subfloat[$M ( t, \sigma ) = \partial_\sigma^2 U ( t, \sigma )$ at $\mu = 0.6$, $T = 0.0625$]{%
				\centering
				\includegraphics[width=\columnwidth]{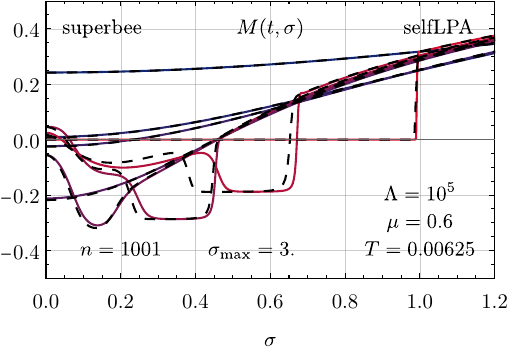}
			}

			\subfloat[$u ( t, \sigma ) = \partial_\sigma U ( t, \sigma )$ at $\mu = 0.6$, $T = 0.0625$]{%
				\centering
				\includegraphics[width=\columnwidth]{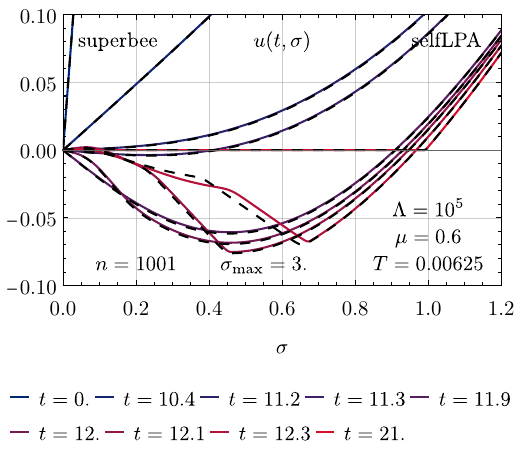}
			}

			\caption{\label{fig:RGFlowGNYsuperbeeSelf}%
				\gls{rg} flow in the \gls{sclpa} of derivative of the effective potential $u ( t, \sigma ) = \partial_\sigma U ( t, \sigma )$, the boson mass $M ( t, \sigma ) = \partial_\sigma^2 U ( t, \sigma )$, and the fermion mass $m ( t, \sigma )$ of the \gls{gny} model with the \gls{kthj} (blue to red) scheme compared to the \gls{lpa} result from the \gls{kt} scheme (black dashed).
				For the \gls{kthj} $u ( t, \sigma )$ is generated by numerical integration of $M ( t, \sigma )$ in $\sigma$-direction.}
		\end{figure}

	Next, we turn to the \gls{sclpa}, where we solve the full system \cref{eq:flow_M_gny,eq:flow_m_gny} including the field-dependent fermion mass $m ( t, \sigma )$.
	Here, of course, there are no longer reference solutions available.
	The only way to cross-check our results is to work at different resolutions and with different flux limiters and to check the independence of the results from these numerical details.
	The results for the \gls{rg} flow of $m ( t, \sigma )$, $M ( t, \sigma ) = \partial_\sigma^2 U ( t, \sigma )$, and $u ( t, \sigma ) = \partial_\sigma U ( t, \sigma )$ are shown in \cref{fig:RGFlowGNYminmodSelf,fig:RGFlowGNYmusclSelf,fig:RGFlowGNYsuperbeeSelf} for the three different flux limiters -- minmod, muscl, and superbee, which all show identical results.
	However, note that the computational time for the minmod and superbee limiters increased so drastically while enlarging the number of grid points that we solely show results for $N = 1001$, while improving the resolution for the muscl limiter was no problem at all.
	We speculate that this might be linked to the fact that minmod and superbee are both on the extreme ends (minimal and maximal diffusion respectively) within the class of second-order accurate total variation diminishing limiters and the problem at hand might require a medium amount of numerical diffusion for optimal performance.
	
	Let us now turn to the explicit results.
	Basically, we find almost the same dynamics for $M ( t, \sigma )$ and $u ( t, \sigma )$ as in the \gls{lpa} case with constant Yukawa coupling.
	Especially the final \gls{ir} shape of $M ( t \to \infty, \sigma )$ and $u ( t \to \infty, \sigma )$ is almost identical to the \gls{lpa} case.
	For direct comparison, we plotted the \gls{lpa} results as black dashed lines in the plots.
	This behavior is not too surprising, if one inspects the flow of the fermion mass $m ( t, \sigma )$.
	Here, we solely observe deviations from the linear behavior in the region of small fields $\sigma < 1$ -- hence at field values smaller than the intermediate minimum of the potential.
	An interesting observation is that the fermion mass $m ( t, \sigma )$ develops a cusp at the position, where the boson mass jumps.
	Furthermore, it seems as if the fermion mass $m ( t, \sigma )$ tries to approach a constant function at small fields that is given by the chemical potential.
	Indeed, we found by studying \gls{rg} flows at different $\mu$ and even lower $T$ that the fermion mass $m ( t, \sigma )$ at small fields approaches $\pm \mu$ in the \gls{ir} limit, which might be linked to the Silver-Blaze property of the model.
	However, a more detailed analysis of this behavior is postponed to future work.

	Instead, let us close our discussion by noting that symmetry restoration -- even at small temperatures -- seems to be a robust feature of the \gls{gny} model in $1 + 1$ dimensions at finite $N$ (non)zero chemical potential.

	Additionally, we remark that our new interpretation and numerical implementation of \gls{rg} flow equations in terms of \gls{hj} systems and conservation laws works perfectly fine for the inclusion of field-dependent Yukawa couplings.

\section{Conclusions and outlook}%
\label{sec:conclusions}

In this work we set out to connect conceptual advances in the understanding of the Wetterich equation with concrete progress on its numerical treatment.
Starting from the observation that the flow of two-point functions can be interpreted as a viscous \gls{hjb}-type equation, we emphasized that this perspective not only clarifies the advective, diffusive, and nonconservative contributions to \gls{frg} flows, but also allows the systematic adaptation of established numerical methods.
The structures identified on a functional level carried through a range of projected \glspl{pde}, which initially prompted the identification on functional level.

Our primary testbed has been the zero-dimensional fermion-boson model, which proved to be structurally extremely rich. 
In this setting we developed and benchmarked the presented \gls{kthj}-hybrid scheme, combining the novel insights with the stability and proven robustness of finite volume methods. 
Using the \gls{kthj} scheme to deal with -- on the level of the \gls{cfd}/finite volume formulation -- nonconservative contributions has proven very promising.
The scheme performed well in zero spacetime dimensions and, within the scope of this work, represents the most reliable approach we were able to construct on the basis of extensive numerical experimentation.
First applications to higher-dimensional models, including a $\mathbb{Z}_2$-symmetric scalar theory in $d=3$ and the Gross-Neveu-Yukawa model in $d=1+1$, further indicate that the method can be successfully generalized to physically relevant systems.
Extending these developments to other models and deepening the analysis of the $\mathbb{Z}_2$ and Gross-Neveu-Yukawa systems are natural next steps, and we expect them to provide new insights into both the method as well as the models.

Despite the promising prospects of our developments, they should be regarded as a first step rather than a definitive solution. 
More advanced or specialized schemes for \gls{hjb} equations, possibly tailored to the specific nonlinear structures encountered in \gls{frg} flows, may lead to even more stable and efficient numerical treatments. 
Exploring these alternatives, and testing them in both toy models and higher-dimensional theories, will be a worthwhile and necessary direction for future work.
We made a similar conclusion and in a sense warnings in our previous works~\cite{Koenigstein:2021syz,Koenigstein:2021rxj,Steil:2021cbu,Stoll:2021ori,Zorbach:2024rre} using adapted \gls{kt}-finite-volume-schemes for the numerical solution of conservative advection-diffusion-source/sink equations. 
In the present context we feel obliged to stress this even more: Conceptually we feel that it should be possible to treat the \gls{pde}-system of this work numerically as coupled \gls{hjb} equations without the need to split off parts
as conservative equations. 
The adapted \gls{kthj}-scheme for the semi-discrete \gls{pde}-systems studied in this work is not ideal for this, which prompted us to split off purely conservative parts.
The coupling of \glspl{pde}, externally driven discontinuities (\eg{} by the chemical potential), or especially the approximation of velocities/wave speeds might be responsible for that.
Our hope is that some ideas and aspects raised in this work will spark further numerical developments.
As with our testcases I-IV for the zero dimensional $O(N)$ model (in this work test 0), we hope that tests 1-3 for our zero dimensional fermion-boson model will be used in the future to benchmark \gls{frg} numerics for coupled systems of scale- and field-dependent couplings.
We would truly be delighted to see a numerical scheme in future works, outperforming our developments made here.
Progress in pushing truncations to ever higher sophistication -- often aided by powerful computer algebra systems and significant human effort -- can only translate into genuine advances if numerical methods keep pace!
At present, we feel that this balance is not yet achieved.

Within the broader landscape of methods for quantum field theory and statistical physics, the \gls{frg} both benefits from and suffers under its inherently nonperturbative character.
When applied to genuinely nonperturbative problems, truncations are a practical necessity, yet their reliability is difficult to judge.
Assessing the quality of a truncation typically requires \aposteriori{} knowledge, accumulated experience, and the study of apparent convergence. 
Such an assessment, however, is only meaningful if based on robust and appropriate numerical schemes -- particularly when the aim is to analyze systems on a fundamental rather than an effective level.

Beyond the numerical aspects, the identification of \gls{hjb} structures in \gls{frg} flow equations may be key for further theoretical developments. 
The notion of viscosity solutions as physical solutions to \gls{grg} flow equations seems very natural and in the spirit of the underlying Wilson's \gls{rg}-approach.
The interpretation of the \gls{frg} as an infinite-dimensional stochastic optimal control problem might open up connections of the \gls{frg} to fields and methods not explored so far.
Given the extreme complexity and practical challenges in solving the explicit \glspl{pde} derived within the \gls{frg}, such structural connections and the insights they provide may become increasingly important for the future of the field.

\acknowledgments%

	The authors are grateful for the support of the following people without whom this work would not have been successful:
	The authors thank J.~Braun, M.~Buballa, F.~Ihssen, D.~H.~Rischke, F.~Sattler, J.~Stoll, and N.~Wink for various discussions during different stages of this project.
	The authors are especially grateful to E.~Grossi and N.~Zorbach for numerous in depth discussions about numerics for \gls{pde} systems and their implementation, critical comments on preliminary results, as well as many inspiring ideas and suggestions for the present work.
	The authors are also grateful to M.~Scherer, R.~Grauer, and K.~Kormann for interesting discussions on \glspl{pde} underlying the \gls{frg} and possible numeric approaches.
	The authors also thank F.~Atteneder, S.~Bernuzzi, H.~Gies, E.~Oevermann, C.~Schmidt, M.~Schröfl, T.~Stötzel, A.~Wipf for comments and discussions on this project at the ITP in Jena.
	The authors are grateful to G.\ Pfeifer for his work on a related project and crosschecking the flow equations of the three-dimensional $\mathbb{Z}_2$ model.
	The authors thank B.~Alt for literature suggestions and discussions on \gls{hjb} equations and optimal control theory.
	The authors thank J.~Stoll and N.~Zorbach for valuable comments and enlightening discussions on the content and the first drafts of the manuscript.

\section*{Data availability}

	The data that support the findings of this work are openly available \cite{koenigstein_2025_18085412}.

\appendix

\section{Correlation and vertex functions of the zero-dimensional model}

In this appendix we present explicit expressions for the correlation and vertex functions at physical point -- here $\varphi = 0 = \tilde{\vartheta} = \vartheta$ -- of the zero-dimensional model introduced in \cref{sec:the_model}.
The relevant governing equations were introduced in \cref{subsec:expval_genfunc}.

\subsection{Connected correlation functions}%
\label{subsec:exp_c}

	In this subsection we provide expressions for the first few relevant connected correlation functions based on the Schwinger function \labelcref{eq:schwinger_functional}.

	For example, the connected two-point correlation function for the bosonic field is given by
		\begin{align}
			\langle \phi^2 \rangle_c = \, & \frac{\delta^2 \W ( \tilde{\eta}, \eta, J )}{\delta J^2} \Bigg|_{\substack{J = 0, \tilde{\eta} = 0, \eta = 0}} =	\Vdistance
			\\
			= \, & \langle \phi^2 \rangle - \langle \phi \rangle^2 = \langle \phi^2 \rangle \, .	\Vdistance	\nonumber
		\end{align}
	Here, we used the fact that the expectation value $\langle \phi \rangle$ vanishes due to the symmetry of the action under $\phi \to - \phi$ (without the additional term $- c \, \phi$ in the potential).
	Similarly, we obtain the connected four-point correlation function for the bosonic field in terms of usual expectation values,
		\begin{align}
			\langle \phi^4 \rangle_c = \, & \frac{\delta^4 \W ( \tilde{\eta}, \eta, J )}{\delta J^4} \Bigg|_{\substack{J = 0, \tilde{\eta} = 0, \eta = 0}} =	\Vdistance
			%\\
			%= \, & \langle \phi^4 \rangle - 3 \, \langle \phi^2 \rangle^2 + 12 \, \langle \phi \rangle^2 \langle \phi^2 \rangle - 4 \, \langle \phi^3 \rangle \langle \phi \rangle - 6 \, \langle \phi \rangle^4 =	\Vdistance	\nonumber
			\\
			= \, & \langle \phi^4 \rangle - 3 \, \langle \phi^2 \rangle^2 \, .	\Vdistance	\nonumber
		\end{align}
	Quantities like these serve as benchmark values for the \gls{frg} calculations in this work.
	In addition, we use the connected two-point correlation function for the fermionic fields
		\begin{align}
			\langle \tilde{\theta} \, \theta \rangle_c = \, & \frac{\delta^2 \W ( \tilde{\eta}, \eta, J )}{\delta( - \eta ) \, \delta \tilde{\eta}} \Bigg|_{\substack{J = 0, \tilde{\eta} = 0, \eta = 0}} = \langle \tilde{\theta} \, \theta \rangle
		\end{align}
	and the connected four-point correlation function of fermionic and bosonic fields
		\begin{align}
			\langle \tilde{\theta} \, \theta \, \phi^2 \rangle_c = \, & \frac{\delta^4 \W ( \tilde{\eta}, \eta, J )}{\delta ( - \eta ) \, \delta \tilde{\eta} \, \delta J^2} \Bigg|_{\substack{J = 0, \tilde{\eta} = 0, \eta = 0}} =	\Vdistance
			\\
			= \, & \langle \tilde{\theta} \, \theta \, \phi^2 \rangle - \langle \tilde{\theta} \, \theta \rangle \langle \phi^2 \rangle \, ,	\Vdistance	\nonumber
		\end{align}
	which can be seen as the expectation value of an interaction/scattering of a fermion-antifermion pair with two bosons.

\subsection{Vertex functions}%
\label{subsec:exp_vf}

	In this subsection we provide expressions for the first few relevant vertex functions based on the \gls{eaa} \labelcref{eq:effective_action}, \ie{} its derivatives \labelcref{eq:effective_action_dIII}.
	
	First, we consider the vertex function for two bosonic fields, which is the inverse of the full propagator.
	In terms of connected correlation functions, it reads
		\begin{align}
			\Gamma^{\varphi \varphi} = \, & \langle \phi^2 \rangle_c^{- 1} \, .	\label{eq:vertex_phi_phi}
		\end{align}
	Analogously, we obtain the fermionic two-point vertex function in terms of connected correlation functions,
		\begin{align}
			\Gamma^{\vartheta \tilde{\vartheta}} = - \Gamma^{\tilde{\vartheta} \vartheta} = \, & - \langle \tilde{\theta} \, \theta \rangle_c^{- 1} \, .	\label{eq:vertex_theta_theta}
		\end{align}
	The other components of the full two-point function vanish.
	For the nonvanishing components of the four-point vertex functions (up to permutations of the fields), we find
		\begin{align}
			\Gamma^{\varphi \varphi \varphi \varphi} = \, & - \langle \phi^4 \rangle_c \, \langle \phi^2 \rangle_c^{- 4} \, ,	\vdistance	\label{eq:vertex_phi_phi_phi_phi}
			\\
			\Gamma^{\varphi \varphi \vartheta \tilde{\vartheta}} = \, & \langle \phi^2 \, \tilde{\theta} \, \theta \rangle_c \, \langle \tilde{\theta} \, \theta \rangle_c^{- 2} \, \langle \phi^2 \rangle_c^{- 2} \, .	\vdistance	\label{eq:vertex_phi_phi_theta_theta}
		\end{align}
	Note that these formulae are valid for arbitrary choices of the functions $H ( \phi )$ and $U ( \phi )$ which obey the conditions mentioned above.
	
\section{Flow equations from exact inversion of the full field-dependent two-point function}%
\label{app:exact_inversion}

	This appendix is devoted to the derivation of the flow equation for the \gls{eaa} with full field dependence and an explicit calculation of the inverse of the full two-point function as well as the trace on the right hand side of the Wetterich equation.
	For the sake of clearness, we recapitulate the Wetterich equation \labelcref{eq:wetterich_equation},
		\begin{align}
			&	\partial_t \bar{\Gamma} ( t, \Phi ) =	\vdistance
			\\
			= \, & \STr \big[ \big( \tfrac{1}{2} \, \partial_t R ( t ) \big) \, \big( \bar{\Gamma}^{(2)} ( t, \Phi ) + R ( t ) \big)^{- 1} \big] \, .	\vdistance	\nonumber
		\end{align}
	Hence, as the first ingredient, we need the full two-point function $\Gamma^{(2)} ( t, \Phi ) = \bar{\Gamma}^{(2)} ( t, \Phi ) + R ( t )$, which has to be inverted.
	From our (exact) ansatz and the matrix representation of the regulator,
		\begin{align}
			R =
			\begin{pmatrix}
				R^{\varphi \varphi}	&	R^{\varphi \vartheta}	&	R^{\varphi \barvartheta}
				\\
				R^{\vartheta \varphi}	&	R^{\vartheta \vartheta}	&	R^{\vartheta \barvartheta}
				\\
				R^{\barvartheta \varphi}	&	R^{\barvartheta \vartheta}	&	R^{\barvartheta \barvartheta}
			\end{pmatrix}
			=
			\begin{pmatrix}
				r_\boson	&	0	&	0
				\\
				0	&	0	&	r_\fermion
				\\
				0	&	- r_\fermion	&	0
			\end{pmatrix} \, ,	\label{eq:regulator_matrix}
		\end{align}
	we find for the single components in field space,
		\begin{align}
			& \big( \bar{\Gamma}^{(2)} + R \big)^{\Phi \Phi} =	\Vdistance	\label{eq:full_2PT_function}
			\\
			= \, &
			\begin{pmatrix}
				\big( \bar{\Gamma}^{(2)} + R \big)^{\varphi \varphi}	&	\big( \bar{\Gamma}^{(2)} + R \big)^{\varphi \vartheta}	&	\big( \bar{\Gamma}^{(2)} + R \big)^{\varphi \barvartheta}
				\\
				\big( \bar{\Gamma}^{(2)} + R \big)^{\vartheta \varphi}	&	\big( \bar{\Gamma}^{(2)} + R \big)^{\vartheta \vartheta}	&	\big( \bar{\Gamma}^{(2)} + R \big)^{\vartheta \barvartheta}
				\\
				\big( \bar{\Gamma}^{(2)} + R \big)^{\barvartheta \varphi}	&	\big( \bar{\Gamma}^{(2)} + R \big)^{\barvartheta \vartheta}	&	\big( \bar{\Gamma}^{(2)} + R \big)^{\barvartheta \barvartheta}
			\end{pmatrix} =	\nonumber
			\\[.1em]
			= \, &
			\begin{pmatrix}
				r_\boson + U^{\prime \prime} + \barvartheta \, H^{\prime \prime} \, \vartheta	&	- \barvartheta \, H^\prime	&	H^\prime \, \vartheta
				\\
				- \barvartheta \, H^\prime	&	0	&	r_\fermion + H
				\\
				H^\prime \, \vartheta	&	- ( r_\fermion + H )	&	0
			\end{pmatrix} =	\nonumber
			\\[.1em]
			= \, &
			\begin{pmatrix}
				\GbInv{}	&	0	&	0
				\\
				0	&	0	&	\GfInv{}
				\\
				0	&	- \GfInv{}	&	0
			\end{pmatrix}
			+	\nonumber
			\\[.1em]
			& + 
			\begin{pmatrix}
				0	&	- H^\prime	&	0
				\\
				- H^\prime	&	0	&	0
				\\
				0	&	0	&	0
			\end{pmatrix} \barvartheta
			+
			\begin{pmatrix}
				0	&	0	&	H^\prime
				\\
				0	&	0	&	0
				\\
				H^\prime	&	0	&	0
			\end{pmatrix} \vartheta
			+	\nonumber
			\\[.1em]
			& + 
			\begin{pmatrix}
				H^{\prime \prime}	&	0	&	0
				\\
				0	&	0	&	0
				\\
				0	&	0	&	0
			\end{pmatrix} \barvartheta \vartheta \, ,	\nonumber
		\end{align}
	where we dropped the $\varphi$ and $t$ arguments of all functions for brevity and introduced
		\begin{align}
			& \GbInv{} \equiv r_\boson + U^{\prime \prime} \, ,	&&	\GfInv{} \equiv r_\fermion + H \, .
		\end{align}
	We note that the matrix is antisymmetric in the purely fermionic sector.
	Furthermore, in the last step, we sorted the expression by powers of Grassmann numbers.
	This structure is crucial for the inversion of the full two-point function, because it allows us to use the most general ansatz
		\begin{align}
			\big( \bar{\Gamma}^{(2)} + R \big)^{- 1} = \, & A + B \, \barvartheta + C \, \vartheta + D \, \barvartheta \, \vartheta \, .	\label{eq:ansatz_inverse_2PT_function}
		\end{align}
	Here, $A$, $B$, $C$, and $D$ are matrices in field space that depend on $t$ and $\varphi$.
	The simple reason, why this ansatz is the most general one, is that a series in Grassmann numbers terminates at the term, which is proportional to the product of all involved Grassmann numbers.
	Now, in order to determine $A$, $B$, $C$, and $D$, we use that
		\begin{align}
			&	\openone \overset{!}{=} \big( \bar{\Gamma}^{(2)} + R \big) \, \big( \bar{\Gamma}^{(2)} + R \big)^{- 1}	\, .
		\end{align}
	We simply expand the right hand side of this equation in powers of Grassmann numbers and compare the coefficients of the resulting series with the left hand side.
	Writing out the product of the matrices, we explicitly find
\begin{widetext}
		\begin{align}
			\begin{pmatrix}
				1	&	0	&	0
				\\
				0	&	1	&	0
				\\
				0	&	0	&	1
			\end{pmatrix}
			\overset{!}{=} \, &
			\begin{pmatrix}
				\GbInv{}	&	0	&	0
				\\
				0	&	0	&	\GfInv{}
				\\
				0	&	- \GfInv{}	&	0
			\end{pmatrix} A +	\label{eq:inversion_2PT_function_line_A}
			\\[.1em]
			& + \left[
				\begin{pmatrix}
					\GbInv{}	&	0	&	0
					\\
					0	&	0	&	\GfInv{}
					\\
					0	&	- \GfInv{}	&	0
				\end{pmatrix} B
				+
				\begin{pmatrix}
					0	&	- H^\prime	&	0
					\\
					- H^\prime	&	0	&	0
					\\
					0	&	0	&	0
				\end{pmatrix} A
			\right] \barvartheta +	\label{eq:inversion_2PT_function_line_B}
			\\[.1em]
			& + \left[
				\begin{pmatrix}
					\GbInv{}	&	0	&	0
					\\
					0	&	0	&	\GfInv{}
					\\
					0	&	- \GfInv{}	&	0
				\end{pmatrix} C
				+
				\begin{pmatrix}
					0	&	0	&	H^\prime
					\\
					0	&	0	&	0
					\\
					H^\prime	&	0	&	0
				\end{pmatrix} A
			\right] \vartheta +	\label{eq:inversion_2PT_function_line_C}
			\\[.1em]
			& + \left[
				\begin{pmatrix}
					\GbInv{}	&	0	&	0
					\\
					0	&	0	&	\GfInv{}
					\\
					0	&	- \GfInv{}	&	0
				\end{pmatrix} D
				+
				\begin{pmatrix}
					0	&	- H^\prime	&	0
					\\
					- H^\prime	&	0	&	0
					\\
					0	&	0	&	0
				\end{pmatrix} C
				-
				\begin{pmatrix}
					0	&	0	&	H^\prime
					\\
					0	&	0	&	0
					\\
					H^\prime	&	0	&	0
				\end{pmatrix} B
				+
				\begin{pmatrix}
					H^{\prime \prime}	&	0	&	0
					\\
					0	&	0	&	0
					\\
					0	&	0	&	0
				\end{pmatrix} A
			\right] \barvartheta \, \vartheta	\label{eq:inversion_2PT_function_line_D}
		\end{align}
	Now, it is straight forward to solve this system by comparing the coefficients.
	From the left hand side and line \labelcref{eq:inversion_2PT_function_line_A}, we find
		\begin{align}
			A = \begin{pmatrix}
				\Gb{}	&	0	&	0
				\\
				0	&	0	&	- \Gf{}
				\\
				0	&	\Gf{}	&	0
			\end{pmatrix} \, .
		\end{align}
	Inserting this in lines \labelcref{eq:inversion_2PT_function_line_B,eq:inversion_2PT_function_line_C} one finds
		\begin{align}
			B = \, & -
			\begin{pmatrix}
				\Gb{}	&	0	&	0
				\\
				0	&	0	&	- \Gf{}
				\\
				0	&	\Gf{}	&	0
			\end{pmatrix}
			\begin{pmatrix}
				0	&	- H^\prime	&	0
				\\
				- H^\prime	&	0	&	0
				\\
				0	&	0	&	0
			\end{pmatrix}
			\begin{pmatrix}
				\Gb{}	&	0	&	0
				\\
				0	&	0	&	- \Gf{}
				\\
				0	&	\Gf{}	&	0
			\end{pmatrix}
			\\[.1em]
			= \, &
			\begin{pmatrix}
				0	&	\Gb{} \, H^\prime	&	0
				\\
				0	&	0	&	0
				\\
				\Gf{} \, H^\prime	&	0	&	0
			\end{pmatrix}
			\begin{pmatrix}
				\Gb{}	&	0	&	0
				\\
				0	&	0	&	- \Gf{}
				\\
				0	&	\Gf{}	&	0
			\end{pmatrix}
			=	\nonumber
			\\[.1em]
			= \, &
			\begin{pmatrix}
				0	&	0	&	- \Gb{} \, H^\prime \, \Gf{}
				\\
				0	&	0	&	0
				\\
				\Gf{} \, H^\prime \, \Gb{}	&	0	&	0
			\end{pmatrix} \, .	\nonumber
		\end{align}
	and
	\begin{align}
		C = \, & -
		\begin{pmatrix}
			\Gb{}	&	0	&	0
			\\
			0	&	0	&	- \Gf{}
			\\
			0	&	\Gf{}	&	0
		\end{pmatrix}
		\begin{pmatrix}
			0	&	0	&	H^\prime
			\\
			0	&	0	&	0
			\\
			H^\prime	&	0	&	0
		\end{pmatrix}
		\begin{pmatrix}
			\Gb{}	&	0	&	0
			\\
			0	&	0	&	- \Gf{}
			\\
			0	&	\Gf{}	&	0
		\end{pmatrix}
		\\[.1em]
		= \, &
		\begin{pmatrix}
			0	&	0	&	- \Gb{} \, H^\prime
			\\
			\Gf{} \, H^\prime	&	0	&	0
			\\
			0	&	0	&	0
		\end{pmatrix}
		\begin{pmatrix}
			\Gb{}	&	0	&	0
			\\
			0	&	0	&	- \Gf{}
			\\
			0	&	\Gf{}	&	0
		\end{pmatrix}
		=	\nonumber
		\\[.1em]
		= \, &
		\begin{pmatrix}
			0	&	- \Gb{} \, H^\prime \, \Gf{}	&	0
			\\
			\Gf{} \, H^\prime \, \Gb{}	&	0	&	0
			\\
			0	&	0	&	0
		\end{pmatrix} \, .	\nonumber
	\end{align}
	Next, we study the coefficient of $\barvartheta \, \vartheta$, thus line \labelcref{eq:inversion_2PT_function_line_D} and insert the previous results.
	Solving for $D$ results in
		\begin{align}
			D = \, &
			\begin{pmatrix}
				\Gb{}	&	0	&	0
				\\
				0	&	0	&	- \Gf{}
				\\
				0	&	\Gf{}	&	0
			\end{pmatrix}
			\left[
			-
			\begin{pmatrix}
				0	&	- H^\prime	&	0
				\\
				- H^\prime	&	0	&	0
				\\
				0	&	0	&	0
			\end{pmatrix} C
			+
			\begin{pmatrix}
				0	&	0	&	H^\prime
				\\
				0	&	0	&	0
				\\
				H^\prime	&	0	&	0
			\end{pmatrix} B
			-
			\begin{pmatrix}
				H^{\prime \prime}	&	0	&	0
				\\
				0	&	0	&	0
				\\
				0	&	0	&	0
			\end{pmatrix} A
			\right] =
			\\[.1em]
			= \, &
			\left[
			\begin{pmatrix}
				0	&	\Gb{} \, H^\prime	&	0
				\\
				0	&	0	&	0
				\\
				\Gf{} \, H^\prime	&	0	&	0
			\end{pmatrix} C
			+
			\begin{pmatrix}
				0	&	0	&	\Gb{} \, H^\prime
				\\
				- \Gf{} \, H^\prime	&	0	&	0
				\\
				0	&	0	&	0
			\end{pmatrix} B
			+
			\begin{pmatrix}
				- \Gb{} \, H^{\prime \prime}	&	0	&	0
				\\
				0	&	0	&	0
				\\
				0	&	0	&	0
			\end{pmatrix} A
			\right] =	\nonumber
			\\[.1em]
			= \, &
			\left[
			\begin{pmatrix}
				0	&	\Gb{} \, H^\prime	&	0
				\\
				0	&	0	&	0
				\\
				\Gf{} \, H^\prime	&	0	&	0
			\end{pmatrix}
			\begin{pmatrix}
				0	&	- \Gb{} \, H^\prime \, \Gf{}	&	0
				\\
				\Gf{} \, H^\prime \, \Gb{}	&	0	&	0
				\\
				0	&	0	&	0
			\end{pmatrix}
			+ \right.	\nonumber
			\\[.1em]
			& +
			\begin{pmatrix}
				0	&	0	&	\Gb{} \, H^\prime
				\\
				- \Gf{} \, H^\prime	&	0	&	0
				\\
				0	&	0	&	0
			\end{pmatrix}
			\begin{pmatrix}
				0	&	0	&	- \Gb{} \, H^\prime \, \Gf{}
				\\
				0	&	0	&	0
				\\
				\Gf{} \, H^\prime \, \Gb{}	&	0	&	0
			\end{pmatrix}	\nonumber
			+
			\\[.1em]
			& \left. +
			\begin{pmatrix}
				- \Gb{} \, H^{\prime \prime}	&	0	&	0
				\\
				0	&	0	&	0
				\\
				0	&	0	&	0
			\end{pmatrix}
			\begin{pmatrix}
				\Gb{}	&	0	&	0
				\\
				0	&	0	&	- \Gf{}
				\\
				0	&	\Gf{}	&	0
			\end{pmatrix}
			\right] =	\nonumber
			\\[.1em]
			= \, &
			\begin{pmatrix}
				2 \, \Gb{} \, H^\prime \, \Gf{} \, H^\prime \, \Gb{} - \Gb{} \, H^{\prime \prime} \, \Gb{}	&	0	&	0
				\\
				0	&	0	&	\Gf{} \, H^\prime \, \Gb{} \, H^\prime \, \Gf{}
				\\
				0	&	- \Gf{} \, H^\prime \, \Gb{} \, H^\prime \, \Gf{}	&	0
			\end{pmatrix}	\nonumber
		\end{align}
	Combing the results in our ansatz \labelcref{eq:ansatz_inverse_2PT_function}, we obtain the full inverse of the two-point function,
		\begin{align}
			\big( \bar{\Gamma}^{(2)} + R \big)^{- 1} = \, &	\label{eq:inverse_2PT_function}
			\begin{pmatrix}
				\Gb{}	&	0	&	0
				\\
				0	&	0	&	- \Gf{}
				\\
				0	&	\Gf{}	&	0
			\end{pmatrix}
			+
			\begin{pmatrix}
				0	&	0	&	- \Gb{} \, H^\prime \, \Gf{}
				\\
				0	&	0	&	0
				\\
				\Gf{} \, H^\prime \, \Gb{}	&	0	&	0
			\end{pmatrix}
			\barvartheta +
			\begin{pmatrix}
				0	&	- \Gb{} \, H^\prime \, \Gf{}	&	0
				\\
				\Gf{} \, H^\prime \, \Gb{}	&	0	&	0
				\\
				0	&	0	&	0
			\end{pmatrix}
			\vartheta +	\nonumber
			\\[.1em]
			&\qquad +
			\begin{pmatrix}
				2 \, \Gb{} \, H^\prime \, \Gf{} \, H^\prime \, \Gb{} - \Gb{} \, H^{\prime \prime} \, \Gb{}	&	0	&	0
				\\
				0	&	0	&	\Gf{} \, H^\prime \, \Gb{} \, H^\prime \, \Gf{}
				\\
				0	&	- \Gf{} \, H^\prime \, \Gb{} \, H^\prime \, \Gf{}	&	0
			\end{pmatrix}
			\barvartheta \, \vartheta \, .	\nonumber
		\end{align}
	Note, that this inversion is exact and does not rely on any approximation!
	However, usually this exact inversion is not possible.
	Next, we we can insert this result together with the derivative of the regulator into the Wetterich equation
		\begin{align}
			\partial_t \bar{\Gamma} ( t, \Phi ) = \STr \big[ \big( \tfrac{1}{2} \, \partial_t R ( t ) \big) \big( \bar{\Gamma}^{(2)} ( t, \Phi ) + R ( t ) \big)^{- 1} \big] \, .
		\end{align}
	Remember, that we also have to replace the left hand side with the derivative of the (exact) ansatz,
		\begin{align}
			\partial_t \bar{\Gamma} ( t, \Phi ) = \barvartheta \, \partial_t H ( t, \varphi ) \, \vartheta + \partial_t U ( t, \varphi ) \, .
		\end{align}
	Again, we compare coefficients of Grassmann numbers on both sides of the equation.
	For the term without Grassmann numbers we find
		\begin{align}
			\partial_t U = \, & \STr
			\left(
			\begin{pmatrix}
				\tfrac{1}{2} \, \partial_t r_\boson	&	0	&	0
				\\
				0	&	0	&	\tfrac{1}{2} \, \partial_t r_\fermion
				\\
				0	&	- \tfrac{1}{2} \, \partial_t r_\fermion	&	0
			\end{pmatrix}
			\begin{pmatrix}
				\Gb{}	&	0	&	0
				\\
				0	&	0	&	- \Gf{}
				\\
				0	&	\Gf{}	&	0
			\end{pmatrix}
			\right)
			= \big( \tfrac{1}{2} \, \partial_t r_\boson \big) \, \Gb{} - ( \partial_t r_\fermion ) \, \Gf{} \, ,	\label{eq:flow_equation_u}
		\end{align}
	which has the well-known structure of a flow equation of the effective potential of a system with Yukawa-type interactions between bosons and fermions.
	For the terms linear in $\barvartheta$ or $\vartheta$ we find that there is no contribution on the left hand side and the supertrace on the right hand side vanishes exactly.
	On the other hand, the coefficients of the $\barvartheta \, \vartheta$-term are nontrivial on both sides.
	Evaluating the supertrace, we find
		\begin{align}
			\partial_t H = \, & \STr
			\left(
			\begin{pmatrix}
				\tfrac{1}{2} \, \partial_t r_\boson	&	0	&	0
				\\
				0	&	0	&	\tfrac{1}{2} \, \partial_t r_\fermion
				\\
				0	&	- \tfrac{1}{2} \, \partial_t r_\fermion	&	0
			\end{pmatrix} \right.
			\cdot	\label{eq:flow_equation_H}
			\\[.1em]
			& \left. \cdot
			\begin{pmatrix}
				2 \, \Gb{} \, H^\prime \, \Gf{} \, H^\prime \, \Gb{} - \Gb{} \, H^{\prime \prime} \, \Gb{}	&	0	&	0
				\\
				0	&	0	&	\Gf{} \, H^\prime \, \Gb{} \, H^\prime \, \Gf{}
				\\
				0	&	- \Gf{} \, H^\prime \, \Gb{} \, H^\prime \, \Gf{}	&	0
			\end{pmatrix}
			\right) =	\nonumber
			\\
			= \, & 2 \, \big( \tfrac{1}{2} \, \partial_t r_\boson \big) \, \Gb{}^2 \, \Gf{} \, ( H^\prime )^2 - \big( \tfrac{1}{2} \, \partial_t r_\boson \big) \, \Gb{}^2 \, H^{\prime \prime} + 2 \, \big( \tfrac{1}{2} \, \partial_t r_\fermion \big) \, \Gb{} \, \Gf{}^2 \, ( H^\prime )^2 \, .	\vdistance	\nonumber
		\end{align}
	Also this equation has a well-known structure: It is the flow equation of a Yukawa-type interaction or the fermionic mass term, respectively.
\end{widetext}

\section{Flow equations via projections of the Wetterich equation}%
\label{app:projection}

	In this appendix, we derive the flow equations for the effective potential and the Yukawa coupling from the Wetterich equation.
	Here, we resort to the standard procedure of projecting the Wetterich equation onto vertices.
	In our case, we have to project the Wetterich equation onto the effective potential and the Yukawa coupling.
	Recapitulating that (exact) ansatz for the \gls{eaa} was given by
		\begin{align}
			\bar{\Gamma} ( t, \Phi ) = \tilde{\vartheta} \, H ( t, \varphi ) \, \vartheta + U ( t, \varphi ) \, ,
		\end{align}
	we can directly project onto the scale-dependent effective potential $U ( t, \varphi )$ by evaluating the Wetterich equation for $\tilde{\vartheta} = \vartheta = 0$.
	Thus, we need to evaluate all objects inside the supertrace on the \gls{rhs} for $\tilde{\vartheta} = \vartheta = 0$.
	Here, it actually does not matter, if one first evaluates the two-point function for vanishing fermion fields and afterwards inverts the matrix or if one proceeds the other way around.\footnote{In our zero-dimensional setup this is explicitly visible from \cref{eq:full_2PT_function,eq:inverse_2PT_function} where in both cases only the $\tilde{\vartheta}$- and $\vartheta$-independent term survives.}
	In any case, one finds exactly \cref{eq:flow_u}.
	However, for the projection onto the Yukawa coupling $H ( t, \varphi )$, we first have to take derivatives with respect to $\tilde{\vartheta}$ and $\vartheta$ first and only afterwards set them to zero.
\begin{widetext}
	One obtains,
		\begin{align}
			\partial_t H = \, & \Big[ \tfrac{\partial^2}{\partial \vartheta \, \partial \tilde{\vartheta}} \STr \big[ \big( \tfrac{1}{2} \, \partial_t R \big) \, \big( \bar{\Gamma}^{(2)} + R \big)^{- 1} \big] \Big]_{\tilde{\vartheta} = \vartheta = 0} =	\Vdistance	\label{eq:flow_H_abstract}
			\\
			= \, & \Big[ \STr \big[ \big( \tfrac{1}{2} \, \partial_t R \big) \, \big( \bar{\Gamma}^{(2)} + R \big)^{- 1} \, \Gamma^{\tilde{\vartheta} \Phi \Phi} \, \big( \bar{\Gamma}^{(2)} + R \big)^{- 1} \, \Gamma^{\vartheta \Phi \Phi} \, \big( \bar{\Gamma}^{(2)} + R \big)^{- 1}\big] -	\Vdistance	\nonumber
			\\
			& - \STr \big[ \big( \tfrac{1}{2} \, \partial_t R \big) \, \big( \bar{\Gamma}^{(2)} + R \big)^{- 1} \, \Gamma^{\vartheta \tilde{\vartheta} \Phi \Phi} \, \big( \bar{\Gamma}^{(2)} + R \big)^{- 1}\big] +	\Vdistance	\nonumber
			\\
			& + \STr \big[ \big( \tfrac{1}{2} \, \partial_t R \big) \, \big( \bar{\Gamma}^{(2)} + R \big)^{- 1} \, \Gamma^{\tilde{\vartheta} \Phi \Phi} \, \big( \bar{\Gamma}^{(2)} + R \big)^{- 1} \, \Gamma^{\vartheta \Phi \Phi} \, \big( \bar{\Gamma}^{(2)} + R \big)^{- 1}\big] \Big]_{\tilde{\vartheta} = \vartheta = 0} \, .	\Vdistance	\nonumber
		\end{align}
\end{widetext}
	For the full propagators (evaluated for $\tilde{\vartheta} = \vartheta = 0$), we can again directly invert the $\tilde{\vartheta}$- and $\vartheta$-independent part of the full two-point function \labelcref{eq:full_2PT_function} and for the derivative of the regulator, we make use of the matrix representation \labelcref{eq:regulator_matrix}.
	It remains to determine the three- and four-point vertices in matrix representation and evaluated for vanishing fermion fields.
	In fact, these are
		\begin{align}
			\Gamma^{\vartheta \Phi \Phi} =
			\begin{pmatrix}
				0	&	- H^\prime	&	0
				\\
				- H^\prime	&	0	&	0
				\\
				0	&	0	&	0
			\end{pmatrix} \, ,
			\\[.1em]
			\Gamma^{\vartheta \Phi \Phi} =
			\begin{pmatrix}
				0	&	0	&	H^\prime
				\\
				0	&	0	&	0
				\\
				H^\prime	&	0	&	0
			\end{pmatrix} \, ,
			\\[.1em]
			\Gamma^{\vartheta \tilde{\vartheta} \Phi \Phi} =
			\begin{pmatrix}
				H^{\prime \prime}	&	0	&	0
				\\
				0	&	0	&	0
				\\
				0	&	0	&	0
			\end{pmatrix} \, .
		\end{align}
	Inserting these matrices into the supertraces of \cref{eq:flow_H_abstract} and evaluating the traces, we find exactly \cref{eq:flow_H_1}.

\section{Flow equations for the \texorpdfstring{$\mathbb{Z}_2$}{Z2}-symmetric model}%
\label{app:flow_equations_z2_three_dimensions}

	In this appendix we include additional expressions related to \cref{sec:field-dependent-wave-function-renormalization} including the explicit expressions for the \gls{frg} flow equations for the $\mathbb{Z}_2$-symmetric model of the single scalar in three Euclidean dimensions.

	The full propagator within our truncation -- evaluated for $\sigma$ -- from \cref{eq:effective_average_action_z2_three_dim} is given by
		\begin{align}
			G_t ( p_2, p_1 ) \big|_{\varphi ( x ) = \sigma} = \, & ( 2 \uppi )^d \, \delta^{(d)} ( p_2 + p_1 ) \, \mathcal{G} ( t, \sigma, p_1^2 ) \, ,	\Vdistance
			\\
			\mathcal{G} ( t, \sigma, p^2 ) = \, & \frac{1}{U^{\prime \prime} ( t, \sigma ) + [ Z ( t, \sigma ) + r_t ( p^2 ) ] \, p^2} \, .	\Vdistance
		\end{align}
	In addition, the three- and four-point functions derived from \cref{eq:effective_average_action_z2_three_dim} read
		\begin{align}
			& \bar{\Gamma}_{t}^{(3)} ( p_3, p_2, p_1 ) \big|_{\varphi ( x ) = \sigma} =	\Vdistance
			\\
			= \, & ( 2 \uppi)^d \, \delta^{(d)} ( p_3 + p_2 + p_1 ) \times	\Vdistance	\nonumber
			\\
			& \times \big[ U^{\prime \prime \prime} ( t, \sigma ) - Z^\prime ( t, \sigma ) \, ( p_3 \cdot p_2 +  p_2 \cdot p_1 + p_1 \cdot p_3 \big) \big] \, ,	\Vdistance	\nonumber
		\end{align}
	as well as
		\begin{align}
			& \bar{\Gamma}_{t}^{(4)} ( p_4, p_3, p_2, p_1 ) \big|_{\varphi ( x ) = \sigma} =	\Vdistance
			\\
			= \, & ( 2 \uppi)^d \, \delta^{(d)} ( p_4 + p_3 + p_2 + p_1 ) \, \big[ U^{\prime \prime \prime \prime} ( t, \sigma ) - Z^{\prime \prime} ( t, \sigma ) \times	\Vdistance	\nonumber
			\\
			& \quad \times [ p_4 \cdot ( p_3 + p_2 + p_1 ) + p_3 \cdot ( p_2 + p_1 ) + p_2 \cdot p_1 ] \big] \, .	\Vdistance	\nonumber
		\end{align}
	Inserting these expressions into \cref{eq:flow_2point_z2_three_dim}, we can evaluate the integrals and find
	\begin{widetext}
		\begin{align}
			\partial_t \bar{\Gamma}^{(2)} ( t, q ) = \, & \int_p \big( \tfrac{1}{2} \, \partial_t R ( t, p^2 ) \big) \, \Big[ - \mathcal{G}^2 ( t, \sigma, p^2 ) \, \big[ U^{\prime \prime \prime \prime} ( t, \sigma ) + Z^{\prime \prime} ( t, \sigma ) \, ( q^2 + p^2 ) \big] +	\Vdistance	\label{eq:flow_2point_z2_three_dim_result}
			\\
			& + 2 \int_p \big( \tfrac{1}{2} \, \partial_t R ( t, p^2 ) \big) \, \mathcal{G}^2 ( t, \sigma, p^2 ) \, \mathcal{G} ( t, \sigma, ( p + q )^2 ) \, \big[ U^{\prime \prime \prime} ( t, \sigma ) + Z^{\prime} ( t, \sigma ) \, [ p^2 + ( p + q ) \cdot q ] \big]^2 \big] \, .	\Vdistance	\nonumber
		\end{align}
	\end{widetext}
	By evaluating this at $q = 0$, we arrive at the flow equation for the boson mass function $M ( t, \sigma )$
		\begin{align}
			\partial_t M = \, & \int_p \big( \tfrac{1}{2} \, \partial_t R \big) \, \Big[ - \mathcal{G}^2 \, \big[ M^{\prime \prime} + Z^{\prime \prime} \, p^2 \big] +	\Vdistance	\label{eq:flow_M_z2_three_dim}
			\\
			& \quad + 2 \int_p \big( \tfrac{1}{2} \, \partial_t R \big) \, \mathcal{G}^3 \, \big[ M^\prime + Z^{\prime} \, p^2 \big]^2 \big] \, .	\Vdistance	\nonumber
		\end{align}
	Here, we stopped indicating the dependences on $t$, $\sigma$, and $p^2$ for the sake of the readability.
	
	For the wave-function renormalization, however, one needs to expand the second contribution in \cref{eq:flow_2point_z2_three_dim_result} up to order $q^2$ and then take the derivative with respect to $q^2$.
	Lastly, one again evaluates the result at $q = 0$ and arrives at the flow equation for the wave-function renormalization $Z ( t, \sigma )$
		\begin{align}
			& \partial_t Z =	\Vdistance	\label{eq:flow_Z_z2_three_dim}
			\\
			= \, & \int_p \big( \tfrac{1}{2} \, \partial_t R \big) \, \Big[ - \mathcal{G}^2 \, Z^{\prime \prime} +	\Vdistance	\nonumber
			\\
			& \quad + 2 \, \mathcal{G}^2 \, \Big( 2 \, \mathcal{G} \, Z^\prime \, \big[ ( M^{\prime \prime} + Z^\prime \, p^2 ) + \tfrac{1}{2d} \, Z^\prime \, p^2 \big] + 	\Vdistance	\nonumber
			\\
			& \qquad + \dot{\mathcal{G}} \, ( M^{\prime \prime} + Z^\prime \, p^2 ) \, \big[ ( M^{\prime \prime} + Z^\prime \, p^2 ) + \tfrac{4}{d} \, Z^\prime \, p^2 \big] +	\Vdistance	\nonumber
			\\
			& \qquad + \ddot{\mathcal{G}} \, ( M^{\prime \prime} + Z^\prime \, p^2 )^2 \, \tfrac{2}{d} \, p^2 \Big) \Big] \, ,	\Vdistance	\nonumber
		\end{align}
	where some propagators $\mathcal{G}$ are retained in the expression for which the $p^2$-derivative is not taken yet, indicated by the dot(s).
	Before we can evaluate these derivatives and the remaining integrals, we need to specify the regulator shape function.
	Explicit expressions for \cref{eq:flow_M_z2_three_dim,eq:flow_Z_z2_three_dim} with Callan-Symanzik regulator and Litim regulator are discussed in the following Appendices~\ref{app:callan_symanzik_regulator} and~\ref{app:litim_regulator} respectively.

\subsection{Callan-Symanzik regulator in three dimensions}%
\label{app:callan_symanzik_regulator}

	Starting with the Callan-Symanzik regulator \labelcref{eq:callan_symanzik_regulator}, we find that the flow equation for $M$, \cref{eq:flow_M_z2_three_dim}, or respectively for the potential $U$ itself,
		\begin{align}
			\partial_t U = \, & \int_p \big( \tfrac{1}{2} \, \partial_t R ) \, \mathcal{G} \, ,	\label{eq:flow_U_z2_three_dim}
		\end{align}
	still contains an \gls{uv} divergence in three Euclidean dimensions, since $\dd^3 p \propto \dd p \, p^2$ and $\mathcal{G} \propto 1/p^2$ for large momenta, while the Callan-Symanzik regulator does not contain any momentum dependence and therefore does not regulate the \gls{uv}.
	For the flow equation of the wave-function renormalization $Z$, \cref{eq:flow_Z_z2_three_dim}, this problem does not occur, because of the higher powers of the propagators -- the integrand always falls off at least like $1/p^2$.

	In any case, we used dimensional regularization with
		\begin{align}
			\int_p \frac{1}{( p^2 + \Delta )^n} = \frac{\Gamma ( n - \frac{d}{2} )}{( 4 \uppi )^{\frac{d}{2}} \, \Gamma ( n )} \, \Delta^{\frac{d}{2} - n} \, ,	\Vdistance
		\end{align}
	for \cref{eq:flow_U_z2_three_dim} and take two derivatives with respect to $\sigma$ to obtain the \gls{uv} renormalized flow equation for $M$.
	
	In total, we find the following flow equations for the effective potential and the wave-function renormalization,
	\begin{align}
		& \partial_t M =	\Vdistance	\label{eq:flow_M_callan_symanzik_conservative}
		\\
		= \, & \frac{\partial}{\partial \sigma} \bigg[ \frac{k^2}{8 \uppi} \, \bigg( \frac{1}{\sqrt{k^2 + M}} \, \frac{M^{\prime}}{Z^{\frac{3}{2}}} - 3 \, \sqrt{k^2 + M} \, \frac{Z^{ \prime}}{Z^{\frac{5}{2}}} \bigg) \bigg]	=	\Vdistance	\nonumber
		\\
		= \, & \frac{k^2}{8 \uppi} \, \bigg[ \frac{1}{\sqrt{k^2 + M}} \, \frac{M^{\prime \prime}}{Z^{\frac{3}{2}}} - \frac{1}{2} \, \frac{1}{\sqrt{k^2 + M}^3} \, \frac{( M^{\prime} )^2}{Z^{\frac{3}{2}}} +	\Vdistance	\nonumber
		\\
		& \quad - 3 \, \frac{1}{\sqrt{k^2 + M}} \, \frac{M^{\prime} Z^\prime}{Z^{\frac{5}{2}}} + \frac{15}{2} \, \sqrt{k^2 + M} \, \frac{( Z^\prime )^2}{Z^{\frac{7}{2}}} +	\Vdistance	\nonumber
		\\
		& \quad - 3 \, \sqrt{k^2 + M} \, \frac{Z^{\prime \prime}}{Z^{\frac{5}{2}}} \bigg] \, ,	\Vdistance	\nonumber
		\\
		& \partial_t Z =	\Vdistance
		\\
		= \, & \frac{k^2}{8 \uppi} \, \bigg[ \frac{1}{\sqrt{k^2 + M}} \, \frac{Z^{\prime \prime}}{Z^\frac{3}{2}} - \frac{49}{24} \, \frac{1}{\sqrt{k^2 + M}} \, \frac{( Z^\prime )^2}{Z^\frac{5}{2}} +	\Vdistance	\nonumber
		\\
		& \quad - \frac{7}{12} \, \frac{1}{\sqrt{k^2 + M}^{\, 3}} \, \frac{Z^\prime \, M^\prime}{Z^\frac{3}{2}} + \frac{1}{8} \, \frac{1}{\sqrt{k^2 + M}^{\, 5}} \, \frac{( M^\prime )^2}{Z^\frac{1}{2}} \bigg] \, .	\Vdistance	\nonumber
	\end{align}

\subsection{Litim regulator in three dimensions}%
\label{app:litim_regulator}
	For the Litim regulator \labelcref{eq:litim_regulator} the problem of \gls{uv} divergences is not present, since the regulator is local in momentum space and therefore cuts off the \gls{uv} divergences via the regulator insertion $\partial_t R$.
	However, there is another problem for the Litim regulator, which is that the regulator is not analytic in the momentum space.
	Hence, especially the terms involving $\dot{\mathcal{G}}$ and $\ddot{\mathcal{G}}$ have to be treated with care.
	In Ref.\ \cite[Appendix D]{Koenigstein:2023wso} one of the authors of this work provided several useful identities for derivatives of the Litim regulator and corresponding integrals, which we also use here.
	The final result for the flow equations of the effective potential and the wave-function renormalization is
\begin{widetext}
	\begin{align}
		& \partial_t M =	\Vdistance	\label{eq:flow_M_litim_conservative}
		\\
		= \, & \frac{\partial}{\partial \sigma} \bigg( \frac{k^3}{8 \uppi^2} \, \bigg[ 2 \, \frac{1}{Z - 1} \, \bigg( - \frac{1}{M + Z \, k^2} + \frac{1}{M + k^2} \, \sqrt{\frac{M + k^2}{( Z - 1) \, k^2}} \, \mathrm{arccot} \bigg( \sqrt{\frac{M + k^2}{( Z - 1) \, k^2}} \bigg) \bigg) \, M^{\prime} +	\Vdistance	\nonumber
		\\
		& \quad + 2 \, \frac{1}{( Z - 1 )^2} \, \bigg( 2 + \frac{M + k^2}{M + Z \, k^2} - 3 \, \sqrt{\frac{M + k^2}{( Z - 1) \, k^2}} \, \mathrm{arccot} \bigg( \sqrt{\frac{M + k^2}{( Z - 1) \, k^2}} \bigg) \bigg) \, Z^{\prime} \bigg] \bigg) =	\Vdistance	\nonumber
		\\
		= \, & \frac{k^3}{8 \uppi^2} \, \bigg[ 2 \, \frac{1}{Z - 1} \, \bigg( - \frac{1}{M + Z \, k^2} + \frac{1}{M + k^2} \, \sqrt{\frac{M + k^2}{( Z - 1) \, k^2}} \, \mathrm{arccot} \bigg( \sqrt{\frac{M + k^2}{( Z - 1) \, k^2}} \bigg) \bigg) \, M^{\prime \prime} +	\Vdistance	\nonumber
		\\
		& \quad + \frac{1}{Z - 1} \, \frac{1}{( M + k^2 )^2} \, \bigg( \frac{( M + k^2 ) \, [ ( M + k^2 ) - ( Z - 1 ) \, k^2 ]}{( M + Z \, k^2 )^2} - \sqrt{\frac{M + k^2}{( Z - 1) \, k^2}} \, \mathrm{arccot} \bigg( \sqrt{\frac{M + k^2}{( Z - 1) \, k^2}} \bigg) \bigg) \, ( M^\prime )^2 +	\Vdistance	\nonumber
		\\
		& \quad + 2 \, \frac{1}{( Z - 1 )^2} \, \bigg( \frac{3 \, ( M + k^2 ) + 5 \, ( Z - 1 ) \, k^2}{( M + Z \, k^2 )^2} - 3 \, \frac{1}{M + k^2} \sqrt{\frac{M + k^2}{( Z - 1) \, k^2}} \, \mathrm{arccot} \bigg( \sqrt{\frac{M + k^2}{( Z - 1) \, k^2}} \bigg) \bigg) \, M^\prime \, Z^\prime +	\Vdistance	\nonumber
		\\
		& \quad - \frac{1}{( Z - 1 )^3} \, \bigg( \frac{15 \, ( M + k^2 )^2 + 25 \, ( M + k^2 )\, ( Z - 1 ) \, k^2 + 8 \, [ ( Z - 1 ) \, k^2 ]^2 }{( M + Z \, k^2 )^2} +	\Vdistance	\nonumber
		\\
		& \qquad - 15 \, \sqrt{\frac{M + k^2}{( Z - 1) \, k^2}} \, \mathrm{arccot} \bigg( \sqrt{\frac{M + k^2}{( Z - 1) \, k^2}} \bigg) \bigg) \, ( Z^\prime )^2 +	\Vdistance	\nonumber
		\\
		& \quad + 2 \, \frac{1}{( Z - 1 )^2} \, \bigg( 2 + \frac{M + k^2}{M + Z \, k^2} - 3 \, \sqrt{\frac{M + k^2}{( Z - 1) \, k^2}} \, \mathrm{arccot} \bigg( \sqrt{\frac{M + k^2}{( Z - 1) \, k^2}} \bigg) \bigg) \, Z^{\prime \prime} \bigg] \, .	\Vdistance	\nonumber
	\end{align}

	\begin{align}
		& \partial_t Z =	\Vdistance
		\\
		= \, & \frac{k^2}{288 \uppi^2} \, \bigg[ - 72 \, \frac{1}{Z - 1} \, \bigg( \frac{k}{M + Z \, k^2} - \frac{1}{\sqrt{M + k^2} \, \sqrt{Z - 1}} \, \mathrm{arccot} \bigg( \sqrt{\frac{M + k^2}{( Z - 1) \, k^2}} \bigg) \bigg) \, Z^{\prime \prime} +	\Vdistance	\nonumber
		\\
		& \quad + \frac{1}{( Z - 1 )^2} \, \bigg( \frac{k}{( M + Z \, k^2 )^4} \, \big[ 48 \, [ ( Z - 1 ) \, k^2 ]^2 \, k^2 + 147 \, ( M + k^2 )^3 + 539 \, ( M + k^2 )^2 \, ( Z - 1 ) \, k^2 +	\Vdistance	\nonumber
		\\
		& \qquad + 581 \, ( M^2 + k^2 ) \, [ ( Z - 1 ) \, k^2 ]^2 + 237 \, [ ( Z - 1 ) \, k^2 ]^3 \big] +	\Vdistance	\nonumber
		\\
		& \qquad - 147 \, \frac{1}{\sqrt{M + k^2} \, \sqrt{Z - 1}} \, \mathrm{arccot} \bigg( \sqrt{\frac{M + k^2}{( Z - 1) \, k^2}} \bigg)  \bigg) \, ( Z^\prime )^2 +	\Vdistance	\nonumber
		\\
		& \quad + \frac{1}{Z - 1} \, \frac{1}{M + k^2} \, \bigg( \frac{k}{( M + Z \, k^2 )^4} \, \big[ 96 \, ( M + k^2 ) \, ( Z - 1 ) \, k^4 + 42 \, ( M + k^2 )^3 - 38 \, ( M + k^2 )^2 \, ( Z - 1 ) \, k^2 +	\Vdistance	\nonumber
		\\
		& \qquad - 26 \, ( M + k^2 ) \, [ ( Z - 1 ) \, k^2 ]^2 - 42 \, [ ( Z - 1 ) \, k^2 ]^3 \big] - 42 \, \frac{1}{\sqrt{M + k^2} \sqrt{Z - 1}} \, \mathrm{arccot} \bigg( \sqrt{\frac{M + k^2}{( Z - 1) \, k^2}} \bigg) \bigg) \, Z^\prime \, M^\prime	+	\Vdistance	\nonumber
		\\
		& \quad + 3 \, \frac{1}{( M + k^2 )^2} \, \bigg( \frac{k}{( M + Z \, k^2 )^4} \, \big[ 16 \, ( M + k^2 )^2 \, k^2 - 3 \, ( M + k^2 )^3 + 21 \, ( M + k^2 )^2 \, ( Z - 1 ) \, k^2 +	\Vdistance	\nonumber	
		\\
		& \qquad + 11 \, ( M + k^2 ) \, [ ( Z - 1 ) \, k^2 ]^2 + 3 \, [ ( Z - 1 ) \, k^2 ]^3 \big] + 3 \, \frac{1}{\sqrt{M + k^2} \, \sqrt{Z - 1}} \, \mathrm{arccot} \bigg( \sqrt{\frac{M + k^2}{( Z - 1) \, k^2}} \bigg) \bigg) \, ( M^\prime )^2 \bigg] \, .	\Vdistance	\nonumber
	\end{align}
\end{widetext}
	Here, however, we have to be careful, when the wave-function renormalization takes the value $Z = 1$, because the flow equations contain terms like $1/(Z - 1)$.
	Nevertheless, by carefully taking the limit $Z \to 1$, one actually finds that the flow equations are well-defined and do not have poles at $Z = 1$.
	In fact, they reduce to
		\begin{align}
			& \lim_{Z \to 1} \partial_t M =	\Vdistance
			\\
			= \, & \frac{k^5}{6 \uppi^2} \, \bigg[ \frac{1}{( M + k^2 )^2} \, M^{\prime \prime} - 2 \, \frac{1}{( M + k^2 )^3} \, ( M^\prime )^2 \bigg] \, ,	\Vdistance	\nonumber
		\end{align}
	and
		\begin{align}
			\lim_{Z \to 1} \partial_t Z = \, & \frac{k^5}{6 \uppi^2} \, \frac{1}{( M + k^2 )^4} \, ( M^\prime )^2 \, .
		\end{align}

\bibliography{main}
% \bibliography{bib/general,bib/gn,bib/inhomo,bib/instanton,bib/lattice,bib/math,bib/numerics,bib/qcd,bib/rg,bib/software,bib/steilPhDaux,bib/symmetries,bib/thermal_qft,bib/thies,bib/virasoro_algebra,bib/zero-dim-qft,bib/time_crystals} % Produces the bibliography via BibTeX -- separated bibfiles for working versions.

\end{document}